\documentclass[pre,amsmath,amssymb,groupedaddress,graphicx,showkeys,showpacs,reprint]{revtex4-1}

\usepackage[colorlinks=true,linkcolor=blue,citecolor=blue,bookmarks]{hyperref}
\usepackage{url}
\usepackage{breakurl}
\usepackage{units}
\usepackage{graphicx}
\usepackage{subfigure}
\usepackage[utf8]{inputenc}
\usepackage{ae}
\usepackage{natbib}
\usepackage{multirow}

\renewcommand{\vec}[1]{\boldsymbol{#1}}
\newcommand\eg{e.g.}
\newcommand\ie{i.e.}

\begin{document}

\title{Characterization of Maximally Random Jammed Sphere Packings:\\
  II. Correlation Functions and Density Fluctuations}

\author{Michael A. Klatt}
\affiliation{Karlsruhe Institute of Technology (KIT), Institute of Stochastics,\\ Englerstraße 2, 76131 Karlsruhe, Germany}
\author{Salvatore Torquato}
\email[Electronic mail: ]{torquato@electron.princeton.edu}
\affiliation{Department of Chemistry, Department of Physics, Princeton
  Institute for the Science and Technology of Materials, and Program
  in Applied and Computational Mathematics, Princeton University, Princeton, New Jersey 08544, USA}
\date{\today}

\begin{abstract}
  In the first paper of this series, we introduced Voronoi correlation functions to characterize the structure of maximally random jammed (MRJ) sphere packings across length scales.
  In the present paper, we determine a variety of different correlation functions that arise in rigorous expressions for the effective physical properties of MRJ sphere packings and compare them to the corresponding statistical descriptors for overlapping spheres and equilibrium hard-sphere systems.
  Such structural descriptors arise in rigorous bounds and formulas for effective transport properties, diffusion and reactions constants, elastic moduli, and electromagnetic characteristics.
  First, we calculate the two-point, surface-void, and surface-surface correlation functions, for which we derive explicit analytical formulas for finite hard-sphere packings.
  We show analytically how the contact Dirac delta function contribution to the pair correlation function $g_2(r)$ for MRJ packings translates into distinct functional behaviors of these two-point correlation functions that do not arise in the other two models examined here.
  Then we show how the spectral density distinguishes the MRJ packings from the other disordered systems in that the spectral density vanishes in the limit of infinite wavelengths; \ie,
  these packings are hyperuniform, which means that density fluctuations on large length scales are anomalously suppressed.
  Moreover, for all model systems, we study and compute exclusion probabilities and pore size distributions, as well as local density fluctuations.
  We conjecture that for general disordered hard-sphere packings, a central limit theorem holds for the number of points within an spherical observation window. 
  Our analysis links problems of interest in material science, chemistry, physics, and mathematics. 
  In the third paper of this series, we will evaluate bounds and estimates of a host of different physical properties of the MRJ sphere packings that are based on the structural characteristics analyzed in this paper.
\end{abstract}

\pacs{}{}
\maketitle

\section{Introduction}
\label{Introduction}

\begin{figure}[t]
  \centering
  \subfigure[][]{\includegraphics[width=0.65\linewidth,angle=90]{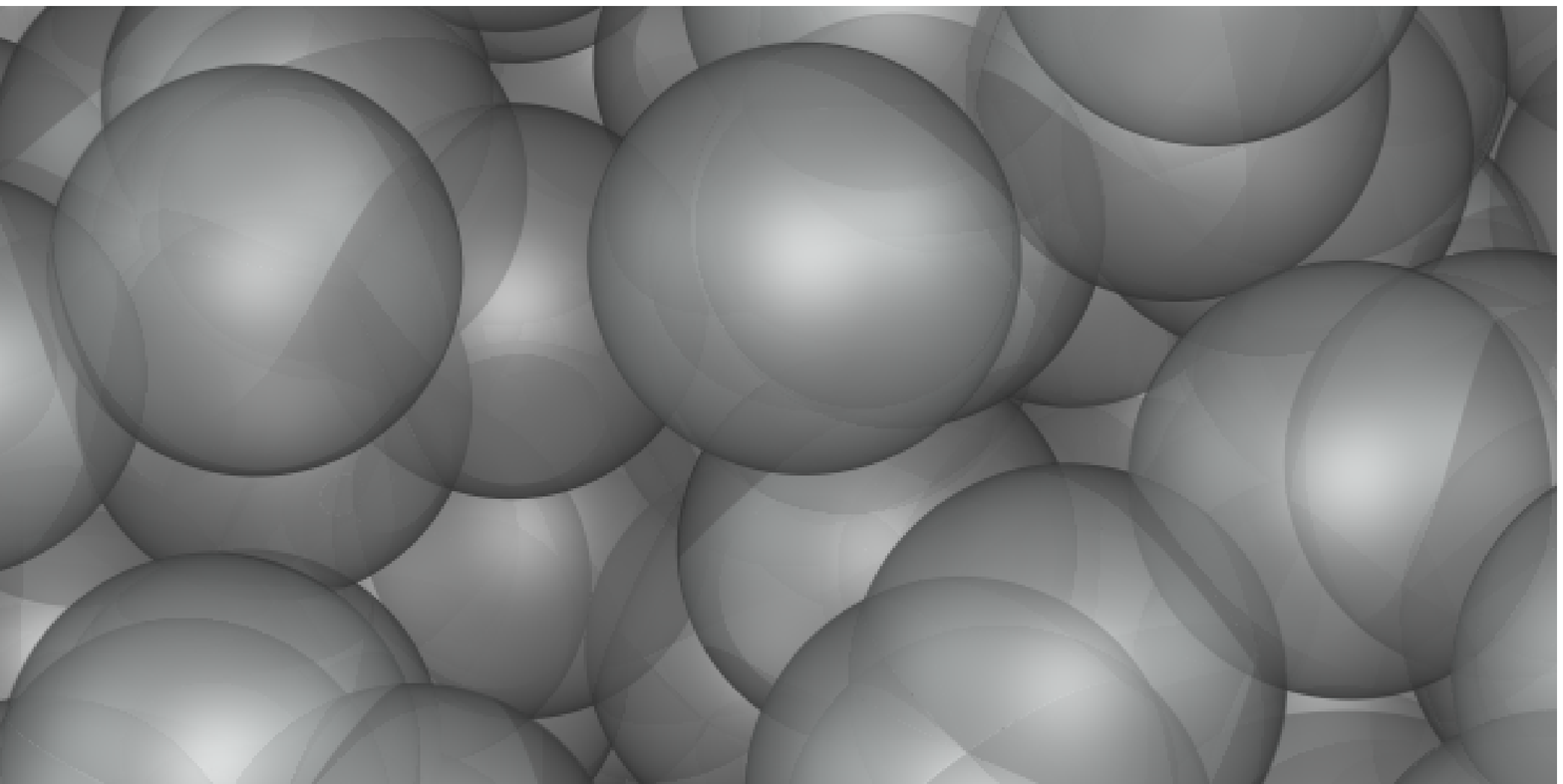}%
  \label{fig:vis-overlap}}
  \hfill
  \subfigure[][]{\includegraphics[width=0.65\linewidth,angle=90]{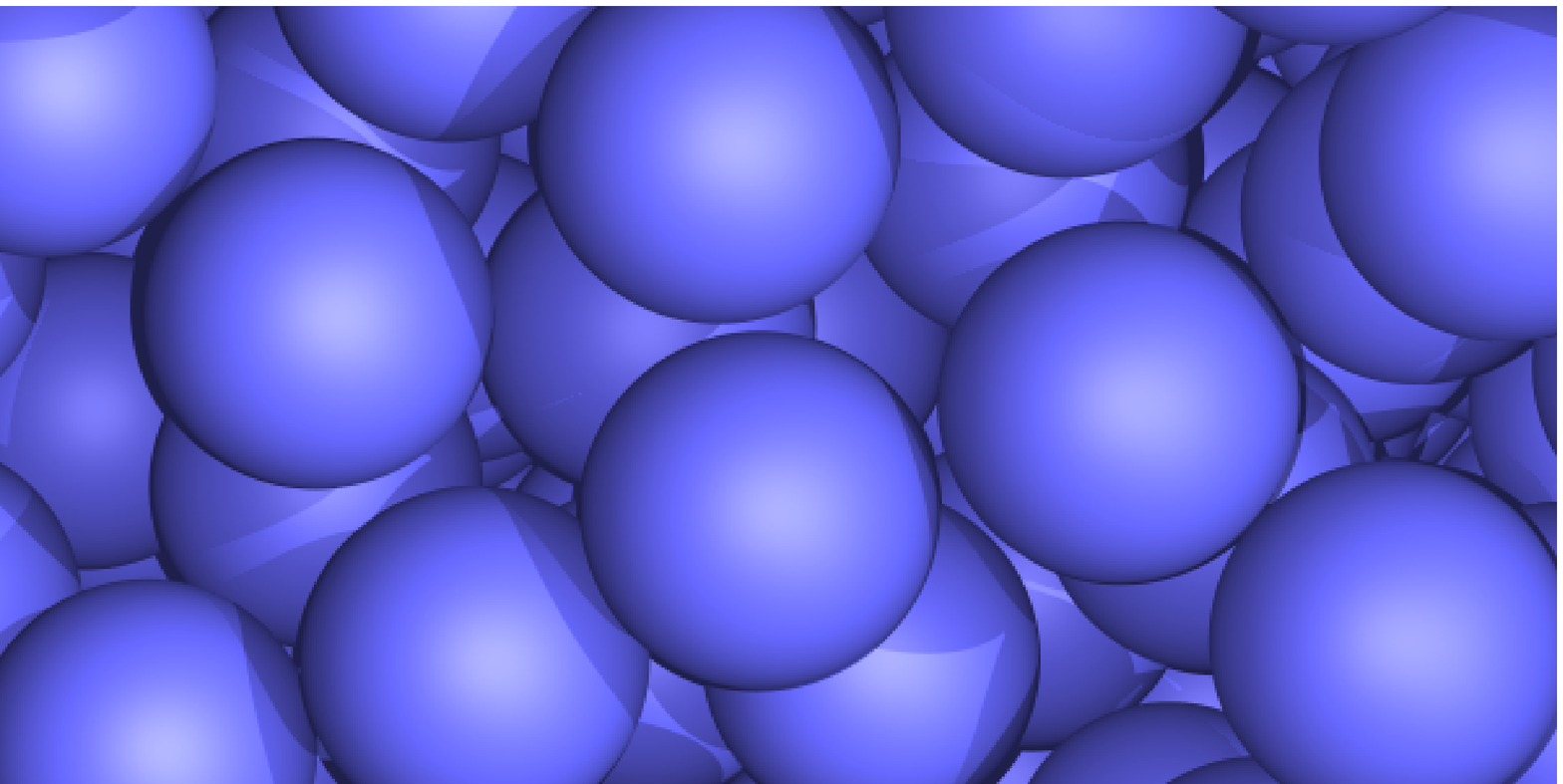}%
  \label{fig:vis-equi}}
  \hfill
  \subfigure[][]{\includegraphics[width=0.65\linewidth,angle=90]{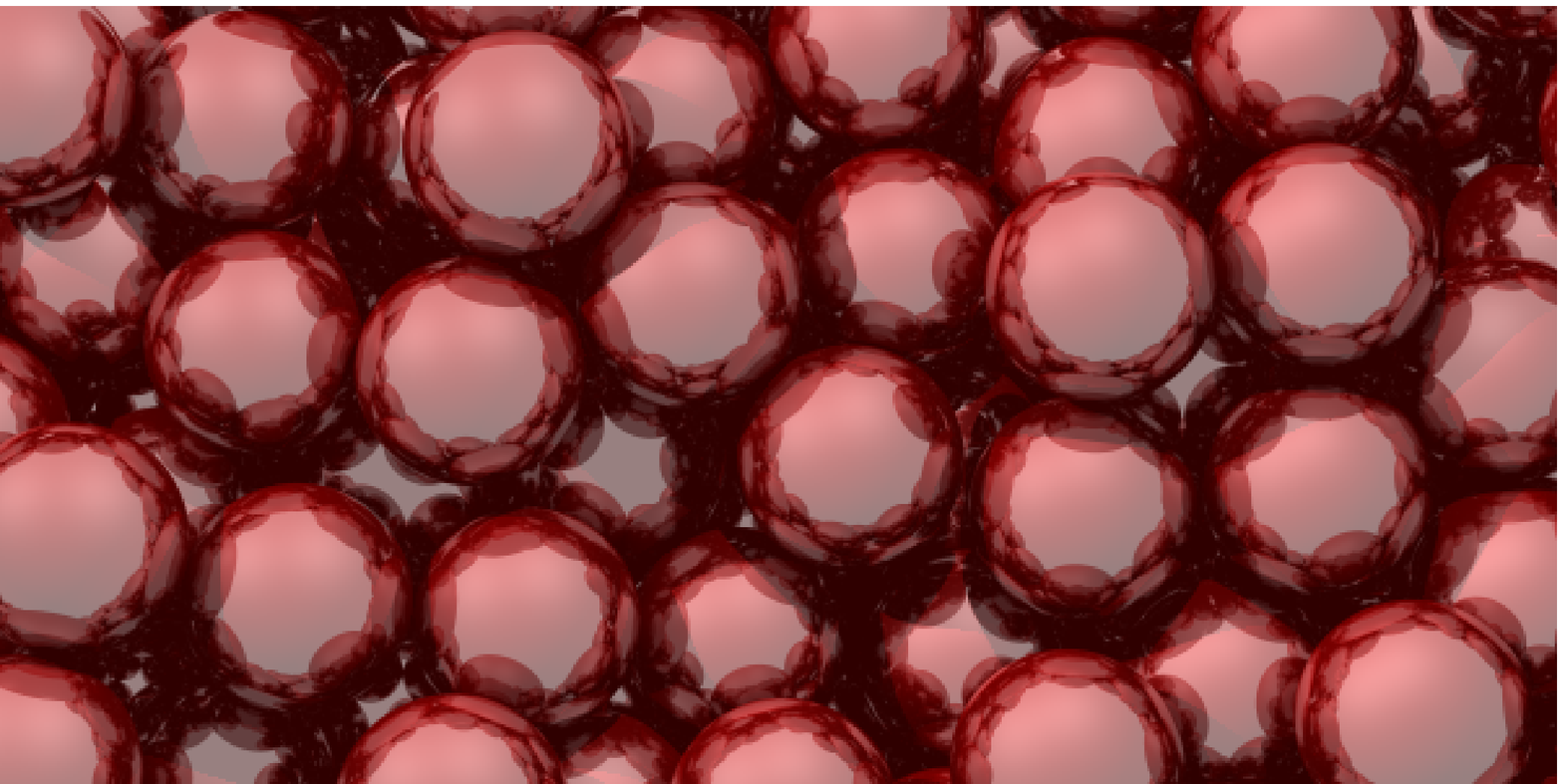}%
  \label{fig:vis-MRJ}}
  \caption{(Color online) Disordered sphere configurations: (a) overlapping spheres, (b) an equilibrium hard-sphere liquid, (c) an MRJ sphere packing}
  \label{fig:vis}
\end{figure}

Among all mechanically stable packings of totally impenetrable spheres in $d$-dimensional Euclidean space $\mathbb{R}^d$, an especially interesting system is the packing that exhibits maximal disorder.
More precisely, among the set of all isotropic, frictionless and statistically homogeneous jammed sphere packings~\cite{Henley1986, Hales2005, Finney1997, Zallen1998, bertrand_protocol_2016},
of particular interest is the state that minimizes some given order metric $\psi$.
This is called the maximally random jammed (MRJ) state~\cite{TorquatoEtAl2000PhysRevLetRCPvsMRJ, TorquatoStillinger2010RevModPhys, ohern_PRL_2002, Karayiannis_PRL_2008, XuRice2011, Ozawa_etal_2012, Baranau_etal_2013, tian_geometric-structure_2015, ramola_disordered_2016}; see Fig.~\ref{fig:vis-MRJ}.

This definition makes mathematically precise the familiar notion of random closed packing (RCP)~\cite{Bernal1960, ScottKilgour1969, Finney1970, Berryman1983, HernEtAl2003, Aste2005, KuritaWeeks2011, Berthier_etal_2011, KapferEtAl2012} in that it can be unambiguously identified for a particular choice of the order metric.
A variety of sensible, positively correlated order metrics produce an MRJ state (minimal order metric) in three dimensions with the same packing fraction 0.64~\cite{TorquatoStillinger2010RevModPhys, Berthier_etal_2011}.
While three-dimensional (3D) RCP and MRJ packings of identical spheres are reported to have similar packing fractions~\cite{TorquatoEtAl2000PhysRevLetRCPvsMRJ, HernEtAl2003, ScottKilgour1969}, other structural attributes can be both subtly and distinctly different~\cite{TorquatoStillinger2010RevModPhys, AtkinsonEtAl2013, KansalEtAl2002}.
Moreover, the packing characteristics of RCP and MRJ packings of two-dimensional identical disks have recently been shown to be dramatically different from one another, including their respective densities, average contact numbers, and degree of order~\cite{AtkinsonEtAl2014}, which serves to punctuate the conceptual differences between RCP and MRJ states.
MRJ packings possess the singular property of hyperuniformity~\cite{TorquatoStillinger2003, ZacharyTorquato2009}, \ie, infinite-wavelength density (volume-fraction) fluctuations are anomalously suppressed.
In MRJ packings in $d$-dimensional Euclidean space $\mathbb{R}^d$, this is manifested as negative quasi-long-range pair correlations that decay asymptotically like $-1/r^{d+1}$~\cite{DonevEtAl2005, ZacharyJiaoTorquato2011}.
Disordered hyperuniformity can be seen as an `inverted critical phenomenon' with a long-ranged direct correlation function, in contrast to thermal critical points in which this function is short ranged~\cite{TorquatoStillinger2003, HopkinsStillingerTorquato2012}.

The purpose of the present series of papers is to delve more deeply into the structure and physical properties of 3D MRJ packings of identical frictionless spheres.
In the first paper of this series~\cite{KlattTorquato2014}, we introduced Voronoi correlation functions to characterize the structure of MRJ sphere packings across length scales.
More precisely, we computed correlation functions associated with the volume and other Minkowski functionals of Voronoi cells.
We investigated similarities and differences in both the local and the global structure of overlapping spheres, equilibrium hard-sphere liquids, and MRJ sphere packings; see Fig.~\ref{fig:vis}.
We demonstrated that although their local structural characteristics appear to be qualitatively similar, their global structure is qualitatively different.
Strong Voronoi anti-correlations that we found in the MRJ state are related to its hyperuniformity.

In this paper, 
we determine a variety of different correlation functions that arise in rigorous bounds on the effective physical properties~\cite{Prager1961, RubinsteinTorquato1988, RubinsteinTorquato1989JFM, TorquatoYeong1997, Prager1963, TorquatoAvellaneda1991, Torquato2002, Beran1968, Torquato1985, *SenTorquato1989, RechtsmanTorquato2008, TorquatoRubinstein1989, Doi1976, Torquato1991Review}
of MRJ sphere packings and compare them to the corresponding statistical descriptors for overlapping spheres and equilibrium hard-sphere systems.
This includes the two-point probability functions, two-point surface correlation functions, and pore-size distributions.
These statistical descriptors arise, \eg, in rigorous bounds for effective transport properties~\cite{Prager1961, TorquatoLado1986, RubinsteinTorquato1988,
RubinsteinTorquato1989JFM, TorquatoYeong1997}, diffusion and reactions constants~\cite{Prager1963, TorquatoAvellaneda1991}, or
mechanical~\cite{Torquato2002} and electromagnetic properties~\cite{Beran1968, Torquato1985, *SenTorquato1989, RechtsmanTorquato2008}.
The surface-void and surface-surface correlation functions allow for improved bounds on the trapping constant~\cite{RubinsteinTorquato1988,
TorquatoRubinstein1989, Doi1976} and the fluid permeability~\cite{RubinsteinTorquato1989JFM, Torquato1991Review}.
All of these bounds will be the topic of the third paper in this series.
Thus, we relate different topics in material science, chemistry, physics, and mathematics.

Moreover, we investigate how the hyperuniformity of the MRJ state affects its global structure (in comparison to the nonhyperuniform equilibrium hard-sphere liquid and overlapping spheres).
Hyperuniformity can, for example, be detected by a vanishing spectral density~\footnote{The spectral density is the Fourier transform of the autocovariance function.} in the limit of large wavelengths.
An alternative equivalent diagnostic is how density fluctuations scale (asymptotically) with the size of the observation window.


In Sec.~\ref{sec:structure-characteristics}, we define and explain the structural descriptors used here to quantify both the two-phase medium formed by the spheres and the point process formed by the sphere centers.
These structural characteristics of the MRJ packings are compared to those of equilibrium hard spheres and overlapping spheres.
In Sec.~\ref{sec:correlation-functions}, we derive, for a given configuration of $N$ hard spheres within a periodic simulation box, MRJ or not, explicit analytical expressions for the two-point correlation function $S_2(r)$, surface-void correlation function $F_{sv}(r)$, and surface-surface correlation function $F_{ss}(r)$ as defined in Ref.~\cite{Torquato2002}, {for example}. 
These formulas allow for a fast and accurate calculation of these correlation functions as well as corresponding integrals that are needed for void and interfacial bounds~\cite{RubinsteinTorquato1988, RubinsteinTorquato1989JFM, TorquatoRubinstein1989, Doi1976, Torquato1991Review}.
We also investigate the behavior of the sphere configurations in reciprocal space in Sec.~\ref{sec:spectral-density}, where we compare different estimators of the spectral density $\tilde{\chi}_{_V}(k)$  associated with
the two-point probability function, see \eg~Refs.~\cite{Torquato2002, DreyfusEtAl2015}.
In Sec.~\ref{sec:pore-size-distribution}, we determine the {complementary} cumulative distribution function $F(\delta)$ of the pore sizes $\delta$ for the MRJ sphere packings and compare them to those of the overlapping and equilibrium hard spheres as well as crystalline structures.
We also obtain the exclusion probability $E_V(r)$, which is trivially related~\cite{Torquato2002}.

Second, we analyze the aforementioned sphere models as point processes.
Therefore, we identify the sphere configurations with the point patterns formed by the sphere centers in order to analyze their local  density fluctuations within a spherical window of radius $R$.
In Sec.~\ref{sec:density-fluctuations}, we first estimate the probability distributions of the number of sphere centers $N$ within a spherical window of radius $R$ for the equilibrium and the MRJ sphere packings for various sizes of the observation window~\cite{quintanilla_local_1997}.
The distributions quickly converge (for increasing window size) to normal distributions, which is consistent with analogous previous results for ``volume-fraction'' fluctuations for particle systems~\cite{quintanilla_local_1997}.
We thus conjecture a central limit theorem for disordered hard-sphere systems.
Then, we study the local number variance as a function of the radius of the observation window.

In Sec.~\ref{sec:Conclusion}, we summarize the results and make concluding remarks.
In Appendix~\ref{sec_analytical}, we analytically derive for a given finite configuration of hard spheres the explicit expressions for the correlation functions mentioned above.

\section{Definitions of the correlation functions and other structural descriptors}
\label{sec:structure-characteristics}

\begin{table*}
  \centering
  \begin{ruledtabular}
  \begin{tabular}{l c c l l l}
    Characteristic & & Unit & \multicolumn{1}{c}{Canonical $n$-point function} & Definition & Results \\
    \hline
    \multicolumn{6}{c}{\rule{0pt}{4ex} Two-phase media} \\
    \rule{0pt}{4ex}Volume or packing fraction & $\phi$ & 1 & $\lim_{a_1\rightarrow R} H_1(\varnothing;\{\vec{x}\};\varnothing)$ & Eq.~\eqref{eq:phi} & Sec.~\ref{sec:one-point-func} \\
    Specific surface & $s$ & $1/l$ & $\lim_{a_1\rightarrow R} H_1(\{\vec{x}\};\varnothing;\varnothing)$ & Eq.~\eqref{eq:s} & Sec.~\ref{sec:one-point-func} \\
    Two-point correlation function& $S_2(r)$ & 1 & $\lim_{a_i\rightarrow R, i=1,2} H_2(\varnothing;\{\vec{x}_1,\vec{x}_2\};\varnothing)+2\phi^2 -1$ & Eq.~\eqref{eq:S2-def} & Fig.~\ref{fig:S2}; Tab.~\ref{tab:correlations-equi},~\ref{tab:correlations}\\
    Surface-void correlation function & $F_{sv}(r)$ & $1/l^{\hphantom{2}}$ & $\lim_{a_i\rightarrow R, i=1,2} H_2(\{\vec{x}_1\};\{\vec{x}_2\};\varnothing)$ & Eq.~\eqref{eq:Fsv-def} & Fig.~\ref{fig:Fsv},~\ref{fig:2bdy-Fsv}; Tab.~\ref{tab:correlations-equi},~\ref{tab:correlations}\\
    Surface-surface correlation function & $F_{ss}(r)$ & $1/l^2$ & $\lim_{a_i\rightarrow R, i=1,2} H_2(\{\vec{x}_1,\vec{x}_2\};\varnothing;\varnothing)$ & Eq.~\eqref{eq:Fss-def} & Fig.~\ref{fig:Fss},~\ref{fig:2bdy-Fss}; Tab.~\ref{tab:correlations-equi},~\ref{tab:correlations}\\
    Spectral density & $\tilde{\chi}_{_V}(k)$ & $l^3$ & & Eq.~\eqref{eq:chi-via-Ft-of-indicator} & Fig.~\ref{fig_spectral_density}; Tab.~\ref{tab:spectral-density-equi},~\ref{tab:spectral-density}\\
    Compl. cumul. pore-size distribution & $F(\delta)$ & 1 & & Eq.~\eqref{eq:def-F} & Fig.~\ref{fig_cum_pore-size_distr}\\
    Mean pore size & $\langle\delta\rangle$ & $l$ & & Eq.~\eqref{eq:def-mean-pore-size} & Sec.~\ref{sec:mean-pore-size} \\
    Second moment of the pore size & $\langle\delta^2\rangle$ & $l^2$ & & Eq.~\eqref{eq:def-2nd-moment} & Sec.~\ref{sec:mean-pore-size} \\
    \multicolumn{6}{c}{\rule{0pt}{4ex} Point processes} \\
    \rule{0pt}{4ex}Exclusion probability & $E_V(r)$ & 1 & $\lim_{a_1\rightarrow r} H_1(\varnothing;\{\vec{x}\};\varnothing)$ & Eq.~\eqref{eq:def-exclusion-probability} & Fig.~\ref{fig:exclusion_probability}; Tab.~\ref{tab:correlations-equi},~\ref{tab:correlations}\\
    Number probability distribution & $f_R(N)$ & 1 &  & Sec.~\ref{sec:def-num-des-fluctuations} & Fig.~\ref{fig_density_probab_distr} \\
    Number variance & $\sigma_N^2(R)$ & 1 & & Eq.~\eqref{eq:def-num-var} & Fig.~\ref{fig_density_fluct} 
  \end{tabular}
  \end{ruledtabular}
  \caption{The structure characteristics used here describe either a random two-phase media, which is formed by the spheres and the surrounding matrix phase, or a point process, which is formed by the sphere centers.
    The unit is denoted by the length $l$, and references to the definition and some results in this paper are collected for each characteristic. 
    If possible, their representation by the canonical $n$-point functions is provided; see Sec.~\ref{sec:canonical}.}
  \label{tab:2phase}
\end{table*}

\begin{figure*}[t]
  \centering
  \subfigure[][]{\includegraphics[width=0.49\linewidth]{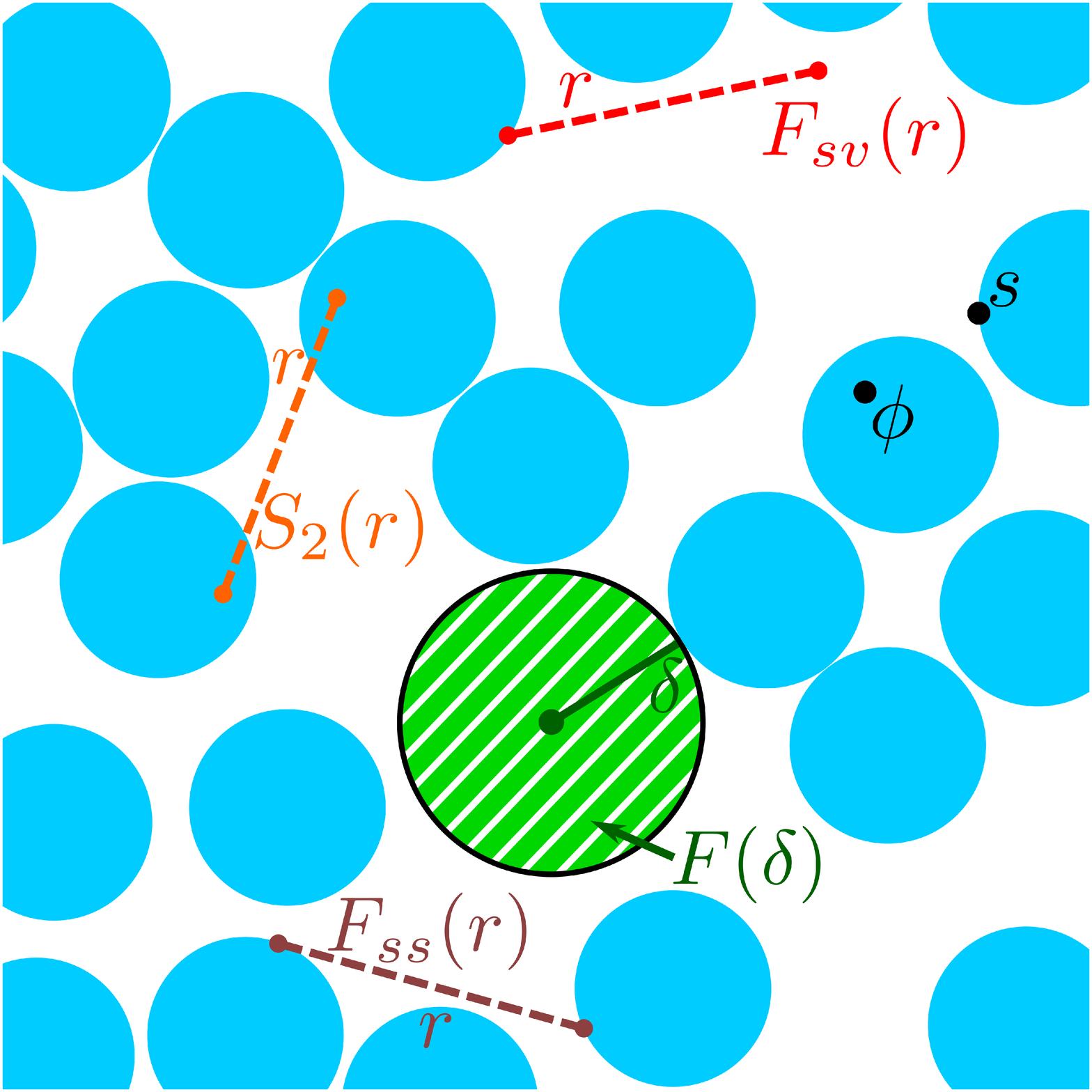}}
  \hfill
  \subfigure[][]{\includegraphics[width=0.49\linewidth]{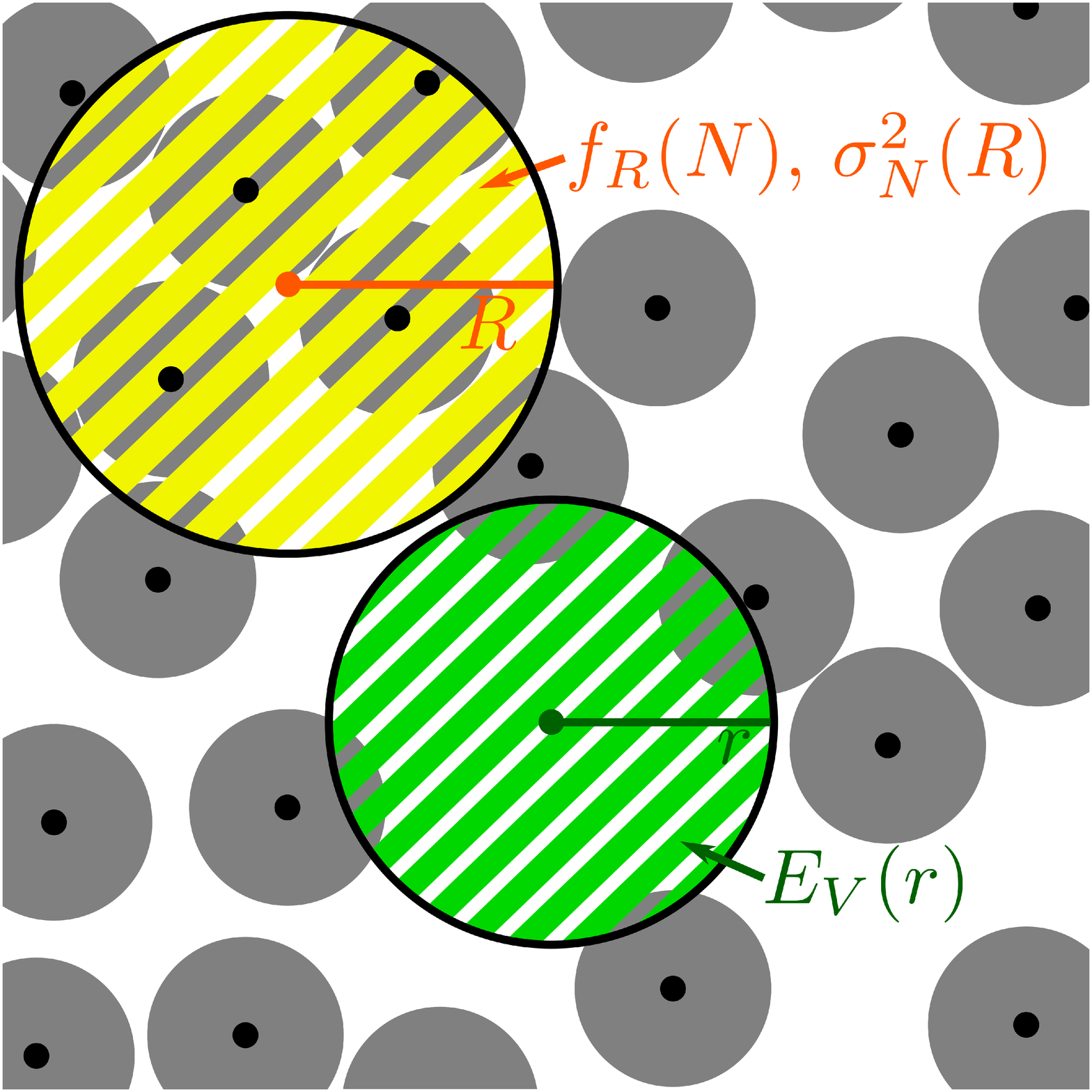}}
  \caption{The schematic depicts events that contribute to the structure characteristics from Tab.~\ref{tab:2phase} when single points, balls, or points at a given distance (dashed lines) are placed randomly in the sample (a) for a packing of hard spheres forming a two-phase random medium or (b) for the point process of the sphere centers.}
  \label{fig:vis-structural-characteristics}
\end{figure*}

A system of hard or overlapping spheres can either be viewed as a medium that consists of two phases,
where the first is formed by the spheres, and the second by the surrounding matrix,
or it can be represented by the point pattern that is formed by the sphere centers.
Here, we analyze both the two-phase random medium and the point process using different structure characteristics, which are summarized in Table~\ref{tab:2phase}; see also Fig.~\ref{fig:vis-structural-characteristics}.

When characterizing the two-phase medium formed by the spheres, we choose the diameter $D$ of the spheres as the unit of length.
In other words, the spheres in different systems have the same diameter.
On the other hand, we compare point patterns at unit density, so that when analyzing a point process, we use $\lambda = \rho^{-1/3}$ as the unit of length, where $\rho$ is the number density (or intensity).
The latter is the mean number of points per unit volume.

\subsection{Two-phase media}

A two-phase medium can be represented by the so-called indicator function $\mathcal{I}^{(j)}(\vec{r})$ for phase $j\in\{1,2\}$~\cite{Torquato2002}:
\begin{align}
  \mathcal{I}^{(j)}(\vec{r}) := \begin{cases}
    1, & \vec{r} \text{ in phase }j,\\
    0, & \text{otherwise},
  \end{cases}
  \label{eq:indicator-phase-def}
\end{align}
which is sometimes also called the characteristic function of phase $j$.
Also the indicator function $\mathcal{M}(\vec{r})$ for the interface
can be defined as a generalized function, \ie, involving Dirac delta functions~\cite{Torquato2002}:
\begin{align}
  \mathcal{M}(\vec{r}) := |\nabla \mathcal{I}^{(1)}(\vec{r})| = |\nabla \mathcal{I}^{(2)}(\vec{r})|.
  \label{eq:indicator-interface-def}
\end{align}
It is nonzero only if $\vec{r}$ is on the interface.

\subsubsection{One-point functions}
\label{sec:one-point-def}

To characterize the two-phase media formed by the different sphere systems, we first consider one-point probability functions~\cite{Torquato2002}.

The \textit{volume (or packing) fraction} $\phi$ of the phase covered by spheres.
    It is equal to the probability that a random point lies within any sphere and thus within the phase $2$ formed by the spheres:
\begin{align}
  \phi := \mathcal{P}\left\{ \mathcal{I}^{(2)}(\vec{r}) = 1 \right\} = \left\langle \mathcal{I}^{(2)}(\vec{r}) \right\rangle.
  \label{eq:phi}
\end{align}
This probability is equal to the expectation of the indicator function because the latter only takes on the values 0 or 1.
The angular brackets denote an ensemble average (over all possible realizations) at a fixed position $\mathbf{r}$.
Because the systems studied here are homogeneous and ergodic, this average does not depend on the position, and it corresponds to a spatial average in the infinite-volume limit.

The \textit{specific surface} $s$ is the ratio of the surface area and the volume of the whole system.
    It can be similarly defined by the expectation of the indicator function for the interface:
\begin{align}
  s := \left\langle \mathcal{M}(\vec{r}) \right\rangle.
  \label{eq:s}
\end{align}
A probably more intuitive interpretation considers the probability that a random point falls within a shell of thickness $\epsilon$ around the interface~\cite{Torquato2002}.
In the limit of vanishing thickness $\epsilon\rightarrow 0$, the ratio of this probability and the thickness $\epsilon$ converges to the specific surface.
For a packing of hard spheres with diameter $D$, the specific surface can easily be related to the volume (or packing) fraction via $s=6\phi/D$. To begin, we consider a one-point probability function, the volume fraction $\phi$ of the phase covered by spheres and the specific surface $s$,
which is (in the limit of infinite system size) the ratio of the surface area and the volume of the whole system~\cite{Torquato2002}.

\subsubsection{Correlation functions}

In contrast to the one-point functions, the two-point functions describe the global structure, i.\,e., they characterize correlations at larger distances.

An intuitive definition of the \textit{two-point correlation function} $S^{(j)}_2(\vec{r}_1,\vec{r}_2)$ for phase $j$ is the probability that the two points $\vec{r}_1$ and $\vec{r}_2$ lie both in phase $j$~\cite{Torquato2002}:
\begin{align}
  \begin{aligned}
  S^{(j)}_2(\vec{r}_1,\vec{r}_2) :=\;& \mathcal{P}\left\{ \mathcal{I}^{(j)}(\vec{r}_1) = 1 \text{ and } \mathcal{I}^{(j)}(\vec{r}_2) = 1 \right\}\\
  =\;& \left\langle \mathcal{I}^{(j)}(\vec{r}_1)\cdot\mathcal{I}^{(j)}(\vec{r}_2) \right\rangle.
  \end{aligned}
  \label{eq:S2-def}
\end{align}
For homogeneous and isotropic systems, this two-point correlation function only depends on the distance $r$ between two points.
$S^{(j)}_2(r)$ can then be interpreted as the probability that two random points at a distance $r$ are both found in phase $j$.
Here, we calculate the two-point correlation function for phase $2$ formed by the spheres.
For convenience, we define $S_2(r):= S^{(2)}_2(r)$.
Note, however, that the difference of the two-point correlation functions of the two phases is simply a constant offset by $(1-2\phi)$.
So, we can easily deduce $S_2(r)$ for the first phase if we know it for the second phase.

In contrast to the two-point correlation function, the \textit{surface-void correlation function} $F_{sv}(\vec{r}_1,\vec{r}_2)$ is not a probability but the limit of a rescaled probability.
It considers the probability that a random point is inside the ``void'' and another random point is in a shell of (vanishing) thickness $\epsilon$ close to the interface between the two phases.
The surface-void correlation function is the limit of the ratio of this probability and the distance $\epsilon$ for $\epsilon\rightarrow 0$~\cite{Torquato2002}.
Therefore, it has units of inverse length.

If the void phase is denoted by $j_v$,
the surface-void correlation function can be defined 
using the indicator functions of the void phase and the interface:
\begin{align}
  \begin{aligned}
  F_{sv}(\vec{r}_1,\vec{r}_2) := \left\langle \mathcal{M}(\vec{r}_1)\cdot\mathcal{I}^{(j_v)}(\vec{r}_2) \right\rangle.
  \end{aligned}
  \label{eq:Fsv-def}
\end{align}
For homogeneous and isotropic systems, it only depends on the distance $r$ between two points,
which is denoted by $F_{sv}(r)$.

For two-phase random media formed by the spheres, $F_{sv}(r)$ depends on the choice of which phase is considered as ``void''.
Either the space exterior to the spheres or the phase formed by the spheres can form the void phase.
However, the correlation functions for both choices can easily be derived from one another.
In the following, $F_{sv}(r)$ denotes the case where the space exterior to the spheres forms the void phase.
We denote by $F_{sv}^{(s)}(r)$ the corresponding function for the complementary system, \ie, where the spheres are the void phase.
The sum of these two correlation functions is a constant and equal to the specific surface:
\begin{align}
  F_{sv}(r) + F_{sv}^{(s)}(r) = s.
  \label{eq:relation-between-Fsv-s-i}
\end{align}

Like the surface-void correlation function, the \textit{surface-surface correlation function} is no probability but the limit of a rescaled probability.
As the name indicates, it considers the probability that both random test points are inside a shell of (vanishing) thickness $\epsilon$ close to the interface between the two phases.
The surface-surface correlation function is the limit of the ratio of this probability and $\epsilon^2$ for $\epsilon\rightarrow 0$~\cite{Torquato2002}.
Therefore, it has units of inverse length squared.

It can be defined using the indicator function of the interface:
\begin{align}
  \begin{aligned}
  F_{ss}(\vec{r}_1,\vec{r}_2) := \left\langle \mathcal{M}(\vec{r}_1)\cdot\mathcal{M}(\vec{r}_2) \right\rangle.
  \end{aligned}
  \label{eq:Fss-def}
\end{align}
For homogeneous and isotropic systems, the correlation function again only depends on the distance $r$ between two points.
This is in the following denoted by $F_{ss}(r)$.

In contrast to the surface-void correlation function, the
surface-surface correlation function does not depend on the choice of
which phase is ``void''. 

\subsubsection{Spectral density}
\label{sec:define-spectral-density}

From the two-point correlation function $S_2(r)$ follows the definition of the autocovariance of a two-phase medium 
\begin{align}
  \chi_{_V}(\vec{r}) := S_2(\vec{r}) - \phi^2,
\end{align}
see, \eg, Ref.~\cite[][Sec.~2.2.5]{Torquato2002}.
Its Fourier representation can be obtained via scattering of radiation~\cite{debye_scattering_1949}.
The Fourier transform of the autocovariance is the \textit{spectral density}
\begin{align}
  \tilde{\chi}_{_V}(\vec{k}) := \mathcal{F}[\chi_{_V}(\vec{r})] = \int\text{d}\vec{r}\,\chi_{_V}(\vec{r})\cdot
  \text{e}^{-i\vec{k}\cdot\vec{r}}.
\end{align}
For a statistically isotropic material, where $S_2(r)$ is only a function of the distance $r$,
also the spectral density only depends on the absolute values $k$ of the wave vector
\begin{align}
  \tilde{\chi}_{_V}(k) := \mathcal{F}[\chi_{_V}(r)] = \frac{4\pi}{k} \int_0^{\infty}\text{d}r\,(S_2(r) - \phi^2) r \cdot \sin(k r).
  \label{eq:chi-FT}
\end{align}
An equivalent definition of the spectral density is given by the Fourier transform of the indicator function $\mathcal{I}^{(2)}(\vec{r})$ of the particle phase, or more precisely, of the function
\begin{align}
  J(\vec{r}) := \mathcal{I}^{(2)}(\vec{r}) - \phi,
\end{align}
where we subtract the mean value of the indicator function.
The spectral density is the absolute square of this Fourier transformation divided by the volume of the system~\cite{DreyfusEtAl2015}:
\begin{align}
  \tilde{\chi}_{_V}(\vec{k})=\frac{1}{V}\left|\tilde{J}(\vec{k})\right|^2.
  \label{eq:chi-via-Ft-of-indicator}
\end{align}
We here consider the spectral density as a function of the wave vector $\vec{k}$ (and not only of its absolute value),
because calculating the spectral density based on this definition we can explicitly take for finite samples nonorthogonal simulation boxes into account.

For monodisperse hard spheres, the Fourier transform $\tilde{J}(\vec{k})$ of the two-phase medium can be rigorously related to the point process formed by the sphere centers.
The absolute value of the first can be expressed by the structure factor $S(k)$ of the latter, where the structure factor can be defined as $S(k) = 1+\rho \tilde{h}(k)$ using the Fourier transform $\tilde{h}(k)$ of the total correlation function $h(r)=g_2(r)-1$ (and $g_2(r)$ is the pair correlation function).
The spectral density is then given by~\cite{ZacharyTorquato2009, Torquato2002, torquato_microstructure_1985}
\begin{align}
  \tilde{\chi}_{_V}(\textbf{k})=\frac{1}{V}\left|\tilde{J}(\textbf{k})\right|^2 = \rho\cdot \tilde{m}^2(k)\cdot S(\textbf{k}),
  \label{eq:chi-Indicator}
\end{align}
where $\rho$ is the number density and $\tilde{m}(k)$ is the Fourier transform of a single sphere.
In $d$-dimensional Euclidean space, it is given by
\begin{align}
  \tilde{m}(k) = \left( \frac{\pi D}{k} \right)^{d/2}\cdot J_{d/2}(\frac{kD}{2}).
  \label{eq:mk}
\end{align}
Here, $J_{d/2}(x)$ is the Bessel function of the first kind of order $d/2$, which is in three dimensions given by
\begin{align}
  J_{3/2}(x)=\sqrt{\frac{2}{\pi\,x^3}}\cdot(\sin(x)-x\cdot\cos(x)).
\end{align}
The structure factor $S(k)$ is not only important because it can be directly measured in scattering experiments.
It can also be used to detect a remarkable property of point processes, hyperuniformity, as discussed in Sec.~\ref{sec:spectral-density}.

\subsubsection{Pore-size distribution}

We also characterize the sphere configurations by the distribution of their pores sizes $\delta$, that is, the maximum radius of a spherical pore that can be assigned to a random point in the matrix phase so that the pore lies wholly in the matrix phase.
The probability density function $P(\delta)$ of the pore sizes is also known as the ``pore-size distribution''~\cite{Prager1963}.
For a point chosen randomly in the matrix (or void) phase, $P(\delta)\text{d}\delta$ is the probability that its shortest distance to the solid-void interface lies between $\delta$ and $\delta +\text{d}\delta$.
Because $P(\delta)$ is a probability density function, it is normalized $\int_0^{\infty}\text{d}\delta\,P(\delta)=1$ and it has the unit of the inverse of length.
Note that the distribution of pore sizes within a phase formed by hard spheres is trivial in the sense that it is independent of the position of the
spheres~\footnote{The pore-size probability density function $P(\delta)$ for pores in a phase formed by hard spheres with radius $R$ is only nonzero for $\delta < R$.
  There, it is always (independent of the arrangement of the spheres) given by $P(\delta)=3(R-\delta)^2/R^3$~\cite{Prager1963}.}.

An equivalent representation is the \textit{complementary cumulative distribution function} $F(\delta)$ of the pore sizes:
\begin{align}
  F(\delta) := \int_{\delta}^{\infty}\text{d}r\,P(r).
  \label{eq:def-F}
\end{align}
It can be interpreted as the fraction of the matrix phase with a pore radius larger than $\delta$.
By definition, $F(0)=1$ and $F(\infty)=0$, and because it is a probability, $F(\delta)$ has no units.
The cumulative distribution function $[1-F(\delta)]$ is also known as the spherical contact distribution function~\cite{ChiuEtAl2013, hug_survey_2002, footnote-correlations}.

The \textit{mean pore size} $\langle \delta \rangle$ and the \textit{second moment} $\langle \delta^2 \rangle$ of $P(\delta)$ can be expressed by $F(\delta)$~\cite{Torquato2002}:
\begin{align}
  \langle \delta \rangle &:= \int_0^{\infty} \text{d}\delta\,F(\delta),
  \label{eq:def-mean-pore-size}\\
  \langle \delta^2 \rangle &:= 2\int_0^{\infty} \text{d}\delta\,F(\delta)\cdot \delta.
  \label{eq:def-2nd-moment}
\end{align}
They can be interpreted as characteristic length scales of the matrix phase. 

\subsection{Point processes}

For a packing of monodisperse spheres, the structure characteristics of the two-phase medium formed by the spheres can be related to those of the point pattern formed by the centers.

\subsubsection{Exclusion probability}
\label{sec:def-EV}

The probability that a test sphere of radius $r$ that is placed randomly in the sample does not contain any point of the point process is called the \textit{exclusion probability} $E_V(r)$.
It is a nonincreasing function, and it can be interpreted as the expected fraction of space available to a test sphere of radius $r$ which is not allowed to contain a point of the point process.
For monodisperse spheres of radius $R$, it is trivially related to the complementary cumulative pore-size distribution $F(\delta)$ via
\begin{align}
  E_V(r) = (1-\phi)F(r-R)\quad \text{for }r > R.
  \label{eq:def-exclusion-probability}
\end{align}
For $r\leq R$, the exclusion probability for hard-sphere centers is simply given by
\begin{align}
  E_V(r) = 1-\frac{4\pi}{3}r^3\cdot\rho.
\end{align}

\subsubsection{Local number density fluctuations}
\label{sec:def-num-des-fluctuations}

The exclusion probability considers whether or not a randomly placed test sphere of radius $R$ contains at least one point of the point process.
This can be generalized to the probability function $f_R(N)$ that there are exactly $N$ points of the point process inside the test sphere.
This \textit{number probability function} $f_R(N)$ includes the exclusion probability $E_V(r) = f_r(0)$.
However, the complete probability function $f_R(N)$ is a more general measure of density fluctuations in the point pattern.
For the example of a Poisson point process, the number probability function $f_R(N)$ is by definition a Poisson distribution~\cite{LastPenrose16}.

The mean value of the number probability function, \ie, the expectation of the number $N$ of points in the test sphere (or ``observation'' window) of radius $R$, is given  by $\langle N \rangle_R := \rho\cdot\frac{4\pi}{3}R^3$ for a statistically homogeneous point process (according to the definition of the number density $\rho$).
The variance of $N$ is known as the \textit{number variance}:
\begin{align}
  \sigma_N^2(R) &:= \langle N^2 \rangle_R - \langle N \rangle_R^2 = \sum_{n=0}^{\infty} f_R(n)\cdot(n - \langle N \rangle_R)^2.
  \label{eq:def-num-var}
\end{align}
For a Poisson distribution, the variance is equal to the mean value.
For a lattice, the number variance scales for large radii like the surface of the spherical observation window, since number fluctuations are concentrated in the vicinity of the window boundary~\cite{TorquatoStillinger2003}.

The number variance is closely related to the structure factor~\cite{TorquatoStillinger2003}:
\begin{align}
  \sigma_N^2(R) &= \frac{\rho}{(2\pi)^3} \int_{\mathbb{R}^3} \text{d}\textbf{k}\,S(\textbf{k})\cdot \tilde{m}^2(k),
  \label{eq:sigma-via-Sk}
\end{align}
where $\tilde{m}(k)$ is the Fourier transform of a single sphere; see Eq.~\eqref{eq:mk}.
Therefore, the number variance can, similar to the structure factor, detect whether or not a point process is hyperuniform.
If the number variance $\sigma_N^2(R)$ grows in the limit of large radii $R\rightarrow\infty$ more slowly than $R^3$, the point process is hyperuniform.
This definition of hyperuniformity based on the scaling of $\sigma_N^2(R)$ is equivalent to the definition via the limit $\lim_{k\rightarrow0}S(k)=0$.

\subsection{Canonical $n$-point functions $H_n$}
\label{sec:canonical}

It is noteworthy that the correlation functions and exclusion probability discussed here are special cases of the more general canonical $n$-point functions $H_n$, which describe higher-order spatial correlations between spheres and test particles~\cite{Torquato2002}.

The canonical $n$-point function $H_n$ statistically characterizes $n$ spherical test particles with radii $b_i$ ($i=1,\ldots, n$).
Before inserting the $i$th test particle, so-called ``exclusion spheres'' with radii $a_i = R + b_i$ are assigned to each of the original sphere centers.
Overlap between exclusion spheres is allowed.
If $b_i = 0$, the exclusion spheres are identical to the original spheres; $b_i > 0$ corresponds to a dilation of the sphere system and $-R < b_i < 0$ to an erosion.
The ``available space'' $D_i$ of the $i$th test particle is defined as the complement of the union of these exclusion spheres.
In other words, the test particle should not fall into any exclusion sphere.
The canonical $n$-point function characterizes these $D_i$.

In the notation of \citet{Torquato2002}, a canonical $n$-point correlation function is denoted by $H_n(\{\vec{x}_1,\ldots,\vec{x}_m\},   \{\vec{x}_{m+1},\ldots,\vec{x}_{p}\};   \{\vec{r}_{p+1},\ldots,\vec{r}_{n}\})$.
It is a very general function that combines
\begin{itemize}
  \item the $m$-point surface correlation function associated with $\partial D_1,\ldots, \partial D_m$, \ie, the surfaces of spaces available to test particles of radii $b_1,\ldots,b_m$, as a function of the positions $\vec{x}_1,\ldots,\vec{x}_m$, respectively,
  \item the $(p-m)$-point correlation function     associated with $D_{m+1},\ldots, D_{p}$, \ie, the spaces available to test particles of radii $b_{m+1},\ldots,b_p$,                     as a function of the positions $\vec{x}_{m+1},\ldots,\vec{x}_{p}$, respectively,
  \item and the $(n-p)$-point correlation function of the sphere centers as a function of the positions $\vec{r}_{p+1},\ldots,\vec{r}_n$, respectively.
\end{itemize}
This huge family of correlation functions includes a wealth of information about the geometry of the point pattern (or the corresponding sphere packings, respectively).
If for a specific $H_n$ we omit one of these three types of correlation functions, the corresponding set of variables is replaced by the symbol $\varnothing$ for the empty set.

For example, the one- and two-point functions discussed here can be expressed by the canonical $n$-point functions $H_n$ in the limit that the radii $a_i$ of the exclusion spheres become equal to the radius $R$ of the (original) spheres.
Using a single test particle with radius $a_1$, we express the occupied volume fraction as
\begin{align}
  \phi = \lim_{a_1\rightarrow R} H_1(\varnothing;\{\vec{x}\};\varnothing),
\end{align}
which actually does not depend for a homogeneous system on the position $\vec{x}$.
Similarly, we express the specific surface $s$ as
\begin{align}
  s = \lim_{a_1\rightarrow R} H_1(\{\vec{x}\};\varnothing;\varnothing).
\end{align}
For the two-point correlation function, we need two test particles with radii $a_1$ and $a_2$:
\begin{align}
  S_2(\vec{x}_1,\vec{x}_2) = 2\phi^2 -1 + \lim_{\substack{a_1\rightarrow R\\a_2\rightarrow R}} H_2(\varnothing;\{\vec{x}_1,\vec{x}_2\};\varnothing).
\end{align}
The surface-void correlation function can be written as
\begin{align}
  F_{sv}(\vec{x}_1,\vec{x}_2) = \lim_{\substack{a_1\rightarrow R\\a_2\rightarrow R}} H_2(\{\vec{x}_1\};\{\vec{x}_2\};\varnothing),
\end{align}
and the surface-surface correlation function as
\begin{align}
  F_{ss}(\vec{x}_1,\vec{x}_2) = \lim_{\substack{a_1\rightarrow R\\a_2\rightarrow R}} H_2(\{\vec{x}_1,\vec{x}_2\};\varnothing;\varnothing).
\end{align}
Instead of directly declaring $a_1,a_2=R$, we explicitly denote the limits to emphasize the generality of these  canonical correlation functions.
It has been shown that the generalizations using test particles with sizes larger than $R$ contain considerably more information than the two-point correlation functions considered here~\cite{ZacharyTorquato2011}
and hence represents an area for future study in the case of MRJ packings.

For example, the exclusion probability $E_V$ can also be represented by a one-point canonical
correlation function like the packing fraction but with a different radius of the test spheres:
\begin{align}
  E_V(r) = \lim_{a_1\rightarrow r} H_1(\varnothing;\{\vec{x}\};\varnothing),
\end{align}
which for a statistically homogeneous system does not depend on the position $\vec{x}$.

\section{Analysis and computation of the correlation functions}
\label{sec:correlation-functions}

Bounds on the trapping constant or permeability can be calculated using the void-void, surface-void, and surface-surface correlation functions of the sphere configurations~\cite{Beran1968, Doi1976, RubinsteinTorquato1988, RubinsteinTorquato1989JFM, Torquato2002, RechtsmanTorquato2008}.
More precisely, the bounds are given in terms of integrals over these correlation functions. 

These integrals can be difficult to estimate by simple Monte Carlo sampling due to statistical fluctuations in the measured volume fraction.
{Any statistical fluctuation in the estimate of the porosity, that is, in the fraction of points hitting the void phase, causes an offset in the long-range limit of the correlation functions.}
This offset can lead to huge errors in the estimates of the bounds which are based on integrals of the correlation functions.

In Appendix~\ref{sec_analytical}, we derive explicit analytical formulas of the two-point, surface-void, and surface-surface correlation functions for a given finite configuration of hard spheres, which heretofore were not put forth.
In the thermodynamic limit, \ie, for infinitely large systems, these correlation functions can be analytically related to the pair-correlation function $g_2(r)$ of the sphere centers.
For example, \citet{torquato_microstructure_1985} and \citet{Torquato1986} used {certain analytical approximations} of the pair-correlation function for equilibrium hard-sphere liquids~\cite{VerletWeis1972} to calculate $S_2(r)$, $F_{sv}(r)$, and $F_{ss}(r)$; see also Ref.~\cite{berryman_computing_1983}, where a similar approach is used to calculate bounds on flow properties.
Here, we provide exact and explicit formulas for {the two-point, surface-void, and surface-surface correlation functions} of finite packings of hard spheres for general ensembles.
These expressions can be viewed as ``discrete versions'' of the formulas in Refs.~\cite{torquato_microstructure_1985,Torquato1986}.
They only depend on the pairwise distances of the spheres.
This allows for the most efficient calculation of $S_2(r)$, $F_{sv}(r)$, and $F_{ss}(r)$ in finite packings {(obtained, \eg, from simulations)} as well as accurate estimates of the bounds on effective properties that depend on these correlation functions.

The samples of the hard-sphere packings that we analyze here were described in detail in the first paper of this series~\cite{KlattTorquato2014}.
For MRJ sphere packings, more than {1000} packings are analyzed each consisting of {2000} spheres.
For equilibrium hard spheres, each of the 100 samples contains {10000} spheres.

We compare the correlation functions of the MRJ sphere packings, as mentioned in the Introduction, to two other systems of spheres with constant diameter $D$:
(i) overlapping spheres that do not interact with each other and 
(ii) an equilibrium hard-sphere liquid at a packing fraction $\phi=0.478$, which is {just below} the freezing transition.
For more details about these systems, the simulations, and the data; see the first paper of this series~\cite{KlattTorquato2014}.

\subsection{One-point functions}
\label{sec:one-point-func}

Before analyzing the two-point functions, we determine the one-point probability functions, namely volume fraction and specific surface.


The average packing fraction of the MRJ sphere packings is $\phi=0.636$ and their specific surface $s=3.81/D$.
The snapshots of the equilibrium hard-sphere liquid have an average packing fraction $\phi=0.478$ and thus a specific surface $s=2.87/D$.
For overlapping spheres, the one-point functions are known analytically as a function of the occupied volume fraction $\phi$.
The specific surface is given by $s=6(1-\phi)\ln(1/(1-\phi))/D$.
For a volume fraction $\phi=0.636$ (equal to the average packing fraction of the MRJ systems),
the specific surface is $s \approx 2.21/D$.

\subsection{Two-point correlation function}
\label{sec_2p_corr}

The two-point correlation function $S_2(r)$  determines bounds on the conductivity~\cite{Beran1968, Torquato1985, *SenTorquato1989, Torquato2002}, the trapping constant~\cite{RubinsteinTorquato1988},
the fluid permeability~\cite{Prager1961, RubinsteinTorquato1989JFM}, and the effective dielectric tensor of electromagnetric waves~\cite{RechtsmanTorquato2008}.

\begin{figure}[t]
  \centering
  \includegraphics[width=\linewidth]{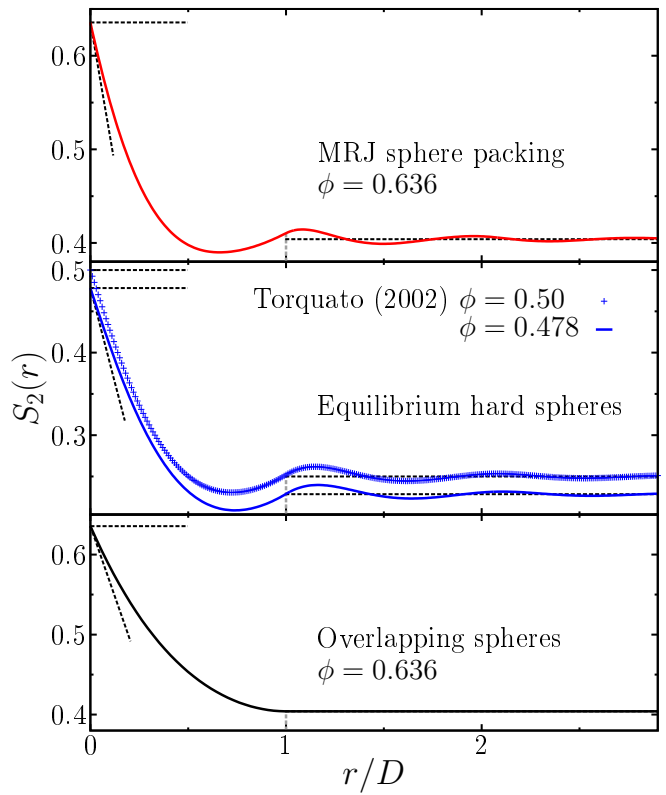}
  \caption{(Color online) Two-point correlation functions $S_2(r)$ for overlapping spheres
    ($\phi = 0.636$), equilibrium hard-sphere liquids ($\phi=0.478$ and
    $\phi=0.50$, data from Ref.~\cite{Torquato2002}), and MRJ sphere
    packings ($\phi=0.636$). The dashed lines indicate the limits of the
    curves. The distance $r$ is rescaled by the diameter of a single sphere $D$.
    The slope at $r=0$ (indicated by a dashed line) is proportional to the specific surface $s$; see Secs.~\ref{sec:one-point-def} and \ref{sec:one-point-func}.}
  \label{fig:S2}
\end{figure}

\begin{figure}[t]
  \centering
  \includegraphics[width=\linewidth]{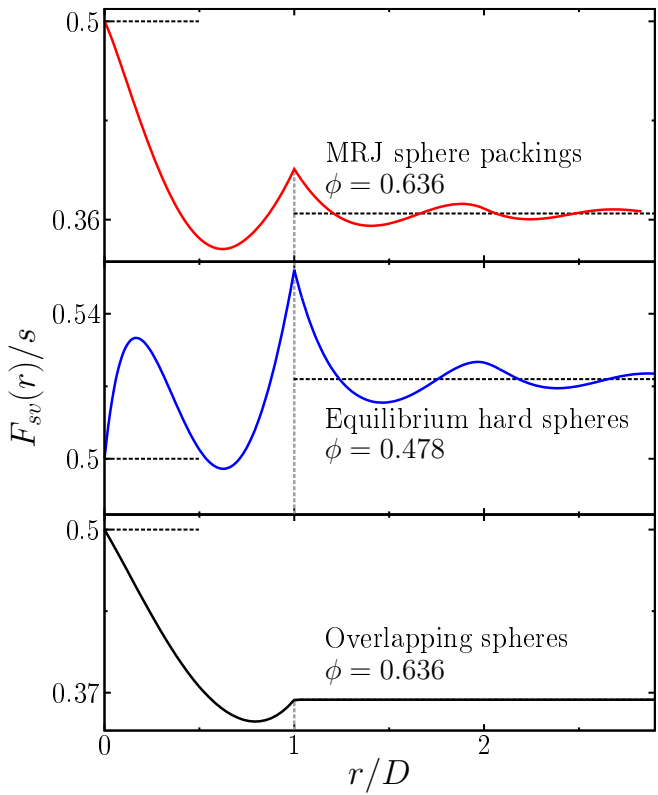}
  \caption{(Color online) Surface-void correlation functions $F_{sv}(r)$ (rescaled by
  the specific surface $s$) for overlapping spheres ($\phi = 0.636$, $s=2.21/D$), an equilibrium
  hard-sphere liquid ($\phi=0.478$, $s=2.87/D$), and MRJ sphere
  packings ($\phi=0.636$, $s=3.81/D$). For details; see Fig.~\ref{fig:S2}.}
  \label{fig:Fsv}
\end{figure}

Figure~\ref{fig:S2} compares the two-point correlation function for the particle phase of the MRJ sphere packings to that of overlapping spheres or two equilibrium hard-sphere liquids at different global packing fractions.
It is well-known analytically for overlapping spheres~\cite[\eg][p. 122]{Torquato2002}.
For both the equilibrium and the MRJ sphere packings, the two-point correlation functions are analytically calculated for each simulated sample according to Eq.~\eqref{eq:S2} and then averaged.
The dashed lines indicate the short- and long-range limits of the two-point correlation function,
$S_2(0)=\phi$ and $\lim_{r\rightarrow \infty} S_2(r)=\phi^2$ (for a homogeneous two-phase medium without long-range interactions),
as well as the slope at $r = 0$.
The latter is proportional to the specific surface.
For isotropic three-dimensional two-phase media, the derivative of $S_2(r)$ in the limit $r\rightarrow 0$ is $-s/4$~\cite{debye_scattering_1957, Torquato2002, footnote-limit-two-point}.

The two-point correlation function $S_2(r)$ appears smooth {for the MRJ state as well as overlapping and equilibrium hard spheres}, and indeed they are continuous and differentiable.
However, the contribution from a single sphere, \ie, the probability that two random points lie in the same sphere, is nonzero only for $r<D$.
For hard spheres, it is proportional to $(D-r)^2$; see~Eq.~\eqref{eq:pii}.
Therefore, the second derivate does not exist at $r=2D$.
(The same can be shown for overlapping spheres using the explicit expressions from Ref.~\cite{Torquato2002}.)

For overlapping spheres, the two-point correlation function is for $r > D$ constant and equal to the long-range limit.
This is because two points at a distance larger than the diameter $D$ of a single sphere cannot belong to the same sphere.
Therefore, the event that one of the test points is inside a sphere is independent of the other point.
There are no anticorrelations in $S_2(r)$ of overlapping spheres.
However, both hard-sphere packings exhibit positive and negative correlations.

For equilibrium hard spheres, Fig.~\ref{fig:S2} compares our results at $\phi=0.478$ to Monte Carlo estimates by \citet{Torquato2002} 
of an equilibrium hard-sphere liquid at $\phi=0.50$ with no detectable crystals.
Their qualitative behavior agrees very well. The functional values from Ref.~\cite{Torquato2002} are slightly larger because of the larger packing fraction.

\subsection{Surface-void correlation function}
\label{sec:Fsv}

The surface-void and surface-surface correlation functions allow for improved bounds on the trapping constant~\cite{RubinsteinTorquato1988, TorquatoRubinstein1989, Doi1976}
and the fluid permeability~\cite{RubinsteinTorquato1989JFM, Torquato1991Review}.

\begin{figure}[t]
  \centering
  \includegraphics[width=\linewidth]{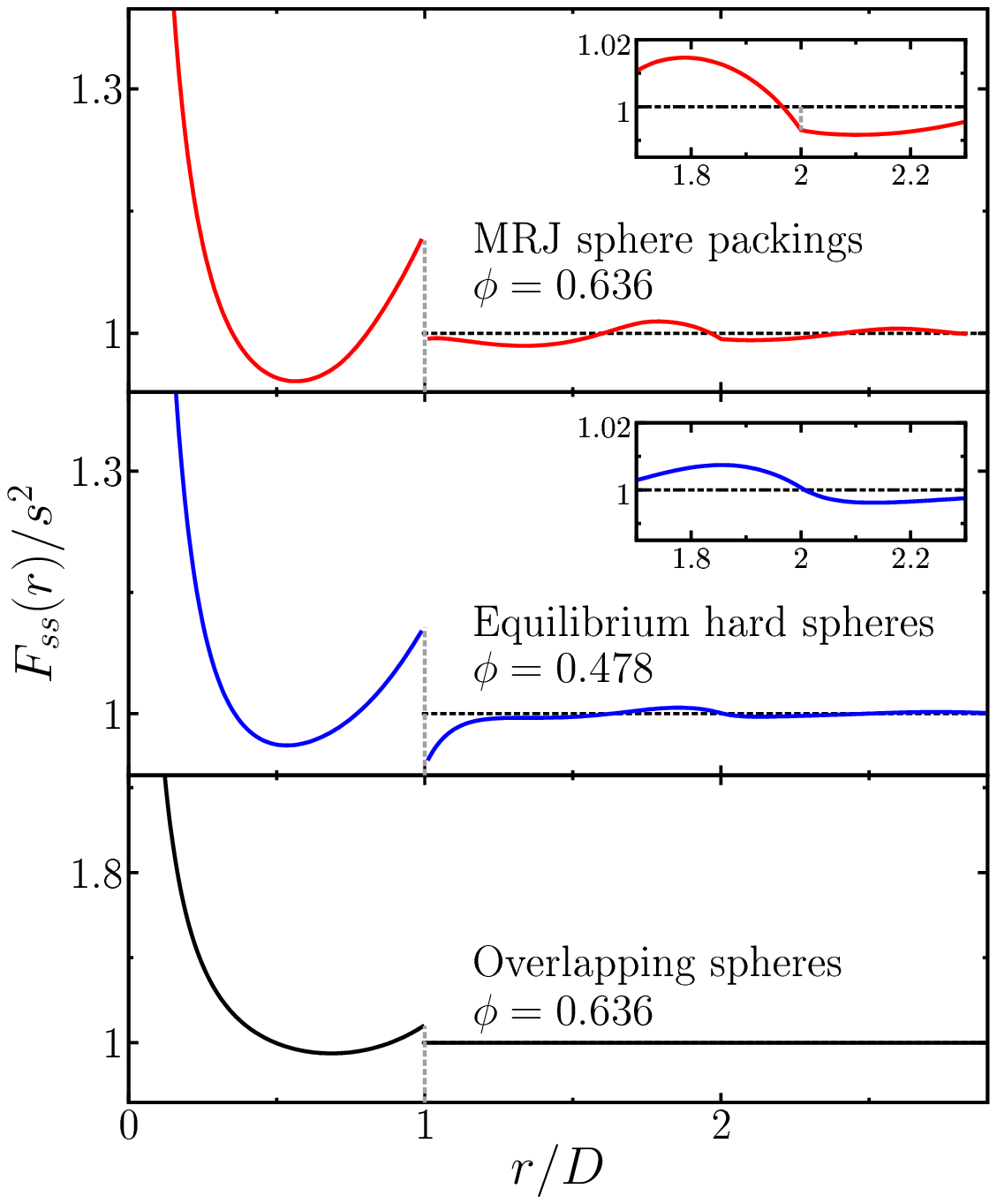}
  \caption{(Color online) Surface-surface correlation functions $F_{ss}(r)$ (rescaled
  by the square of the specific surface $s$) for overlapping spheres ($\phi = 0.636$, $s=2.21/D$), an
  equilibrium hard-sphere liquid ($\phi=0.478$, $s=2.87/D$), and
  MRJ sphere packings ($\phi=0.636$, $s=3.81/D$). For details; see Fig.~\ref{fig:S2}.
  The insets magnify $F_{ss}(r)/s^2$ at $r/D = 2$,
  where the derivative of $F_{ss}$ is discontinuous for the MRJ sphere packings in contrast to the equilibrium hard spheres.}
  \label{fig:Fss}
\end{figure}

\begin{table}[t]%
  \caption{\label{tab:correlations-equi}%
    Probability and correlation functions of the equilibrium hard-sphere liquid (with $\phi=0.4780$)
    corresponding to Figs.~\ref{fig:S2}, \ref{fig:Fsv}, \ref{fig:Fss}, and \ref{fig:exclusion_probability}.
    The specific surface is $s=2.868/D$.
    The statistical errors are smaller than the accuracy of the here presented data.
    For the correlation functions, they are mostly below $3\cdot 10^{-4}$.
    For the exclusion probability, they mainly range between $2\cdot 10^{-5}$ and $10^{-7}$.}
  \centering
  \begin{ruledtabular}
    \begin{tabular}{c c c c c}
       {$r/D$} & {$S_2(r)$} & {$F_{sv}(r)/s$} & {$F_{ss}(r)/s^2$} & {$E_V(r)$} \\
       \hline \\[-1.9ex]
      0    & 0.4780 & $1/2$ & {$\infty$} & 1 \\
      0.09 & 0.4148 & 0.5279 & 2.135 & 0.9972 \\
      0.18 & 0.3577 & 0.5332 & 1.303 & 0.9777 \\
      0.27 & 0.3096 & 0.5274 & 1.081 & 0.9247 \\
      0.36 & 0.2716 & 0.5172 & 0.999 & 0.8216 \\
      0.45 & 0.2435 & 0.5071 & 0.968 & 0.6515 \\
      0.54 & 0.2245 & 0.4998 & 0.961 & 0.4014 \\
      0.56 & 0.2213 & 0.4988 & 0.961 & 0.3401 \\
      0.58 & 0.2186 & 0.4980 & 0.962 & 0.2809 \\
      0.60 & 0.2163 & 0.4975 & 0.964 & $2.26\cdot 10^{-1}$ \\ 
      0.62 & 0.2143 & 0.4973 & 0.966 & $1.76\cdot 10^{-1}$ \\ 
      0.63 & 0.2134 & 0.4972 & 0.968 & $1.54\cdot 10^{-1}$ \\ 
      0.72 & 0.2092 & 0.5006 & 0.985 & $3.05\cdot 10^{-2}$ \\ 
      0.81 & 0.2107 & 0.5104 & 1.013 & $2.63\cdot 10^{-3}$ \\ 
      0.90 & 0.2170 & 0.5266 & 1.053 &  \\ 
      0.99 & 0.2275 & 0.5492 & 1.103 &  \\
      1.08 & 0.2372 & 0.5380 & 0.976 &  \\
      1.17 & 0.2397 & 0.5273 & 0.991 &  \\
      1.26 & 0.2374 & 0.5207 & 0.994 &  \\
      1.35 & 0.2328 & 0.5170 & 0.995 &  \\
      1.44 & 0.2282 & 0.5156 & 0.995 &  \\
      1.53 & 0.2249 & 0.5159 & 0.997 &  \\
      1.62 & 0.2234 & 0.5177 & 0.999 &  \\
      1.71 & 0.2239 & 0.5204 & 1.003 &  \\
      1.80 & 0.2257 & 0.5234 & 1.007 &  \\
      1.89 & 0.2281 & 0.5259 & 1.007 &  \\
      1.98 & 0.2303 & 0.5267 & 1.003 &  \\
      2.07 & 0.2316 & 0.5251 & 0.997 &  \\
      2.16 & 0.2315 & 0.5225 & 0.996 &  \\
      2.25 & 0.2303 & 0.5206 & 0.997 &  \\
      2.34 & 0.2288 & 0.5196 & 0.998 &  \\
      2.43 & 0.2274 & 0.5196 & 0.999 &  \\
      2.52 & 0.2267 & 0.5203 & 1.000 &  \\
      2.61 & 0.2268 & 0.5214 & 1.001 &  \\
      2.70 & 0.2274 & 0.5225 & 1.002 &  \\
      2.79 & 0.2283 & 0.5233 & 1.002 &  \\
      {$\infty$} & 0.2285 & 0.5220 & 1 & 
    \end{tabular}
  \end{ruledtabular}
\end{table}

\begin{table}[t]%
  \caption{\label{tab:correlations}%
    Probability and correlation functions the MRJ hard-sphere packings (with $\phi=0.6356$)
    corresponding to Figs.~\ref{fig:S2}, \ref{fig:Fsv}, \ref{fig:Fss}, and \ref{fig:exclusion_probability}.
    The specific surface is $s=3.814/D$.
    The statistical errors are smaller than the accuracy of the here presented data.
    For the correlation functions, they are mostly below $2\cdot 10^{-4}$.
    For the exclusion probability, they mainly range between $5\cdot 10^{-6}$ and $10^{-7}$.}
  \centering
  \begin{ruledtabular}
    \begin{tabular}{c c c c c}
       {$r/D$} & {$S_2(r)$} & {$F_{sv}(r)/s$} & {$F_{ss}(r)/s^2$} & {$E_V(r)$} \\
       \hline \\[-1.9ex]
      0    & 0.6356 & $1/2$ & {$\infty$} & 1 \\
      0.09 & 0.5583 & 0.4680 & 1.958 & 0.9963 \\
      0.18 & 0.4983 & 0.4328 & 1.282 & 0.9703 \\
      0.27 & 0.4542 & 0.4005 & 1.080 & 0.8999 \\
      0.36 & 0.4238 & 0.3739 & 0.994 & 0.7628 \\
      0.45 & 0.4046 & 0.3543 & 0.955 & 0.5367 \\
      0.54 & 0.3942 & 0.3427 & 0.942 & 0.2225 \\
      0.56 & 0.3928 & 0.3412 & 0.941 & 0.1630 \\
      0.58 & 0.3917 & 0.3402 & 0.942 & 0.1138 \\
      0.60 & 0.3909 & 0.3395 & 0.943 & $7.60\cdot 10^{-2}$ \\ 
      0.62 & 0.3904 & 0.3393 & 0.944 & $4.83\cdot 10^{-2}$ \\ 
      0.63 & 0.3902 & 0.3393 & 0.945 & $3.77\cdot 10^{-2}$ \\ 
      0.72 & 0.3907 & 0.3438 & 0.965 & $1.99\cdot 10^{-3}$ \\ 
      0.81 & 0.3945 & 0.3547 & 1.002 & $2.45\cdot 10^{-5}$ \\ 
      0.90 & 0.4009 & 0.3710 & 1.053 &  \\ 
      0.99 & 0.4094 & 0.3930 & 1.116 &  \\
      1.08 & 0.4144 & 0.3807 & 0.993 &  \\
      1.17 & 0.4118 & 0.3682 & 0.989 &  \\
      1.26 & 0.4067 & 0.3602 & 0.986 &  \\
      1.35 & 0.4021 & 0.3563 & 0.985 &  \\
      1.44 & 0.3995 & 0.3559 & 0.987 &  \\
      1.53 & 0.3992 & 0.3583 & 0.993 &  \\
      1.62 & 0.4006 & 0.3623 & 1.002 &  \\
      1.71 & 0.4029 & 0.3666 & 1.012 &  \\
      1.80 & 0.4052 & 0.3699 & 1.015 &  \\
      1.89 & 0.4068 & 0.3711 & 1.010 &  \\
      1.98 & 0.4072 & 0.3689 & 0.997 &  \\
      2.07 & 0.4060 & 0.3640 & 0.992 &  \\
      2.16 & 0.4040 & 0.3611 & 0.992 &  \\
      2.25 & 0.4024 & 0.3603 & 0.994 &  \\
      2.34 & 0.4018 & 0.3612 & 0.997 &  \\
      2.43 & 0.4022 & 0.3632 & 1.001 &  \\
      2.52 & 0.4032 & 0.3653 & 1.005 &  \\
      2.61 & 0.4043 & 0.3668 & 1.006 &  \\
      2.70 & 0.4051 & 0.3672 & 1.004 &  \\
      2.79 & 0.4053 & 0.3664 & 1.000 &  \\
      {$\infty$} & 0.4040 & 0.3644 & 1 & 
    \end{tabular}
  \end{ruledtabular}
\end{table}

Figure~\ref{fig:Fsv} compares the surface-void correlation function $F_{sv}(r)$ for MRJ sphere packings, see Eq.~\eqref{eq:Fsv}, to those of overlapping and equilibrium hard spheres.
For overlapping spheres, the surface-void correlation function is known analytically~\cite[\eg][p. 125]{Torquato2002}.
The curve for the equilibrium hard-sphere liquid agrees with previous findings in Refs.~\cite{Torquato1986, SeatonGlandt1986}.
If the surface-void correlation function is divided by the specific surface, the ratio takes on only values between zero and one like a probability.
For a two-phase medium with a sufficiently smooth boundary,
the surface-void correlation function converges for $r\rightarrow 0$ to $F_{sv}(0) = s/2$~\cite{Torquato2002}.

For a homogeneous random two-phase media without long-range correlations, it converges in the limit of large {distances} to
$\lim_{r\rightarrow\infty}F_{sv}(r) = s(1-\phi)$, where $\phi$ is the volume fraction of the solid phase.
Both limits are depicted by dashed lines in Fig.~\ref{fig:Fsv}.

The surface-void correlation functions $F_{sv}(r)$ are continuous.
{However, in contrast to $S_2(r)$, they are not smooth at $r=D$.
The discontinuity in the first derivative stems from the contributions of events where the interior and surface of the same sphere are hit.
For example, for hard spheres these contributions to $F_{sv}(r)$ are proportional to $(D-r)$; see Eq.~\eqref{eq:qii}.}

As for the two-point correlation function, the surface-void correlation function of overlapping spheres takes on the value of the long-range limit for all $r > D$.
This is again because two different spheres are independent of each other.

{The derivative of $F_{sv}(r)$ for small distances $r\rightarrow 0$ has a different sign for the MRJ and equilibrium hard spheres, which is mainly due to the different global packing fraction (above or below 0.5).
However, there is also a more interesting and subtle difference in the slope at $r=0$.
A distinct signature of contacts between spheres can be found in the two-body contribution $F^*_{sv}(r)$ (see Appendix~\ref{sec:2body}) because at least in finite packings, the slope of $F^*_{sv}(r)$ at $r=0$ can be related to the mean contact number.
Therefore, this slope vanishes for the equilibrium but not for the MRJ sphere packings.}

\subsection{Surface-surface correlation function}
\label{sec:Fss}

Figure~\ref{fig:Fss} compares the surface-surface correlation function $F_{ss}(r)$ for MRJ sphere packings [cf. Eq.~\eqref{eq:Fss}] to those of overlapping spheres and an equilibrium hard-sphere liquid.
For overlapping spheres, also the surface-surface correlation function is known analytically~\cite[\eg,][p. 125]{Torquato2002}.
Like $S_2(r)$ and $F_{sv}(r)$, $F_{ss}(r)$ is for $r>D$ constant and equal to the long-range limit.
The surface-surface correlation functions for the equilibrium hard-sphere liquid also agree with previous findings in Refs.~\cite{Torquato1986, SeatonGlandt1986}.
{The long-range limit is $\lim_{r\rightarrow \infty} F_{ss}(r) = s^2$ (indicated by a dashed line in Fig.~\ref{fig:Fss}).
For $r\rightarrow 0$, the surface-surface correlation function diverges
because the probability to find a single point in the shell of thickness
$\epsilon$ only vanished like $\epsilon$ but it is rescaled by
$\epsilon^2$.}
{The surface-surface correlation functions are discontinuous at $r=D$, because the single-body contribution is discontinuous; for hard spheres, see Eq.~\eqref{eq:aii}.}

{Interestingly, it is only for the MRJ sphere packings that the surface-surface correlation of the MRJ sphere packings is not smooth at $r=2D$,
which is caused by the contacts between the spheres. 
More precisely, the first derivative is discontinuous at $r=2D$, which we can rigorously relate to the Dirac delta contribution of $g_2(r)$ at $r=D$; see Appendix~\ref{sec:2body}.
The spheres at contact with each other also cause at $r=D$ a discontinuity in the derivative of two-body contributions $F^*_{ss}(r)$ to the surface-surface correlation functions.
Moreover, the functional value of the two-body contributions at $r=0$ can be related to the mean contact number.
It therefore only vanishes for the equilibrium liquid but not for the MRJ sphere packings; see Appendix~\ref{sec:2body}.}

Tables~\ref{tab:correlations-equi} and \ref{tab:correlations} list numerical values of the two-point, surface-void, and surface-surface correlation functions for both the equilibrium and MRJ sphere packings;
see also the Supplemental Material for estimates of the correlation functions at more radial distances~\cite{supplementary_material}.

\section{Spectral density}
\label{sec:spectral-density}

\begin{figure}[t]
  \centering
  \includegraphics[width=\linewidth]{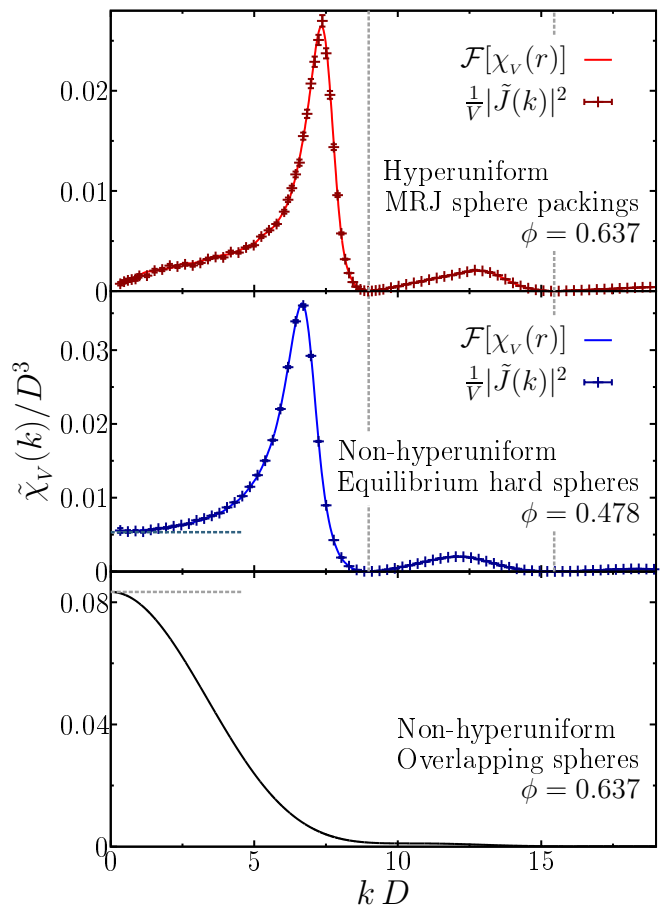}
  \caption{(Color online) Spectral densities of the overlapping spheres (bottom), equilibrium hard sphere liquid (center) and MRJ
    sphere packings (top): for the hard spheres, they are calculated by a Fourier transformation of either the autocovariance (solid line), see
    Eq.~\eqref{eq:chi-FT}, or directly of the sphere packings themselves (crosses), see
    Eq.~\eqref{eq:chi-Indicator}. For the overlapping and equilibrium hard spheres, the dashed horizontal lines indicate the value in the infinite wavelength limit.
    The MRJ state is hyperuniform, therefore the structure factor vanishes for $k\rightarrow 0$.
    For both hard-sphere systems, each analyzed packing contains {10000} spheres.
    The dashed vertical lines indicate the zeros of the spectral densities that are universal for all disordered hard-sphere packings.}
  \label{fig_spectral_density}
\end{figure}

As mentioned in the Introduction, MRJ packings possess---in contrast to the equilibrium hard-sphere liquid below the freezing transition---the singular property of hyperuniformity~\cite{TorquatoStillinger2003, ZacharyTorquato2009};
for detailed discussion of this exotic state of matter, see Refs.~\cite{TorquatoStillinger2003, DonevEtAl2005, ZacharyTorquato2009, TorquatoStillinger2010RevModPhys, DreyfusEtAl2015, Torquato2016b}.
Large-scale density fluctuations or volume-fraction fluctuations are anomalously suppressed~\cite{TorquatoStillinger2003, ZacharyTorquato2009, DonevEtAl2005, ZacharyJiaoTorquato2011}.
Therefore, not only are MRJ packings characterized by short-range order, but they can be regarded to possess a ``hidden long-range order'' due to the global hyperuniformity property.

In a hyperuniform point process, the structure factor $S(k)$ vanishes as the wavenumber $k$ tends to zero,
\begin{align}
  \lim_{k\rightarrow0}S(k)=0.
\end{align}
For a monodisperse packing of hard spheres, $S(k)$ of the sphere centers is directly proportional to the spectral density as explained in Sec.~\ref{sec:define-spectral-density}; see Eq.~\eqref{eq:chi-Indicator}.
Because the Fourier transform of a single sphere $\tilde{m}(k)$ converges for $k\rightarrow 0$ to a constant~\footnote{For three dimensions, $\lim_{k\rightarrow 0}\tilde{m}(k) = \pi D^3/6$.} that is strictly greater than zero,
the spectral density vanishes for $k\rightarrow 0$ if and only if the structure factor vanishes $\lim_{k\rightarrow0}S(k)=0$.
Hyperuniformity can therefore also be detected by a vanishing spectral density in the limit of short wave vectors (\ie, long wavelengths):
\begin{align}
  \lim_{k\rightarrow 0} \tilde{\chi}_{_V}(k) = 0.
  \label{eq:hyperuniform-def-by-vanishing-spectral-density}
\end{align}
This latter definition of hyperuniformity can also be applied to polydisperse packings and even more general two-phase media~\cite{zachary_hyperuniformity_2011}.
It is equivalent to a quasi-long-range asymptotic behavior of the variance $\sigma_V^2(R)$ of the packing (or volume) fraction within a spherical window of radius $R$ that is placed randomly into the sample.
For hyperuniform heterogeneous materials, this variance goes for large $R$ faster to zero than the inverse of the volume of the observation window, \ie, faster than $1/R^{d}$.
This in contrast to, \eg, overlapping spheres or equilibrium hard spheres.

Here, we determine the spectral density not only to examine the hyperuniformity of the MRJ sphere packings
but also to obtain the Fourier representation of the two-point correlation function, which is useful for evaluating rigorous bounds on physical properties.
We compare the spectral density of the hyperuniform MRJ packings to the nonhyperuniform overlapping spheres and equilibrium hard spheres.

For overlapping spheres, the spectral density can easily be calculated by numerical integration using the explicit analytical expressions for the two-point correlation function given, \eg, in Ref.~\cite[][p. 122]{Torquato2002}.
For hard-sphere packings, there are two different approaches to compute the spectral density, as described in Ref.~\cite{DreyfusEtAl2015}:
first, by an explicit calculation of the Fourier transform of the autocovariance function, cf. Eq.~\eqref{eq:chi-FT};
second, by a direct Fourier transformation of the indicator function of the particle phase, cf. Eq.~\eqref{eq:chi-via-Ft-of-indicator}.

In the first approach, the Fourier transform is calculated by a numerical integration of the curves in Fig.~\ref{fig:S2} (after subtracting the long-range limit).
Because the simulation boxes are finite, a cutoff is assumed for the autocovariance $\chi_{_V}(r)$, which induces a minimal absolute value $k$ of the wave vector that is reliable~\footnote{Below
  this minimal value of $k$, the Fourier transform strongly depends on the cutoff.}.  

In the second approach, we use the relation between the Fourier transform of the indicator function and the structure factor; see Eq.~\eqref{eq:chi-Indicator}.
Therefore, no binning or discretization of the sample is needed.
Moreover, we consider the spectral density as a function of the wave vector $\vec{k}$ and explicitly take the nonorthogonal simulation boxes into account.
We evaluate for each sample the structure factor for all wave vectors that are allowed in a simulation box with periodic boundary conditions.
These are integer multiples of the reciprocal lattice vectors.
Collecting the data for all samples, we finally bin the results for the spectral density w.r.t. the absolute value $k$ of the wave vector with a bin width of $\Delta k \approx 0.133$.

\begin{table}[t]%
  \caption{\label{tab:spectral-density-equi}%
    Spectral density derived from the direct Fourier transformation of the periodic simulation box (corresponding to the crosses in Fig.~\ref{fig_spectral_density}) of the equilibrium hard-sphere liquid.
    The allowed values of the wave vector are binned w.r.t. their absolute value $k$ and the resulting spectral densities are then averaged over 100 different packings.
    The third and sixth column display the standard error of the mean.}
  \centering
  \begin{ruledtabular}
    \begin{tabular}{c c c c c c}
      {$k\cdot D$} & {$\tilde{\chi}_{_V}(k)/D^3$} & {$\sigma[\tilde{\chi}_{_V}(k)]/D^3$} & {$k\cdot D$} & {$\tilde{\chi}_{_V}(k)/D^3$} & {$\sigma[\tilde{\chi}_{_V}(k)]/D^3$} \\
      \cline{1-3}\cline{4-6} \\[-1.9ex]
      0.6 & $5.40\cdot 10^{-3}$ & $1\cdot 10^{-4}$ &       9.6 & $2.36\cdot 10^{-4}$ & $6\cdot 10^{-7}$  \\
      1.1 & $5.60\cdot 10^{-3}$ & $1\cdot 10^{-4}$ &       10.2 & $6.47\cdot 10^{-4}$ & $2\cdot 10^{-6}$ \\
      1.7 & $5.88\cdot 10^{-3}$ & $7\cdot 10^{-5}$ &       10.8 & $1.14\cdot 10^{-3}$ & $2\cdot 10^{-6}$ \\
      2.3 & $6.17\cdot 10^{-3}$ & $6\cdot 10^{-5}$ &       11.3 & $1.63\cdot 10^{-3}$ & $3\cdot 10^{-6}$ \\
      2.8 & $6.68\cdot 10^{-3}$ & $5\cdot 10^{-5}$ &       11.9 & $1.98\cdot 10^{-3}$ & $4\cdot 10^{-6}$ \\
      3.4 & $7.37\cdot 10^{-3}$ & $4\cdot 10^{-5}$ &       12.5 & $1.93\cdot 10^{-3}$ & $4\cdot 10^{-6}$ \\
      4.0 & $8.42\cdot 10^{-3}$ & $4\cdot 10^{-5}$ &       13.0 & $1.46\cdot 10^{-3}$ & $3\cdot 10^{-6}$ \\
      4.5 & $1.00\cdot 10^{-2}$ & $5\cdot 10^{-5}$ &       13.6 & $8.60\cdot 10^{-4}$ & $2\cdot 10^{-6}$ \\
      5.1 & $1.31\cdot 10^{-2}$ & $6\cdot 10^{-5}$ &       14.2 & $3.66\cdot 10^{-4}$ & $1\cdot 10^{-6}$ \\
      5.7 & $1.83\cdot 10^{-2}$ & $6\cdot 10^{-5}$ &       14.7 & $9.56\cdot 10^{-5}$ & $3\cdot 10^{-7}$ \\
      6.2 & $2.90\cdot 10^{-2}$ & $1\cdot 10^{-4}$ &       15.3 & $5.09\cdot 10^{-6}$ & $4\cdot 10^{-8}$ \\
      6.8 & $3.46\cdot 10^{-2}$ & $1\cdot 10^{-4}$ &       15.9 & $1.74\cdot 10^{-5}$ & $1\cdot 10^{-7}$ \\
      7.4 & $1.32\cdot 10^{-2}$ & $4\cdot 10^{-5}$ &       16.4 & $8.61\cdot 10^{-5}$ & $5\cdot 10^{-7}$ \\
      7.9 & $2.68\cdot 10^{-3}$ & $8\cdot 10^{-6}$ &       17.0 & $1.75\cdot 10^{-4}$ & $1\cdot 10^{-6}$ \\
      8.5 & $3.23\cdot 10^{-4}$ & $1\cdot 10^{-6}$ &       17.6 & $2.71\cdot 10^{-4}$ & $2\cdot 10^{-6}$ \\
      9.1 & $8.60\cdot 10^{-6}$ & $5\cdot 10^{-8}$ &            &                     & 
    \end{tabular}
  \end{ruledtabular}
\end{table}

\begin{table}[t]%
  \caption{\label{tab:spectral-density}%
    Spectral density derived from the direct Fourier transformation of the periodic simulation box (corresponding to the crosses in Fig.~\ref{fig_spectral_density}) of the MRJ hard-sphere packings.
    The allowed values of the wave vector are binned w.r.t. their absolute value $k$ and the resulting spectral densities are then averaged over 14 different packings.
    The third and sixth column display the standard error of the mean.}
  \centering
  \begin{ruledtabular}
    \begin{tabular}{c c c c c c}
      {$k\cdot D$} & {$\tilde{\chi}_{_V}(k)/D^3$} & {$\sigma[\tilde{\chi}_{_V}(k)]/D^3$} & {$k\cdot D$} & {$\tilde{\chi}_{_V}(k)/D^3$} & {$\sigma[\tilde{\chi}_{_V}(k)]/D^3$} \\
      \cline{1-3}\cline{4-6} \\[-1.9ex]
      0.6 & $9.36\cdot 10^{-4}$ & $9\cdot 10^{-5}$ &	      10.2 & $6.03\cdot 10^{-4}$ & $1\cdot 10^{-5}$ \\
      1.2 & $1.63\cdot 10^{-3}$ & $2\cdot 10^{-4}$ &	      10.8 & $9.67\cdot 10^{-4}$ & $2\cdot 10^{-5}$ \\
      1.8 & $2.29\cdot 10^{-3}$ & $1\cdot 10^{-4}$ &	      11.4 & $1.35\cdot 10^{-3}$ & $2\cdot 10^{-5}$ \\
      2.4 & $2.43\cdot 10^{-3}$ & $2\cdot 10^{-4}$ &	      12.0 & $1.69\cdot 10^{-3}$ & $3\cdot 10^{-5}$ \\
      3.0 & $2.96\cdot 10^{-3}$ & $2\cdot 10^{-4}$ &	      12.6 & $2.02\cdot 10^{-3}$ & $2\cdot 10^{-5}$ \\
      3.6 & $3.39\cdot 10^{-3}$ & $9\cdot 10^{-5}$ &	      13.2 & $1.86\cdot 10^{-3}$ & $2\cdot 10^{-5}$ \\
      4.2 & $3.83\cdot 10^{-3}$ & $1\cdot 10^{-4}$ &	      13.8 & $1.11\cdot 10^{-3}$ & $1\cdot 10^{-5}$ \\
      4.8 & $4.37\cdot 10^{-3}$ & $9\cdot 10^{-5}$ &	      14.4 & $4.13\cdot 10^{-4}$ & $3\cdot 10^{-6}$ \\
      5.4 & $5.55\cdot 10^{-3}$ & $2\cdot 10^{-4}$ &	      15.0 & $6.08\cdot 10^{-5}$ & $1\cdot 10^{-6}$ \\
      6.0 & $8.10\cdot 10^{-3}$ & $2\cdot 10^{-4}$ &	      15.6 & $5.31\cdot 10^{-6}$ & $2\cdot 10^{-7}$ \\
      6.6 & $1.38\cdot 10^{-2}$ & $4\cdot 10^{-4}$ &	      16.2 & $6.41\cdot 10^{-5}$ & $7\cdot 10^{-7}$ \\
      7.2 & $2.35\cdot 10^{-2}$ & $7\cdot 10^{-4}$ &	      16.8 & $1.41\cdot 10^{-4}$ & $1\cdot 10^{-6}$ \\
      7.8 & $1.34\cdot 10^{-2}$ & $4\cdot 10^{-4}$ &	      17.4 & $2.17\cdot 10^{-4}$ & $2\cdot 10^{-6}$ \\
      8.4 & $1.10\cdot 10^{-3}$ & $4\cdot 10^{-5}$ &	      18.0 & $2.99\cdot 10^{-4}$ & $5\cdot 10^{-6}$ \\
      9.0 & $9.28\cdot 10^{-6}$ & $5\cdot 10^{-7}$ &	      18.6 & $3.77\cdot 10^{-4}$ & $4\cdot 10^{-6}$ \\
      9.6 & $2.37\cdot 10^{-4}$ & $6\cdot 10^{-6}$ &	      19.2 & $4.08\cdot 10^{-4}$ & $8\cdot 10^{-6}$
    \end{tabular}
  \end{ruledtabular}
\end{table}

Figure~\ref{fig_spectral_density} compares the results for the spectral density from the two different approaches.
They are in excellent agreement with each other for both the MRJ (top) and equilibrium hard spheres (center).
Tables~\ref{tab:spectral-density-equi} and \ref{tab:spectral-density} provide estimates of the spectral density via the second approach for equilibrium or MRJ sphere packings, respectively;
see also the Supplemental Material for estimates of the spectral density for further absolute values of the wave vector~\cite{supplementary_material}.

A test for the accuracy of our data is given by the zeros of the spectral density.
According to Eq.~\eqref{eq:chi-Indicator}, the zeros of the spectral density are given by the zeros of the structure factor and the zeros of $\tilde{m}(k)$.
The latter are given by the zeros of the Bessel function $J_{d/2}(\frac{kD}{2})$; see Eq.~\eqref{eq:mk}. 
Therefore, they are universal for all disordered packings free of any Dirac delta (or Bragg) peaks in their spectral density~\cite{Torquato2016a, Torquato2016b}.
The dashed vertical lines in Fig.~\ref{fig_spectral_density} indicate these exact positions.
They are in excellent agreement with the simulation results.

In the limit of small wave vectors $k\rightarrow 0$, the structure factor (and thus the spectral density) for equilibrium hard-sphere liquids can be related to the isothermal
compressibility $\kappa_T=\rho^{-1}(\partial \rho/\partial p)_T$ (with pressure $p$, temperature $T$, and number density $\rho$);
\begin{align}
  S(0) = \rho k_B T \kappa_T,
\end{align}
where $k_B$ is the Boltzmann constant. 
The right-hand side can be well estimated by using accurate analytical approximate formulas for the pressure of equilibrium hard spheres~\cite{HansenMcDonald2006}.
This is indicated by the dashed horizontal line in Fig.~\ref{fig_spectral_density}.
Because the hard-sphere liquid has a positive compressibility, the spectral density does not vanish in
the limit of infinite wavelength $\tilde{\chi}_{_V}(0)>0$ and hence is not hyperuniform, which translates into a volume-fraction variance that asymptotically decreases like $R^{-3}$.
This is qualitatively the same behavior as for overlapping spheres (see bottom of Fig.~\ref{fig_spectral_density}, where the dashed gray line indicates $\tilde{\chi}_{_V}(0)/D^3$).

In contrast to this, the spectral density for the hyperuniform MRJ packings should vanish for short wave vectors according to the definition in Eq.~\eqref{eq:hyperuniform-def-by-vanishing-spectral-density}.
Of course, for any finite packing derived from simulations there is a smallest accessible wave vector at which reliable estimates of the spectral density can be measured.
For a careful extrapolation of the spectral density to $k=0$ as well as a detailed discussion of binning effects, noise at the smallest wavenumbers, and numerical and protocol-dependent errors, see Ref.~\cite{AtkinsonEtAl2016}.

Within the scope of this paper, we only compare the binned spectral densities of the MRJ packings to those of the equilibrium hard spheres.
We use simulations with the same system size and therefore analyze in this section 14 MRJ packings with {10000}\ spheres.
For small wave vectors, the spectral density of the MRJ state is distinctly smaller than that of the equilibrium hard-sphere liquid.
The former vanishes at least approximately as $k\rightarrow 0$.

\section{Pore-size distribution}
\label{sec:pore-size-distribution}

The pore-size distribution contains at least a coarse level of connectedness information about the matrix phase~\cite{Torquato2002}.
Its lower-order moments arise in bounds on the mean survival and principal relaxation times~\cite{Prager1963,TorquatoAvellaneda1991}.

We estimate the complementary cumulative pore-size distribution $F(\delta)$ and the exclusion probability $E_V$ as an equivalent representation.
We compare the results for MRJ sphere packings not only to those for overlapping and equilibrium hard spheres but also for crystalline sphere packings.
Moreover, we directly estimate the first and second moments of the pore-size distribution.

\subsection{The complementary cumulative pore-size distribution}
\label{sec:compl-cum-pore-size}

\begin{figure}[t]
  \centering
  \includegraphics[width=\linewidth]{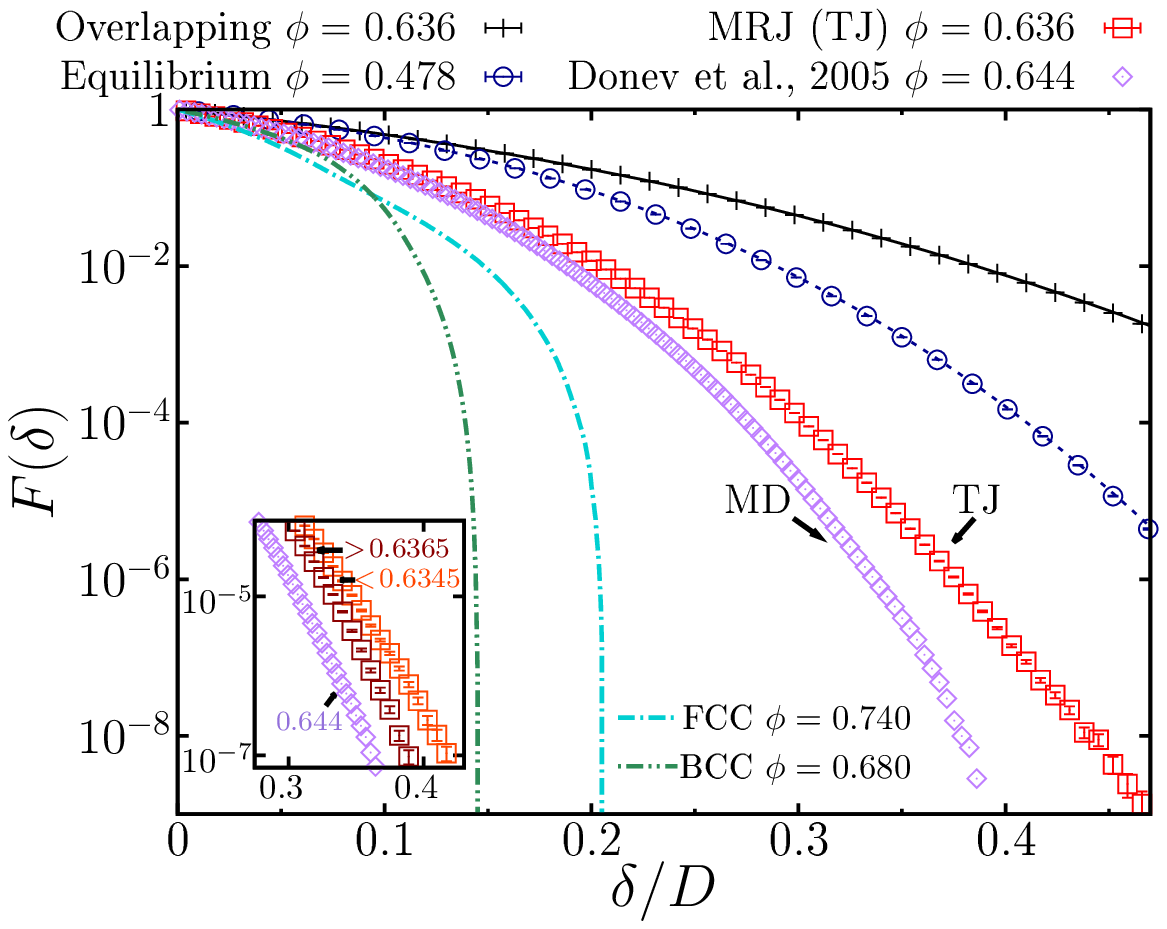}
  \caption{(Color online) Complementary cumulative pore-size distribution $F(\delta)$ in MRJ sphere packings compared to those in overlapping and equilibrium hard spheres as well as face-centered-cubic (FCC) and body-centered-cubic (BCC) lattices:
    the solid (black) and dashed (blue) lines show the analytical curves for overlapping and equilibrium hard spheres, respectively.
    They are in excellent agreement with simulation results (black and blue points).
    The MRJ packings produced by the Torquato-Jiao (TJ) sphere packing algorithm are also compared to results from molecular dynamics (MD) simulations; see Ref.~\cite{DonevEtAl2005}.
    The (quantitative) difference can be explained by slightly different global packing fractions.
    The inset shows that there is a strong variation in $F(\delta)$ if the TJ results are restricted to packings with either slightly larger or smaller global packing fractions than the average global packing fraction.}
  \label{fig_cum_pore-size_distr}
\end{figure}

Figure~\ref{fig_cum_pore-size_distr} compares $F(\delta)$ for the disordered systems of MRJ, equilibrium, and overlapping spheres to those of crystalline sphere packings, \eg, see Ref.~\cite{Torquato2010reformulation}.
These are perfectly ordered packings of monodisperse spheres that are arranged either on a face-centered cubic (FCC) or on a body-centered cubic (BCC) lattice.
The radius of the spheres is chosen such that neighboring spheres touch each other.
The FCC packing corresponds to the densest possible sphere packing~\cite{Hales2005}.
The BCC packing has the smallest known covering radius, see also the discussion in Sec.~\ref{sec:exclusion-probability}.

For the MRJ, equilibrium, and overlapping spheres, we estimate $F(\delta)$ by a simple Monte Carlo sampling.
Random points are placed in the matrix phase uniformly distributed ($10^7$ points per sample).
For each point, the smallest distance to a sphere is determined and recorded.
We determine the empirical histogram weighted by the total number of samples and the bin width as an estimate of the pore-size distribution.
The complementary cumulative pore-size distribution follows immediately according to Eq.~\eqref{eq:def-F}.

For the overlapping spheres, we compare in Fig.~\ref{fig_cum_pore-size_distr} these numerical results to the analytic curve.
The latter follows immediately from the definition of a Poisson point process, that is, the point process that describes the positions of the sphere centers~\cite{LastPenrose16}. 
For the equilibrium hard-sphere liquid, we compare the numerical results to an accurate analytical approximation~\cite{torquato_nearest-neighbor_1995, rintoul_hard-sphere_1998, Torquato2002}; see also Refs.~\cite{torquato_nearest-neighbor_1990, TorquatoAvellaneda1991}.
The numerical estimates agree very well with the analytical predictions.

As expected $F(\delta)$ decreases faster for the crystalline sphere packings than for the disordered systems.
Moreover, the complementary cumulative pore-size distribution decreases faster for the MRJ than for the equilibrium or overlapping spheres.
Note that in the latter systems the spheres occupy the same volume fraction as in the MRJ sphere packing.

For a two-phase medium, the pore-size distribution $P(\delta)$ always vanishes for $\delta\rightarrow\infty$.
For the equilibrium hard-sphere liquid below the freezing transition and for the overlapping spheres, $P(\delta)>0$ vanishes exponentially for large pore sizes $\delta$.
However, $P(\delta)>0$ and thus $F(\delta)>0$ for all finite values of $\delta$.
This is in contrast to the MRJ sphere packings and also to the crystalline sphere packings studied here.
These packings are saturated, that is, no additional sphere can be inserted in the system without intersecting any other sphere.
This implies that $P(\delta)$ is zero at least for all $\delta > R$, and thus, $F(\delta)=0$ at least for all $\delta>R$.

In Fig.~\ref{fig_cum_pore-size_distr}, we also compare our results for MRJ packings produced by the Torquato-Jiao (TJ) sphere packing algorithm~\cite{TorquatoJiao2010}
to those from Ref.~\cite{DonevEtAl2005}, which used molecular dynamics (MD) simulations.
The latter packings are also strictly jammed and saturated, for more details see Ref.~\cite{DonevEtAl2005}.
The complementary cumulative pore-size distribution decreases faster for the latter system, which can be expected for two reasons.
The MD simulations were carried out at a slightly larger packing fraction than that of the MRJ packings produced by the TJ algorithm.
The inset in Fig.~\ref{fig_cum_pore-size_distr} shows how a slight change in the packing fraction can strongly affect $F(\delta)$. 
The complementary cumulative pore-size distribution is shown for only those packings produced by the TJ algorithm where the final packing fraction is either slightly below or slightly above the average packing fraction of all samples.
However, the decrease of $F(\delta)$ is not only determined by the packing fraction as can be seen from the comparison of the FCC and BCC lattices.
The arrangement of spheres is crucial, and $F(\delta)$ decreases faster for a more ordered packing at the same packing fraction.
Because the TJ algorithm is, in contrast to the MD simulation, explicitly designed to find a maximally disordered sample,
the latter is expected to be more ordered which results in a faster decrease of $F(\delta)$.

\subsection{Mean pore size and second moment}
\label{sec:mean-pore-size}

For diffusion-controlled reactions among static traps, the mean survival
time $\tau$ and the principal relaxation time $T$ are intimately related
to the first and second moments of the pore-size probability density
function~\cite{TorquatoAvellaneda1991}. 

In particular, the mean survival time, {which is the mean time that a Brownian particle can diffuse in the void phase before it hits the solid phase}, is bounded from below by the mean
pore size~\cite{Prager1963}.  Moreover, if the mean survival time $\tau$
is rescaled by the diffusion constant $\mathcal{D}$ and the diameter $D$
of a single sphere, it can be very accurately predicted by the mean pore
size via a universal scaling law~\cite{TorquatoYeong1997}.

The principal diffusion relaxation time, {which is the largest diffusion relaxation time}, is bounded from below by the
second moment of pore-size function~\cite{TorquatoAvellaneda1991,
Torquato2002}.
We evaluate both the prediction of the mean survival time and the bound
on the principle relaxation time in the third paper of this series.

Moreover, the second moment is proportional to the so-called ``quantizer
error''~\cite{Torquato2010reformulation}. The latter is defined as the
mean squared distance from a random point in space to the nearest point
in the point process. Minimizing this quantizer error is, \eg,
important for an optimal meshing of space for numerical
applications~\cite{DuEtAl1999}, coding and cryptography~\cite{Bose2008},
and digital communications~\cite{Conway1998}.

We estimate the first and second moments of the pore-size distribution by the sample mean and sample variance of 
the pore sizes found in a MC sampling as described above in Sec.~\ref{sec:compl-cum-pore-size}.
For the overlapping spheres ($\phi = 0.636$), we estimate $\langle\delta\rangle \approx 0.115\,D$ and $\langle\delta^2\rangle \approx 0.021\,D^2$;
for the equilibrium liquid ($\phi=0.478$), $\langle\delta\rangle \approx 0.098\,D$ and $\langle\delta^2\rangle \approx 0.014\,D^2$,
and for the MRJ sphere packings ($\phi=0.636$),  $\langle\delta\rangle \approx 0.063\,D$ and $\langle\delta^2\rangle \approx 0.006\,D^2$.
The statistical errors in units of $D$ or $D^2$ are less than $10^{-3}$. 
The values for the overlapping spheres agree with the analytical results and those for equilibrium spheres agree with the corresponding analytical approximation.

\subsection{Exclusion probability}
\label{sec:exclusion-probability}

\begin{figure}[t]
  \centering
  \includegraphics[width=\linewidth]{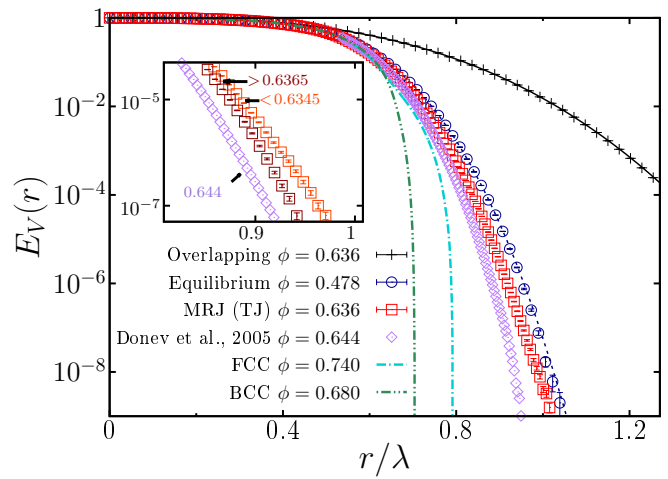}
  \caption{(Color online) The exclusion probability $E_V(r)$ of the sphere centers in MRJ packings is compared to those in overlapping spheres, equilibrium hard spheres, and crystalline sphere packings (FCC or BCC).
    The point patterns are compared at unit density (the unit of length is given by $\rho^{-1/3}$ where $\rho$ is the number density);
    for details, see Fig.~\ref{fig_cum_pore-size_distr}.}
  \label{fig:exclusion_probability}
\end{figure}

\begin{figure*}[t]
  \centering
  \subfigure[][]{\includegraphics[width=0.31\textwidth]{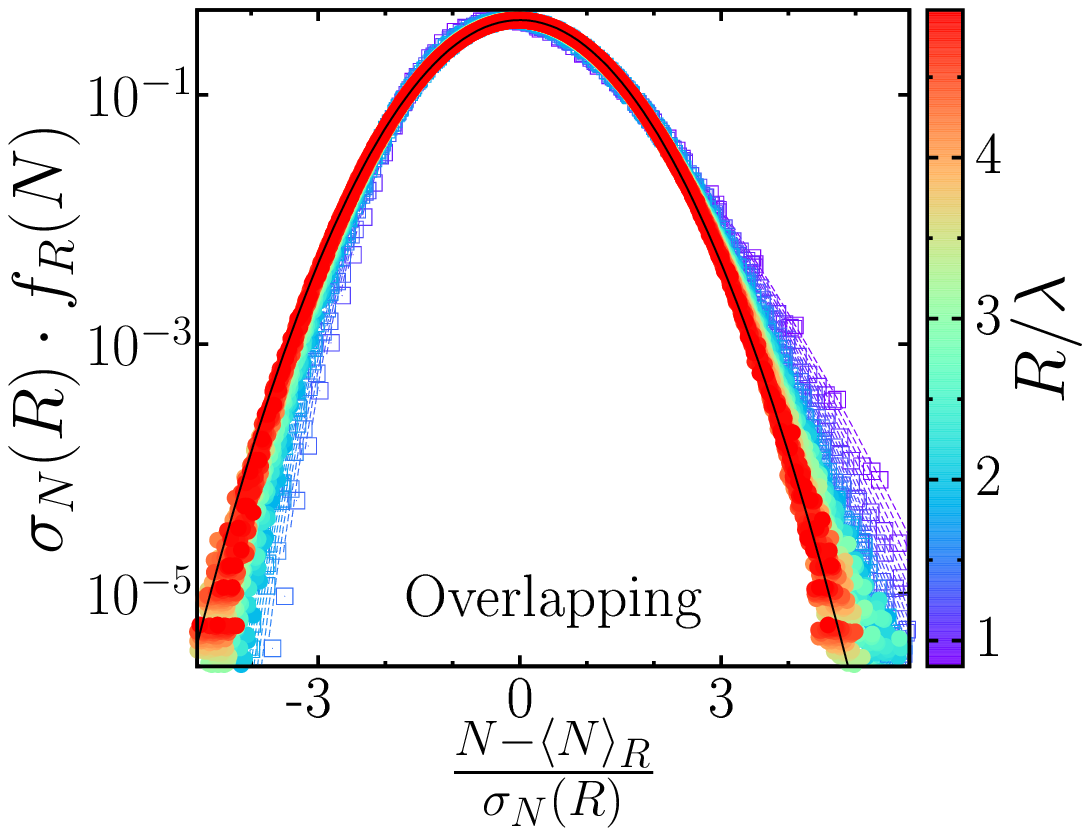}}
  \hfill
  \subfigure[][]{\includegraphics[width=0.31\textwidth]{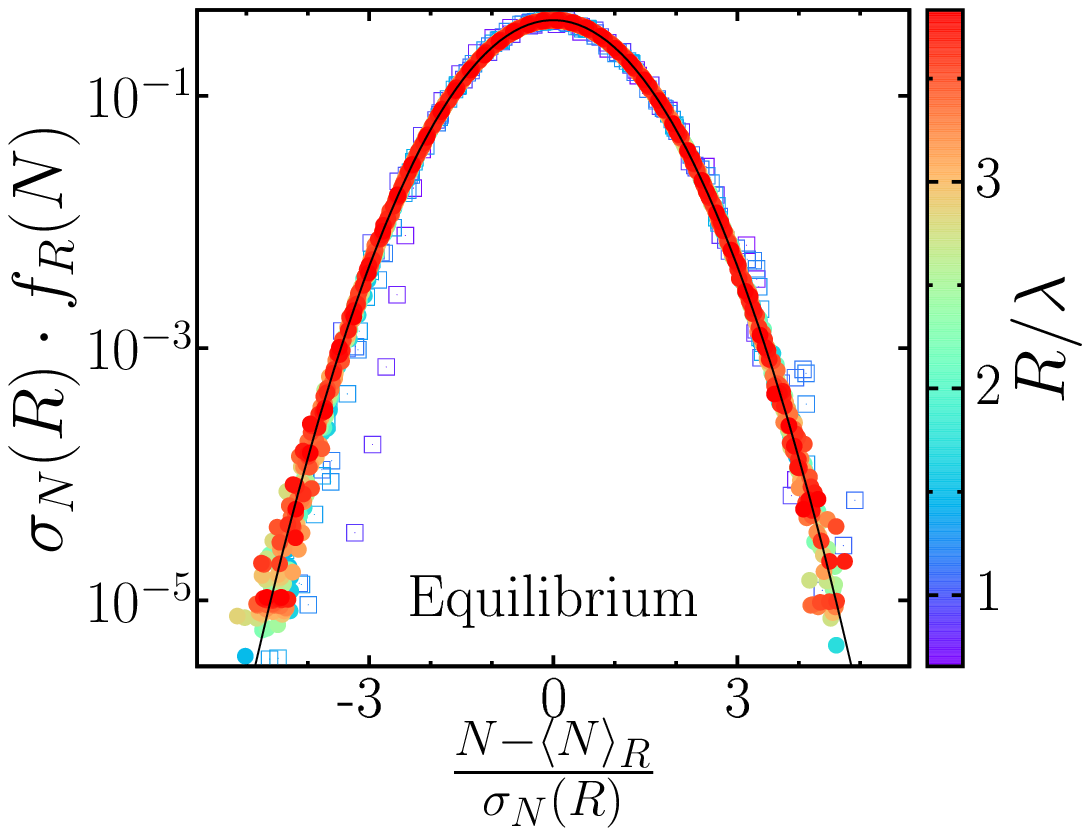}}
  \hfill
  \subfigure[][]{\includegraphics[width=0.31\textwidth]{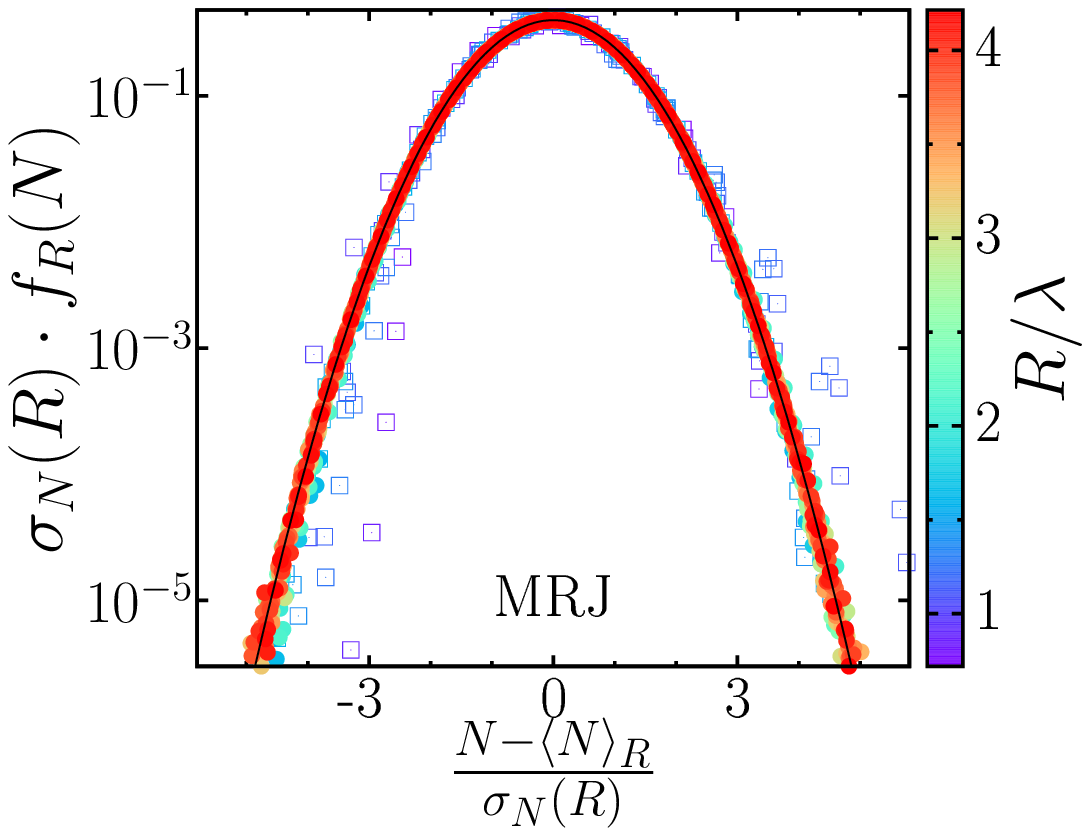}}
  \caption{(Color online) The number probability function $f_R(N)$ is rescaled by the square root of the number variance $\sigma_R^2(N)$ and plotted as a function of the normalized number of points $N$ inside a spherical observation window of radius $R$ that is randomly placed in the sample.
    For (a) overlapping spheres, (b) equilibrium hard spheres, and (c) MRJ sphere
    packings, the rescaled number probability functions are compared to the probability density function of the normal distribution (solid black line).
    For overlapping spheres, $f_R(N)$ corresponds to a Poisson distribution (dashed colored lines).
    It converges for an increasing test sphere radius $R$ to a normal distribution, but slowly compared to the equilibrium and MRJ sphere packings.
    For the hard sphere systems, $f_R(N)$ can well be approximated by Gaussian probability density functions already for $R > 1.5\, D$ (represented by filled circles instead of open squares),
    where $\lambda/D\approx 1.03$ and $\lambda/D\approx 0.937$ for the equilibrium or MRJ sphere packings, respectively.}
  \label{fig_density_probab_distr}
\end{figure*}

So far, we have considered the MRJ packings, as well as the equilibrium liquid and the overlapping spheres, as a two-phase medium formed by the spheres and the surrounding matrix.
Now, we analyze the point processes that is formed by the sphere centers.
As explained in Eq.~\eqref{eq:def-exclusion-probability} in Sec.~\ref{sec:def-EV}, the complementary cumulative pore-size distribution $F(\delta)$, which characterizes the two-phase medium, is trivially related to the exclusion probability $E_V(r)$, which analyzes the point process.
For the equilibrium and MRJ sphere packings, Tab.~\ref{tab:correlations-equi} and \ref{tab:correlations} list numerical values of the exclusion probability;
see also the Supplemental Material for estimates of the exclusion probability at more radial distances~\cite{supplementary_material}.

Figure~\ref{fig:exclusion_probability} shows $E_V(r)$ using the data for $F(\delta)$ in Fig.~\ref{fig_cum_pore-size_distr}.
However, the unit of length is different. Now, we compare point processes with unit number density by choosing $\rho^{-1/3}$ as the unit of length, where $\rho$ is the number density.

Interestingly, the value of $r$ at which $E_V(r)$ first ceases to have support defines the {\it covering radius}~\cite{Torquato2010reformulation}.
If to each point in a point process a sphere of the same radius is assigned, the covering radius $\mathcal{R}_c$ is the minimal radius that is needed to cover the entire space.
In other words, no point in $\mathbb{R}^3$ is further away from a point in the point process than $\mathcal{R}_c$;
therefore, $E_V(r) = 0$ for all $r\geq\mathcal{R}_c$ if ${\cal R}_c$ is bounded.

In three dimensions, the BCC lattice has the smallest known covering radius at unit density.
Therefore, its exclusion probability decreases faster than that of the FCC lattice, although the latter has a higher packing fraction.
\citet{Torquato2010reformulation} provides the exact values for the covering radii of both lattices:
for BCC, $\mathcal{R}_c/\lambda = \sqrt{5}/2^{5/3}\approx 0.7043$ (which corresponds to $\mathcal{R}_c/D = \sqrt{5/3}/2\approx 0.6455$),
and for FCC, $\mathcal{R}_c/\lambda = 1/2^{1/3}\approx 0.7937$ (which corresponds to $\mathcal{R}_c/D = 1/\sqrt{2}\approx 0.7071$).

For the MRJ sphere packings, there appears to be a cutoff at $r\gtrsim\lambda$, which is related to the saturation as explained in Sec.~\ref{sec:compl-cum-pore-size}.
However, for $E_V(r) < 10^{-8}$ the statistical errors become too large for a numerical precise estimate of the covering radius only via the exclusion probability $E_V(r)$.

\section{Local number density fluctuations}
\label{sec:density-fluctuations}

The analysis of the spectral density $\tilde{\chi}_{_V}(k)$ for $k\rightarrow0$ showed that
the random two-phase medium formed by the spheres in the MRJ packings is hyperuniform in contrast to the equilibrium and overlapping spheres; see Sec.~\ref{sec:spectral-density}.
Therefore, the point pattern formed by the sphere centers in the monodisperse MRJ packings must also be hyperuniform; see Sec.~\ref{sec:def-num-des-fluctuations}.
The hyperuniformity of a point pattern can be shown by studying the number density fluctuations,
that is, the fluctuations of the number $N$ of points within a spherical observation window that is randomly placed in the systems, and showing that it decays for large $R$ more slowly than $R^3$ in three dimensions.

First, we study for large radii $R$ the qualitative behavior of the probability functions $f_R(N)$ of the number of points $N$.
Then, we analyze the scaling of the number variance $\sigma_N^2(R)$ with the radius $R$.
The latter again reveals the hyperuniformity of the MRJ sphere packings.

\subsection{Number probability distribution}
\label{sec:nmbr-probab-distribution}

First, we estimate the probability function of the normalized number of points, that is, we subtract from $N$ the mean number of points $\langle N \rangle_R$ and divide by the square root of the number variance $\sigma_N^2(R)$,
where the expectation and the variance are also estimated by the sample mean and sample variance.
Therefore, we determine the estimated probability density function (PDF)~\footnote{The estimated probability density function is the empirical histogram weighted by the total number of samples and the bin width. In other words, the PDF is a relative frequency histogram weighted by the size of each bin.}
$f_R$ of the number $N$ of points of the pattern that lie within a test ball of radius $R$ placed randomly in the system.

First, we randomly place a point uniformly distributed in the simulation box.
It serves as the center of a score of test balls with different radii $R$ ranging from the maximal radius~\footnote{In a finite simulation box with periodic boundary conditions, the maximal, allowed radius is half of the minimum width of the box. Otherwise different representatives of the same sphere might be included in one test ball.} to a fraction of the diameter of a sphere in the sample.
For each radius, the number of sphere centers inside the test ball is recorded.
The PDF can be estimated by repeating this numerical experiment not only for the different samples but also by distributing several random centers inside a single sample.
Note that the estimates of $f_R(N)$ at different radii $R$ are correlated with each other.

Figure~\ref{fig_density_probab_distr} shows the normalized PDFs for the overlapping, equilibrium, and MRJ spheres.
As mentioned in Sec.~\ref{sec:compl-cum-pore-size}, the centers of the overlapping spheres form a Poisson point process~\cite{LastPenrose16}.
Therefore, $f_R(N)$ is by definition a Poisson distribution with mean value $\langle N \rangle_R = \rho\cdot\frac{4\pi}{3}R^3$, which is depicted in Fig.~\ref{fig_density_probab_distr} by dashed lines.
For large radii $R\rightarrow\infty$, the distribution of the normalized number of points inside the test ball can be approximated by the normal distribution (depicted as a black line).
However, even for large radii of about three times the diameter of a single sphere, there are significant deviations from a normal distribution.

In contrast to this the rescaled probability distributions for the equilibrium and MRJ sphere packings can be very well approximated by a normal distribution even for relatively small radii $R$.
Only for radii $R < 1.5\, D$ (denoted in Fig.~\ref{fig_density_probab_distr} by open squares), there are deviations because of the nonoverlap constraint.
So, from the simulations we can conjecture that for the equilibrium hard sphere liquid and MRJ state a central limit theorem holds for the number of points in a test ball.

A Gaussian distribution is determined by its first and second moments.
Therefore, the number density $\rho$ and number variance $\sigma_N^2(R)$ are the main parameters of the number probability function $f_R(N)$.

\subsection{Number variance}
\label{sec:nmbr-variance}

From the MC sampling of the PDF of the number of points, the sample variance provides an estimate for the number variance $\sigma_N^2(R)$ (as mentioned in the previous Sec.~\ref{sec:nmbr-probab-distribution}).
Recall that the estimates at different radii $R$ are correlated.

An important choice is that of the number of throws in a finite simulation box for a fixed radius $R$~\cite{AtkinsonEtAl2016}.
A too small number of throws leads to large statistical errors.
If there were too many throws, a systematic bias could arise because the same data is sampled several times but the throws are assumed to be independent.

\begin{figure}[t]
  \centering
  \includegraphics[width=\linewidth]{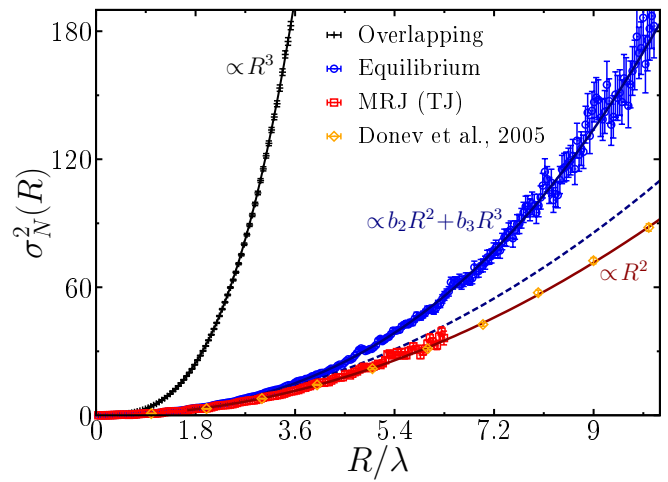}
  \caption{(Color online) The number variance $\sigma_N^2(R)$ for MRJ sphere packings as a function of
    the radius $R$ of the test sphere is compared to those of overlapping and equilibrium hard spheres.
    The solid black line shows the analytical curve for the overlapping spheres, which is proportional to $R^3$. 
    The data for the equilibrium hard spheres agrees well with a polynomial $b_2R^2+b_3R^3$ (solid blue line) but can clearly not be described by the fit of only a parabola (dashed line).
    This is in contrast to the MRJ data, which agrees well with a parabola (solid red line).
    Moreover, the extrapolation of the parabola is in perfect agreement with the results from MD simulations of MRJ packings with up to $10^6$ spheres; see Ref.~\cite{DonevEtAl2005}.}
  \label{fig_density_fluct}
\end{figure}

A benchmark test to check for such a bias is the comparison of numerical estimates of $\sigma_N^2(R)$ for overlapping spheres to the corresponding analytic curve.
As mentioned above, their sphere centers form a Poisson point process.
By definition, the number variance is in this case equal to the mean number of points $\sigma_N^2(R) = \langle N \rangle_R = \rho\cdot\frac{4\pi}{3}R^3$.

If for each radius $R$ a different number of throws is chosen, the maximum possible statistics can be used without introducing a systematic bias at large radii.
Here, we choose as a simple and efficient criterion for the number of throws that the expected volume fraction of the sample covered by the test balls remains below a fixed value $\phi_B$.
This choice is robust in the sense that similar results are obtained for reasonable values of $\phi_B$.
We here choose $\phi_B=0.8$.
For our choice of the number of throws, the numerical estimates agree for the benchmark test of overlapping spheres very well with the analytical curve, see Fig.~\ref{fig_density_fluct}.

Figure~\ref{fig_density_fluct} compares the number variance for the hyperuniform MRJ packings to that of the nonhyperuniform overlapping and equilibrium spheres.
The number variance for the MRJ packings is for radii $R>2\lambda$ not only smaller than for the other two more disordered systems, but what is most important, the scaling is qualitatively different.
According to Eq.~\eqref{eq:sigma-via-Sk}, the behavior of the structure factor (and thus of the spectral density) in the limit $k\rightarrow 0$ is related to the asymptotic behavior of density fluctuations in spherical observation windows in the limit of infinite radius.
The number variance for the hyperuniform MRJ state grows for large radii $R$ slower than the volume of the test ball, in contrast to the nonhyperuniform sphere systems.
When we fit a polynomial to the numerical estimates to study the scaling behavior,
the fit is not applied to very small radii $R$ but only to those radii for which the number probability functions can well be approximated by a Gaussian probability density function, see Sec.~\ref{sec:nmbr-probab-distribution}.

As explained above, the number variance for the overlapping spheres is analytically known to be proportional to $R^3$.
The data for the hard-sphere liquid is well approximated by the fit of a polynomial $b_2R^2+b_3R^3$ (solid blue line in Fig.~\ref{fig_density_fluct}).
However, a fit of only a parabola $\propto R^2$ (dashed line) is not sufficient to describe the data.
Therefore, the leading asymptotic behavior of the number variance for the hard-sphere liquid is also $R^3$.
It is clearly nonhyperuniform.

This is in contrast to the MRJ packings.
The number variance $\sigma_N^2(R)$ for to the TJ data of hyperuniform MRJ spheres agrees well with the fit of a parabola (solid red line in Fig.~\ref{fig_density_fluct}).
Moreover, the results for the TJ data with {2000} spheres per packing are in very good agreement with those from MD simulations with up to $10^6$ spheres~\cite{DonevEtAl2005}.
This holds not only in the range accessible by the TJ samples, but also the extrapolation of the quadratic fit to larger radii is in excellent agreement with the number variance obtained from the MD simulations of MRJ packings.

\section{Conclusions and Outlook}
\label{sec:Conclusion}

We have studied in detail the global and local structure of maximally random jammed (MRJ) sphere packings.
We considered it both as a two-phase random medium and as a point pattern that is formed by the sphere centers and evaluated certain structural characteristics accordingly.
In the first case, we have determined the two-point, surface-void, and surface-surface correlation functions, the spectral density and the pore-size distribution.
In the second case, we have estimated the number probability function and number variance.
These structural characteristics were then compared to those of equilibrium hard-sphere liquids as well as completely uncorrelated overlapping spheres.

The correlation functions and pore-size distribution are related to effective physical properties of the two-phase random medium, which we will evaluate in the third paper of this series.
Our results, for example, allow predictions of the effective transport~\cite{Prager1961, RubinsteinTorquato1988,
RubinsteinTorquato1989JFM, TorquatoYeong1997}, diffusion and reactions constants~\cite{Prager1963, TorquatoAvellaneda1991} as well as
mechanical~\cite{Torquato2002} and electromagnetic properties~\cite{Beran1968, Torquato1985, *SenTorquato1989, RechtsmanTorquato2008}.
Thereby, we can compare the (physical) behavior of a hyperuniform system, like the MRJ sphere packings, to that of nonhyperuniform disordered systems.
From the novel and unique structural properties of the hyperuniform materials can follow interesting physical properties, like isotropic band gaps~\cite{florescu_designer_2009}.


We have derived explicit expressions of the correlation functions of finite packings, \eg, from simulations. 
They are only functions of the radial distances between the spheres, which allows for a both accurate and fast calculation of these correlation functions.

By comparing the two-point, surface-void, and surface-surface correlation functions of the MRJ packings to those of the overlapping and equilibrium spheres,
we have found distinctive signatures of the contacts between spheres in the MRJ state.
For example, there are additional discontinuities in the derivatives of the correlation functions, which we have rigorously related to the contact Dirac delta contribution to the pair correlation function for MRJ packings.

As described in Sec.~\ref{sec:canonical}, the correlation functions evaluated here are special cases of the far more general canonical $n$-point correlation functions.
Future studies of MRJ packings could include generalizations of these functions for test particles with a variable size, which have been shown to contain considerably more information~\cite{ZacharyTorquato2011}.


The Fourier transform of the autocovariance, which follows from the two-point correlation function, reveals the hyperuniformity of the MRJ sphere packings.
The spectral density vanishes in the limit of infinite wavelengths (which is here equivalent to a vanishing structure factor).
This is in contrast to the equilibrium hard spheres because of their positive compressibility.

For a rigorous test of hyperuniformity given only a finite sample of the MRJ state, the spectral density would have to be extrapolated to $k\rightarrow0$~\cite{AtkinsonEtAl2016}.
It exceeds the scope of this article, but a statistical test could easily be developed to select the appropriate model of the vanishing structure factor or estimate a remaining finite value of the structure factor at $k=0$ as well as the statistical error.
Because the functional values of the spectral density at a given wave vector $\mathbf{k}$ are exponentially distributed, a maximum likelihood fit corresponds to an iterated weighted least square fit~\cite{charnes_equivalence_1976}.
Such an approach could detect hyperuniformity possibly even from relatively small samples.


In the pore-size distribution (or more precisely in the complementary cumulative distribution), we find a distinctive difference in the 
structure of MRJ packings that are either created by the TJ algorithm or by MD simulations.
The complementary cumulative distribution function decreases slower for the samples of the first than for the latter algorithm.
This is because the MD simulations have a slightly larger packing fraction and because they tend to be more ordered.

The pore-size distribution, or equivalently the exclusion probability, is also related to the covering problem~\cite{Torquato2010reformulation}.
Therefore, we compare the numerical estimates of the MRJ packings not only to the equilibrium and the overlapping spheres but also to perfectly regular lattices.
A more regular system often tends to exhibit a faster decrease of the complementary cumulative distribution function.

An open question is whether there are nontrivial necessary and/or sufficient conditions for hyperuniformity based on the pore-size distribution (besides the trivial observation that a completely empty or filled system is hyperuniform).
There are probably no sufficient conditions; for example, a finite covering radius is not a sufficient condition for a hyperuniform point process.
A counterexample would be a superposition of a nonuniform Poisson point process and a BCC lattice.
However, there might be necessary conditions on the asymptotic behavior or covering radius of the point process.


Concerning the number density fluctuations, we conjecture a central limit theorem for the equilibrium and MRJ hard sphere systems.
Already for relatively small radii $R$, the distribution of the normalized number of points inside a test ball of radius $R$ can be well approximated by a Gaussian distribution.
These observations are consistent with previous results in which it was shown that the distribution of local volume fraction for various particle systems tends to the normal distribution for sufficiently large windows~\cite{quintanilla_local_1997}.

Therefore, besides the number density the only nontrivial moment of the number probability function is the number variance.
Its scaling with the radius $R$ can be related to the structure factor.
If it grows slower than the volume of the test ball, the point process is hyperuniform.
We compare the scaling for the overlapping and equilibrium spheres to that of the hyperuniform MRJ packings.
A prediction from relatively small samples is difficult.
Nevertheless, we can demonstrate that a fit of a quadratic function is in good agreement with the results for the MRJ packings in contrast to the equilibrium hard-sphere liquid.

A crucial step in this analysis is the choice of the number of throws of the test ball.
Too many throws in a finite sample can lead to a systematic bias.
Here, we have chosen a conservative estimate based on a comparison of the numerical results for the overlapping spheres to the analytic curve.
In a future study, either a detailed analysis of the allowed number of throws or of the potential bias could help to significantly improve the statics that can be derived from a relatively small sample.
This would be very valuable to detect hyperuniformity, \eg, in experimental observations.
Very interesting would be also a rigorous hypothesis test or model selection that compares different scalings and takes the correlation between different radii $R$ into account.

In future work on the number variance, it would be interesting to compare the performance of the direct estimate via the sample variance, which is used here, to an estimate based on the so-called excess coordination $\Delta Z$, which is the average excess number of points compared to the ideal-gas expectation~\cite{DonevEtAl2005}.
For a square lattice, the excess coordination is connected to the so-called Gauss circle problem.


Recently, hyperuniformity was generalized to interfacial area fluctuations, random scalar fields, and statistically anisotropic many-particle systems and heterogeneous media~\cite{Torquato2016b}.
These concepts combined with the observations from this article call for further extensions and pose new questions.
For example, valuable insights might be gained by comparing the spectral density of MRJ sphere packings (see Sec.~\ref{sec:spectral-density}) to the corresponding spectral density of the surface defined in Ref.~\cite{Torquato2016b}.
(The latter should also vanish for the hyperuniform MRJ sphere packings in the limit of infinite wavelengths.)

Moreover, the variance $\sigma_S^2(R)$ of fluctuations in the surface area (similar to the number variance in Sec.~\ref{sec:density-fluctuations}) is related to the surface-surface correlation function $F_{ss}(r)$.
The explicit expressions for finite packings of hard spheres, which we have derived here, can help to efficiently compute $\sigma_S^2(R)$.
Even another generalization could be introduced by relating also the surface-void correlation function $F_{sv}(r)$ (studied in Sec.~\ref{sec:Fsv}) to a variance of fluctuations in finite observation window similar to $\sigma_S^2(R)$.

Our analysis provides insight into and links different problems of interest in various fields of research like material science, chemistry, physics, and mathematics.
The structural descriptors studied here determine a host of different effective properties of random two-phase or particulate medium, including transport, mechanical, electromagnetic, and chemical characteristics.
For point patterns, some of them are linked to the quantizer error or covering problem as well as the Gauss circle problem. 
The singular property of hyperuniformity is of special fundamental interest and has already seen surprising applications.

\begin{acknowledgments}
  We thank Steven Atkinson for his simulated samples of MRJ packings and hard-sphere liquids.
  This work was supported in part by the National Science Foundation under Grant No. DMS-1211087.
  We also thank the German Research Foundation (DFG) for the Grants No. HU1874/3-2, No. LA965/6-2, No. SCHR1148/3 and No. ME1361/11 awarded as part of the DFG-Forschergruppe FOR 1548 ``Geometry and Physics of Spatial Random Systems''.
\end{acknowledgments}

\appendix

\section{Analytical correlation functions of finite hard-sphere packings}
\label{sec_analytical}

Given a specific configuration of a finite packing of $N$ hard spheres, more precisely all pairwise distances $\rho_{ij}$, we here derive explicit analytical formula for the two-point correlation function~$S_2(r)$, the surface-void correlation function $F_{sv}(r)$, and the surface-surface correlation function $F_{ss}(r)$.
For convenience, here we only show the calculation for monodisperse sphere packings, but the calculations can easily and straightforwardly be generalized to any polydisperse packing of hard spheres.
Moreover, the approach can be easily adapted to various edge-corrections.
Here, we only consider periodic boundary conditions.

\subsection{Analytical two-point correlation function of finite hard-sphere packings}
\label{sec_analytical_2point}

The analytical calculation of the two-point correlation function follows closely the concept of the Monte Carlo sampling as described, \eg, in Refs.~\cite{SmithTorquato1988, Torquato2002}.
There, a test pattern that consists of points on the boundary of a sphere of radius $r$ is repeatedly and randomly placed onto the sample.
As described in Sec.~\ref{sec_2p_corr}, the two-point correlation function $S_2(r)$ is the probability that two points at a distance $r$ are found in the particle phase, \ie, within one of the spheres.
The Monte Carlo sampling estimates this probably by the frequency with which a point on the outside of the test pattern and its center both fall inside the particle phase.
For a hard-sphere packing, this hitting probability is here calculated analytically (given the pair distances $\rho_{ij}$ of the spheres).

\begin{figure}[t]
  \centering
  \includegraphics[width=0.7\linewidth]{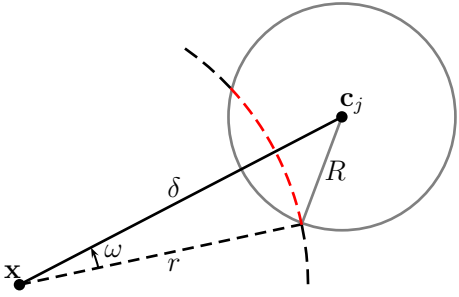}
  \caption{(Color online) A two-dimensional section through the spheres $B_R(\vec{c_j})$ (solid
  line) and $B_r(\vec{x})$ (dashed line): 
  the dashed red line indicates points at a distance $r$
  from $\vec{x}$ that lie inside sphere $j$; if the radii $r$ and $R$
  and the distance $\delta:=\|\vec{x}-\vec{c}_j\|$ are given, the cosine
  of the angle $\omega$ follows from the law of cosines; this in turn
  allows for the computation of the area $A\left(\partial B_r(\vec{x})\cap
  B_R(\vec{c_j})\right)$.}
  \label{fig:2dsection}
\end{figure}

For a system of hard-spheres the probability that a random point falls
inside sphere $i$ is $\phi/N$ where $\phi$ is the fraction of space
occupied by the spheres and $N$ the number of spheres. Given a point
inside sphere $i$, the conditional probability that another random point
at a distance $r$ is inside sphere $j$ is denoted by $p_{ij}(r)$.
The two-point correlation function can then by expressed by
\begin{align}
\begin{aligned}
  S_2(r) &= \sum_{i=1}^{N} \frac{\phi}{N} \sum_{j=1}^{N}p_{ij}(r)\\
  &= \frac{\phi}{N} \sum_{i=1}^{N} p_{ii}(r) + \frac{\phi}{N} \sum_{i=1}^{N} \sum_{j\neq i}p_{ij}(r)\\
  &= \phi \cdot p_{ii}(r) + \frac{2\phi}{N} \sum_{i=1}^{N}\sum_{j>i}p_{ij}(r),
\end{aligned}
\label{eq:S2}
\end{align}
because $p_{ii}$ is the same for all spheres and $p_{ij}=p_{ji}$.

For the calculation of $p_{ij}(r)$, we must first determine the
probability that for a given point $\vec{x}$ (in sphere $i$) at a
distance $\delta$ of the center $\vec{c}_j$ of sphere $j$ (with radius
$R$), another point, which is at a distance $r$ of $\vec{x}$, is inside
the sphere $j$, see Fig.~\ref{fig:2dsection}. This probability is
denoted by $f_{r,R}(\delta)$. It is simply the fraction of the surface
area of a sphere with radius $r$ centered at $\vec{x}$ that lies inside
sphere $j$:
\begin{align}
  f_{r,R}(\delta) = \frac{A\left(\partial B_r(\vec{x})\cap
  B_R(\vec{c_j})\right)}{4\pi r^2},
\end{align}
where $B_r(\vec{x})$ denotes, as usual, a ball of radius $r$ with center
$\vec{x}$, and the argument of the function indicates that it only
depends on the distance $\delta:=\|\vec{x}-\vec{c}_j\|$.

There are two cases where $f_{r,R}(\delta) \neq 0$.
First, if the sum of radius $r$ and distance $\delta$ is less than $R$,
$B_r(\vec{x}) \subset B_R(\vec{c_j})$ and $f_{r,R}(\delta) = 1$. Second, if
$|R-\delta| < r \leq R+\delta$, the fraction depends on an angle
$\omega$ between $(\vec{x}-\vec{c}_j)$ and the intersection line of the
two spheres $B_R(\vec{c_j})$ and $B_r(\vec{x})$, see
Fig.~\ref{fig:2dsection}. The cosine of this angle can be expressed by
$\delta$, $r$, and $R$ using the law of cosines:
\begin{align}
  \cos\omega = \frac{\delta^2+r^2-R^2}{2r\delta}.
  \label{eq:cosineslaw}
\end{align}
The corresponding surface area of $\partial B_r(\vec{x})\cap
  B_R(\vec{c_j})$ is then given by
\begin{align}
  \begin{aligned}
      A_{r,R}(\delta) &:= 2\pi r^2 \int_0^{\omega}\mathrm{d}\theta \,
  \sin\theta = 2\pi r^2 (1-\cos\omega)  \\
&= 2\pi r^2 \frac{R^2-(\delta^2-2r\delta+r^2)}{2r\delta}  \\
&= \pi r \frac{R^2-(\delta-r)^2}{\delta} 
  \end{aligned}
  \label{eq:ArRdelta}
\end{align}
using the rotational symmetry around the axis $\overline{\vec{x}-\vec{c}_j}$ and
Eq.~\eqref{eq:cosineslaw}.
Therefore,
\begin{align}
f_{r,R}(\delta) = \begin{cases}
    1 & \text{if } r + \delta < R\\
    \frac{R^2-(\delta-r)^2}{4r\delta} & \text{if } |R-\delta| \leq r \leq R+\delta\\
    0 & \text{else}
  \end{cases} .
  \label{eq:frRdelta}
\end{align}

For the conditional probability $p_{ij}(r)$, we only assume that the
initial random point $\vec{x}$ is any point in sphere $i$. So, it is the
integral of $f_{r,R}(\delta)$ over all positions $\vec{x}$ in sphere $i$
divided by the volume $v_R:=\nicefrac{4\pi}{3}R^3$ of the sphere. A
case-by-case analysis is needed. For $p_{ii}(r)$ (and thus $\delta \leq
R$), this integration of
Eq.~\eqref{eq:frRdelta} results in
\begin{align}
  \begin{aligned}
      p_{ii}(r) &= \frac{1}{v_R}\int_0^{R}\mathrm{d}\delta\,4\pi\delta^2f_{r,R}(\delta)\\
      &= \frac{1}{16R^3}(2R-r)^{2}(r+4R)\cdot\Theta(2R-r),
  \end{aligned}
\label{eq:pii}
\end{align}
where $\Theta(2R-r)$ is the Heaviside step function.

The calculation for $p_{ij}(r)$ (and thus $\delta > R$) for two
different spheres $i\neq j$ at a distance
$\rho_{ij}:=\|\vec{c}_i-\vec{c}_j\|$ can be tremendously simplified by
using suitable coordinates. To integrate the sphere $B_R(\vec{c_i})$,
spherical coordinates should be used. However, $\vec{c}_j$ should be
chosen as the origin instead of $\vec{c_i}$. 
Then, the sphere $i$ is foliated in shells with a constant distance
$\delta$ to $\vec{c}_j$ (the center of sphere $j$). (This distance can
of course only take on values between $\rho_{ij}-R$ and $\rho_{ij}+R$.)
On each sheet, the function $f_{r,R}(\delta)$ is constant, and the
integral over the sheet is simply its surface area, which was already
calculated in Eq.~\eqref{eq:ArRdelta} (only the parameters must be
exchanged). A straightforward case-by-case analysis then provides the
result. The probability $p_{ij}(r)$ can be expressed using the indicator
function $\mathbf{1}_{I_{\rho_{ij},R}}(r)$, which takes on the value unity
on $I_{\rho_{ij},R}:=[\rho_{ij} -2R,\rho_{ij} + 2R)$ and zero otherwise:
\begin{align}
  p_{ij}(r) =& \frac{1}{v_R}\int_{\rho_{ij}-R}^{\rho_{ij}+R}\mathrm{d}\delta
  \, A_{\delta,R}(\rho_{ij}) \cdot f_{r,R}(\delta) \nonumber\\
  =& \frac{1}{v_R}\int_{\rho_{ij}-R}^{\rho_{ij}+R}\mathrm{d}\delta
  \, \pi\delta \frac{R^2-(\rho_{ij}-\delta)^2}{\rho_{ij}} \cdot f_{r,R}(\delta) \nonumber\\
  \begin{split}
    =& \frac{\mathbf{1}_{I_{\rho_{ij},R}}(r)}{160\cdot R^3
    \rho_{ij}r}\cdot
    \left(2R-|r-\rho_{ij}|\right)^3\cdot\\&\cdot\left[|r-\rho_{ij}|^2+2R\cdot\left(3|r-\rho_{ij}|+2R\right)\right] .
  \end{split}
\label{eq:pij}
\end{align}
Inserting Eqs.~\eqref{eq:pii} and \eqref{eq:pij} in Eq.~\eqref{eq:S2}
yields the final result.

In a finite simulation box with periodic boundary conditions, there is
of course a maximal radius beyond which the two-point
correlation function cannot be calculated because different
representations of the same sphere would contribute. The here
presented Eqs.~\eqref{eq:pii} and \eqref{eq:pij} in Eq.~\eqref{eq:S2}
are only valid for values of $r$ smaller than half of the minimum
width of the simulation box $h_w$ minus the diameter of a sphere $D$
\begin{align}
  r < h_w - D .
\end{align}
If necessary, the calculation could be modified to make it possible to
calculate $S_2(r)$ also for $h_w-D<r<h_w$. Therefore, different
representations of the same sphere must be taken into account, such that
the minimal distance of the points in two different spheres is used.

\subsection{Analytical surface-void correlation functions of finite hard-sphere packings}
\label{sec_analytical_surfacevoid}

The surface-void correlation function $F_{sv}(r)$ of hard-sphere packings can be derived analytically in a very similar calculation.

For convenience and a better comparison to the calculation in
Appendix~\ref{sec_analytical_2point}, we calculate in this appendix the
correlation of the particle phase and the interface $F_{sv}^{(s)}(r)$.
In other words, the ``void'' phase is formed by the spheres. As
discussed in Sec.~\ref{sec:Fsv}, the corresponding correlation function
$F_{sv}(r)$ of the intermediate space between the spheres and the
interface can easily be derived from $F_{sv}^{(s)}(r)$ and the specific
surface area $s$ according to Eq.~\eqref{eq:relation-between-Fsv-s-i}.

If the surface-void correlation function $F_{sv}^{(s)}(r)$ is estimated from
Monte Carlo simulations~\cite{SeatonGlandt1986, Torquato2002}, a finite
shell of thickness $\epsilon$ is defined for each sphere $i$ with center
$\vec{c}_i$ and radius $R$:
\begin{align}
   S_{R,\epsilon}(\vec{c}_i) := B_{R}(\vec{c}_i)\setminus B_{R-\epsilon}(\vec{c}_i) .
\end{align}
Then, the frequency is estimated that a random point is inside such a
spherical shell and that another random point at a distance $r$ from the
first point is inside of any particle.
In the limit $\epsilon\rightarrow 0$, the ratio of this hitting
probability and the shell thickness $\epsilon$ converges to the
surface-void correlation function $F_{sv}^{(s)}(r)$ (and
$S_{R,\epsilon}(\vec{c}_i)$ converges to $\partial B_R(\vec{c}_i)$ of
sphere $i$).

Because the probability of a point hitting the spherical shell vanishes
and the ratio needs to be extrapolated, this procedure is numerically
rather expensive. As mentioned above, even small statistical errors can lead to huge
errors in the bounds on physical parameters.

Here, we derive the surface-void correlation function analytically for a
monodisperse hard-sphere packing given the pair distances $\rho_{ij}$ of
the spheres. As for the two-point correlation function in
Appendix~\ref{sec_analytical_2point}, the calculation can easily be
generalized to any polydisperse hard-sphere packing.

The derivation is very similar to that of the two-point correlation
function. Only, the conditional probabilities $p_{ij}$ have to be
replaced, and the limit of vanishing shell thickness is carried
out.
Conditional on the first point lying in sphere $i$, we define
$p^{(\epsilon)}_{ij}(r)$ as the probability that this first point lies
inside the shell $S_{R,\epsilon}(\vec{c}_i)$ and that the second point
at a distance $r$ hits sphere $j$.
The surface-void correlation function is then given by
\begin{align}
\begin{aligned}
  F_{sv}^{(s)}(r) &= \lim_{\epsilon\rightarrow 0} \frac{1}{\epsilon} \sum_{i=1}^{N} \frac{\phi}{N} \sum_{j=1}^{N}p^{(\epsilon)}_{ij}(r)\\
  &= \lim_{\epsilon\rightarrow 0} \frac{1}{\epsilon} \left[ \phi \cdot p^{(\epsilon)}_{ii}(r) +
  \frac{2\phi}{N} \sum_{i=1}^{N}\sum_{j>i}p^{(\epsilon)}_{ij}(r)\right] \\
  &= s \frac{v_R}{a_R} \lim_{\epsilon\rightarrow 0}
  \frac{1}{\epsilon} p^{(\epsilon)}_{ii}(r) + \frac{2s}{N} \frac{v_R}{a_R} \sum_{i=1}^{N}\sum_{j>i}\lim_{\epsilon\rightarrow 0} \frac{1}{\epsilon} p^{(\epsilon)}_{ij}(r),
\end{aligned}
\end{align}
where we use that for a hard-sphere packing the ratio of the packing fraction
$\phi$ and specific surface area $s$ is equal to the ratio of the
surface area $a_R:=4\pi R^2$ and volume $v_R:=\frac{4\pi}{3}R^3$ of a
single sphere.
Using this ratio, we define the unit-free limit
\begin{align}
  q_{ij}(r) :=  \frac{v_R}{a_R} \cdot \lim_{\epsilon\rightarrow 0} \frac{1}{\epsilon} p^{(\epsilon)}_{ij}(r).
\end{align}
The surface correlation function is then given by
\begin{align}
\begin{aligned}
  F_{sv}^{(s)}(r) &= s \cdot q_{ii}(r) + \frac{2s}{N} \sum_{i=1}^{N}\sum_{j>i} q_{ij}(r),
\end{aligned}
\label{eq:Fsv}
\end{align}
which is very similar to Eq.~\eqref{eq:S2} but the volume fraction
$\phi$ is replaced by the specific surface area $s$, and the conditional
probabilities $p_{ij}(r)$ are replaced by the limit $q_{ij}(r)$.

The calculation of this limit is very similar to the derivation of
$p_{ij}$ using the same auxiliary function $f_{r,R}(\delta)$. The main
difference is that the integral over the sphere $i$ is restricted to the
spherical shell $S_{R,\epsilon}(\vec{c}_i)$. In the case of both test
points lying in the same sphere, we derive in accordance with
Refs.~\cite{berryman_computing_1983, Torquato1986}
\begin{align}
  q_{ii}(r) = \frac{2R-r}{4R}\cdot\Theta(2R-r),
  \label{eq:qii}
\end{align}
where $\Theta(2R-r)$ is again the Heaviside step function. If the points
lie in two different spheres, we derive
\begin{align}
  \begin{aligned}
    q_{ij}(r) =& \frac{\mathbf{1}_{I_{\rho_{ij},R}}(r)}{24\cdot R \rho_{ij}r}\cdot
    \left(2R-|r-\rho_{ij}|\right)^2\cdot\\&\cdot\left(|r-\rho_{ij}|+R\right),
  \end{aligned}
    \label{eq:qij}
\end{align}
where $\mathbf{1}_{I_{\rho_{ij},R}}(r)$ is again the indicator function of the interval $I_{\rho_{ij},R}:=[\rho_{ij} -2R,\rho_{ij} + 2R)$.
Inserting Eqs.~\eqref{eq:qii} and \eqref{eq:qij} in Eq.~\eqref{eq:Fsv}
yields the final result.
In a finite simulation box with periodic boundary conditions, the same
restriction $r < h_w - D$ holds, where $h_w$ is half of the minimum
width.

\subsection{Analytical surface-surface correlation functions of finite hard-sphere packings}
\label{sec_analytical_surfacesurface}

The surface-surface correlation function $F_{ss}(r)$ of a finite packing of monodisperse hard spheres is derived similarly to the two-point and void-surface correlation functions.

The definition of the surface-surface correlation function $F_{ss}(r)$ uses the same limit of vanishing shell thickness $\epsilon$ as in Appendix~\ref{sec_analytical_surfacevoid}.
It is the limit of the probability that both random points at distance $r$ lie inside a spherical shell~\cite{SeatonGlandt1986, Torquato2002}.
Therefore, the Monte Carlo estimates are even more difficult, and our analytical solution for hard spheres avoids strong statistical errors.

To express the surface-surface correlation function analogously to Eqs.~\eqref{eq:S2} and \eqref{eq:Fsv},
we define the conditional probability $b^{(\epsilon)}_{ij}(r)$: based on the condition that $\vec{x}$ lies inside sphere $i$,
$b^{(\epsilon)}_{ij}(r)$ is the probability that the point $\vec{x}$ lies in the spherical shell $S_{R,\epsilon}(\vec{c}_i)$
and that simultaneously another point at distance $r$ from $\vec{x}$ lies in the spherical shell $S_{R,\epsilon}(\vec{c}_j)$ (of sphere $j$).
We also define the limit
\begin{align}
  a_{ij}(r):=\frac{v_R}{a_R}\cdot\lim_{\epsilon\rightarrow 0}\frac{1}{\epsilon^2}b^{(\epsilon)}_{ij}(r).
\end{align}
Then, the surface-surface correlation function can be expressed as
\begin{align}
  F_{ss}(r) &= \lim_{\epsilon\rightarrow 0} \frac{1}{\epsilon^2} \sum_{i=1}^{N} \frac{\phi}{N} \sum_{j=1}^{N}b^{(\epsilon)}_{ij}(r)\nonumber\\
  &= s \frac{v_R}{a_R} \lim_{\epsilon\rightarrow 0}
  \frac{1}{\epsilon^2} b^{(\epsilon)}_{ii}(r) + \frac{2s}{N}
  \frac{v_R}{a_R} \sum_{i=1}^{N}\sum_{j>i}\lim_{\epsilon\rightarrow 0}
  \frac{1}{\epsilon^2} b^{(\epsilon)}_{ij}(r)\nonumber\\
  &= s \cdot a_{ii}(r) + \frac{2s}{N} \sum_{i=1}^{N}\sum_{j>i}a_{ij}(r).
  \label{eq:Fss}
\end{align}

The calculation is again very similar to
Appendices~\ref{sec_analytical_2point} and \ref{sec_analytical_surfacevoid}.
There, $f_{r,R}(\delta)$ is needed to describe the probability that a
random point at distance $r$ of a given point $\vec{x}$ hits a sphere
with center $\vec{c}_i$ at a distance $\delta:=\|\vec{x}-\vec{c}_j\|$.
Here, we analogously define $g^{(\epsilon)}_{r,R}(\delta)$ as the
probability that the random point hits the spherical shell
$S_{R,\epsilon}(\vec{c}_j)$ (and not only the sphere $B_R(\vec{c}_j)$). 
More precisely, it is the fraction of the surface area of a sphere with radius $r$
centered at $\vec{x}$ that lies inside the spherical shell
$S_{R,\epsilon}(\vec{c}_j)$, where $\delta:=\|\vec{x}-\vec{c}_j\|$:
\begin{align}
  g^{(\epsilon)}_{r,R}(\delta) := \frac{A\left(\partial B_r(\vec{x})\cap
  S_{R,\epsilon}(\vec{c}_j)\right)}{4\pi r^2}.
  \label{eq:gerRdelta}
\end{align}
Only first order terms do not vanish in the limit. Therefore, we only
need to consider the case $|R-\delta| < r < R+\delta$, \ie, the
intersection $\partial B_r(\vec{x})\cap \partial B_R(\vec{c_j})$
contains more than a single point. All other cases lead to second or
smaller terms. For these values of $r$, we can choose $\epsilon$ small
enough so that $|R-\epsilon-\delta| < r < R-\epsilon+\delta$, \ie, that
the test sphere $\partial B_r(\vec{x})$ also intersects
$B_{R-\epsilon}(\vec{c_j})$ (the inner sphere of the shell) in more than
a single point. Then, Equation~\eqref{eq:gerRdelta} can easily be calculated
using Eq.~\eqref{eq:ArRdelta}:
\begin{align}
  g^{(\epsilon)}_{r,R}(\delta) = \frac{A_{r,R}(\delta)-A_{r,R-\epsilon}(\delta)}{4\pi r^2}.
\end{align}
Using the indicator function $\mathbf{1}_{J_{R,\delta}}(r)$ with
$J_{R,\delta}:=(|R-\delta|,R+\delta)$, we find
\begin{align}
  g^{(\epsilon)}_{r,R}(\delta) = \mathbf{1}_{J_{R,\delta}}(r)\cdot\frac{R}{2r\delta}\cdot \epsilon + \mathcal{O}(\epsilon^2).
  \label{eq:gerRdelta_final}
\end{align}

The limits $a_{ij}$ are then calculated analogously to $q_{ij}(r)$
by integration over the shell $S_{R,\epsilon}(\vec{c}_i)$.
In the case that both test points lie on the same sphere, we derive in accordance with
Refs.~\cite{berryman_computing_1983, Torquato1986}
\begin{align}
  a_{ii}(r) = \frac{1}{2r}\cdot\Theta(2R-r),
  \label{eq:aii}
\end{align}
where $\Theta(2R-r)$ is again the Heaviside step function.
If the points lie on two different spheres, we derive
\begin{align}
  a_{ij}(r) = \frac{\mathbf{1}_{I_{\rho_{ij},R}}(r)}{4
    \rho_{ij}r}\cdot\left(2R-|r-\rho_{ij}|\right).
  \label{eq:aij}
\end{align}
where $\mathbf{1}_{I_{\rho_{ij},R}}(r)$ is again the indicator function of the interval $I_{\rho_{ij},R}:=[\rho_{ij} -2R,\rho_{ij} + 2R)$.
Inserting Eqs.~\eqref{eq:aii} and \eqref{eq:aij} in Eq.~\eqref{eq:Fss}
yields the final result.
Again, the radius must be restricted to $r < h_w - D$ for a finite
simulation box with periodic boundary conditions, where $h_w$ is half of
the minimum width.

\section{Two-body contributions}
\label{sec:2body}

\begin{figure}[t]
  \centering
  \includegraphics[width=\linewidth]{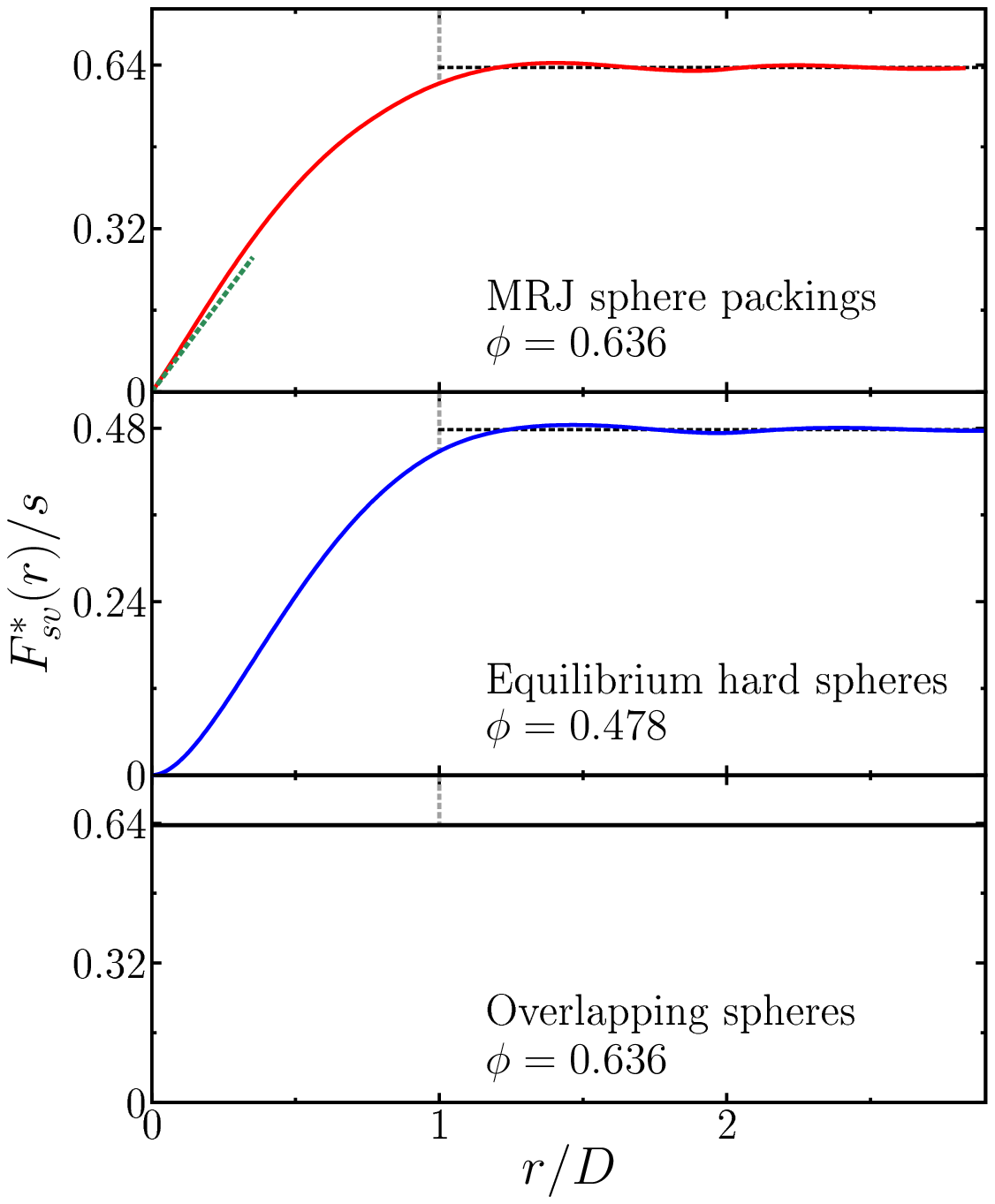}
  \caption{Two-body contributions $F^*_{sv}(r)$ to the surface-void correlation functions (rescaled by
  the specific surface $s$) for overlapping spheres (for which it is constant and equal to $\phi = 0.636$, the specific surface area is $s=2.21/D$), an equilibrium
  hard-sphere liquid ($\phi=0.478$, $s=2.87/D$), and MRJ sphere
  packings ($\phi=0.636$, $s=3.81/D$).
  For the MRJ packings, the dashed (green) line indicates the slope at $r=0$, which is strictly positive for the jammed sphere packings in contrast to the other two systems.
  The slope can be related to the mean number of contacts $\bar{z}$, see Eq.~\eqref{eq:slope-Fsv}.}
  \label{fig:2bdy-Fsv}
\end{figure}

\begin{figure}[t]
  \centering
  \includegraphics[width=\linewidth]{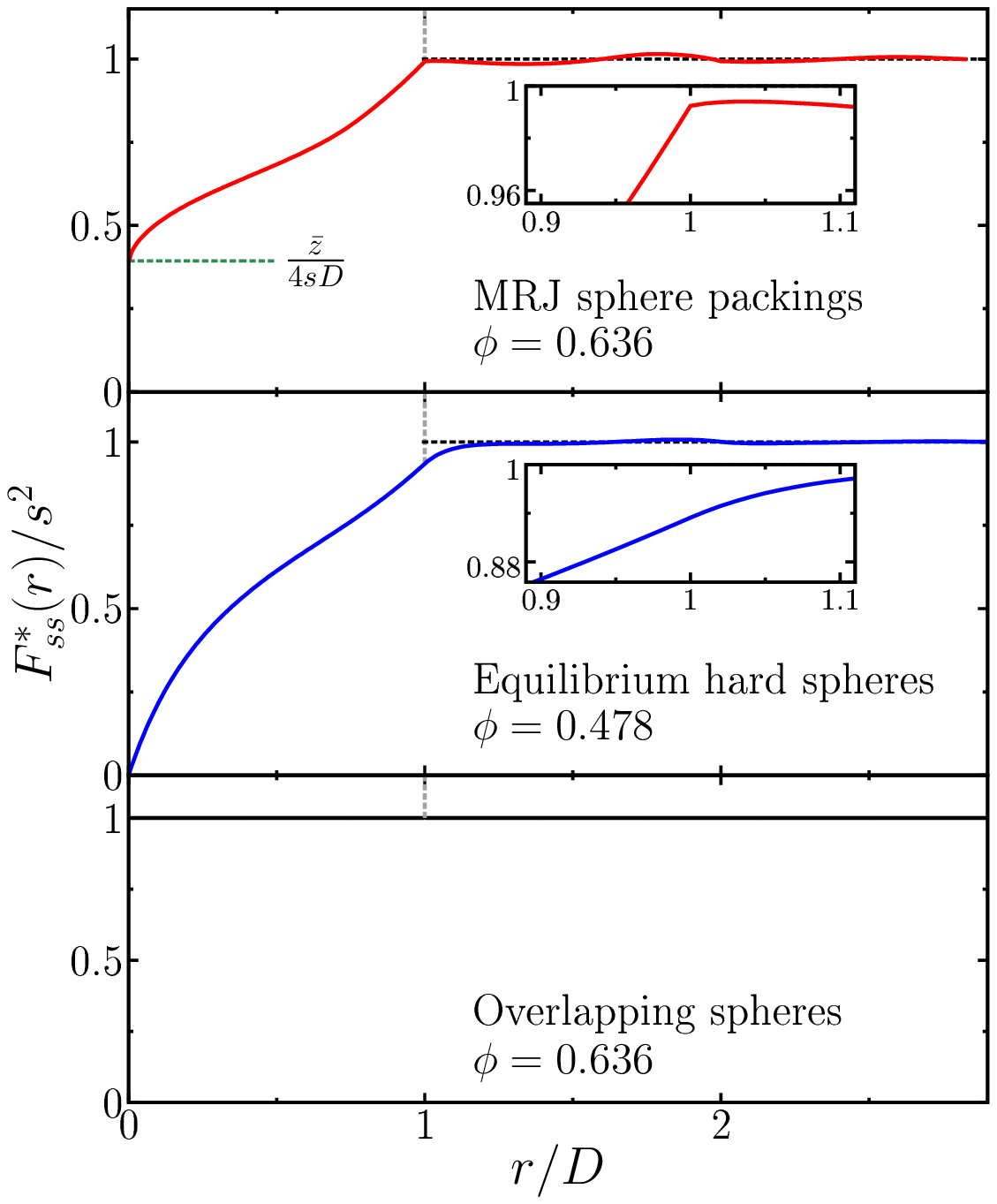}
  \caption{Two-body contributions $F^*_{ss}(r)$ to the surface-surface correlation functions (rescaled
  by the square of the specific surface $s$) for overlapping spheres (for which it is constant unity; $\phi = 0.636$, $s=2.21/D$), an
  equilibrium hard-sphere liquid ($\phi=0.478$, $s=2.87/D$), and
  MRJ sphere packings ($\phi=0.636$, $s=3.81/D$).
  For the MRJ packings, the dashed (green) line shows the functional value at $r=0$, which is strictly positive for the jammed sphere packings in contrast to the equilibrium hard spheres.
  The value can be related to the mean number of contacts $\bar{z}$, see Eq.~\eqref{eq:value-Fss}.
  Moreover, for the MRJ packings, the value $F^*_{ss}(D)/s^2=0.9924(1)$ is close to unity (in contrast to the equilibrium hard-sphere liquid).
  The insets magnify $F^*_{ss}(r)/s^2$ at $r = D$,
  where the derivative of $F^*_{ss}$ is discontinuous for the MRJ sphere packings in contrast to the equilibrium hard spheres;
  see also the insets of Fig.~\ref{fig:Fss} for the same finding at $r = 2D$.
  Note that $F^*_{ss}(r)=F_{ss}(r)$ for $r > D$.}
  \label{fig:2bdy-Fss}
\end{figure}

The correlation functions are sums of contributions from either a single sphere (see Eqs.~\eqref{eq:qii} and \eqref{eq:aii}) or from two spheres (see Eqs.~\eqref{eq:qij} and \eqref{eq:aij}).
The contributions from a single sphere are by definition the same for all possible arrangements of hard spheres.
To learn more about a specific system and to compare different packings, it is helpful to calculate the contributions from two spheres separately from those of a single sphere.

For finite packings of hard spheres, we define the two-body contributions to the correlation functions $F_{sv}$ and $F_{ss}$ by
\begin{align}
  F^*_{sv}(r) & :=  \frac{2s}{N} \sum_{i=1}^{N}\sum_{j>i} q_{ij}(r)\\
  F^*_{ss}(r) & :=  \frac{2s}{N} \sum_{i=1}^{N}\sum_{j>i} a_{ij}(r),
  \label{eq:def_2body}
\end{align}
where $q_{ij}(r)$ and $a_{ij}(r)$ are given in Eqs.~\eqref{eq:qij} and \eqref{eq:aij}, respectively.
The two-body contributions have also already been defined in the thermodynamic limit (\ie, infinite system size), where they can be connected to the pair-correlation function~$g_2(r)$, as discussed for the correlation functions in Sec.~\ref{sec_2p_corr}~\cite{Torquato1986, Torquato2002}.

Figures~\ref{fig:2bdy-Fsv} and \ref{fig:2bdy-Fss} compare the two-body contributions $F^*_{sv}(r)$ and $F^*_{ss}(r)$ for the MRJ sphere packings to those of the equilibrium hard spheres and the overlapping spheres.
Because the single-body contributions are nonzero only for $r\leq D$, the curves in Fig.~\ref{fig:2bdy-Fss} deviate from the curves in Fig.~\ref{fig:Fss} only for $r\leq D$.
For the same reason, $F_{sv}(r)$ in Fig.~\ref{fig:Fsv} is for $r>D$ identical to $s-F^*_{sv}(r)$, see Fig.~\ref{fig:2bdy-Fsv}.

Because the overlapping spheres are completely independent from each other, their two-body contributions are trivial, that is, constant and equal to the long-range limit, see Sec.~\ref{sec_2p_corr}.
The results for the equilibrium hard spheres are in good agreement with the previous findings in Ref.~\cite{Torquato1986} at similar global packing fractions.
There, an approximation of the pair-correlation function by \citet{VerletWeis1972} was used to calculate the correlation functions.
The curves are continuous and differentiable.

Comparing the MRJ sphere packings to the equilibrium hard-sphere liquid, we find some distinct signatures produced by the contacts between the spheres.

At least for finite packings of hard spheres, the slope of $F^*_{sv}(r)$ for $r \rightarrow 0$ can be related to the mean number of contacts $\bar{z}$ using the explicit expression from Eq.~\eqref{eq:qij} (because only spheres at contact contribute in a finite packing to $F^*_{sv}(r)$ for $r \rightarrow 0$):
\begin{align}
  \frac{\text{d}F^*_{sv}}{\text{d}r}(0) = \frac{s}{8D} \cdot \bar{z}.
  \label{eq:slope-Fsv}
\end{align}
The mean contact number $\bar{z}$ vanishes for the equilibrium hard spheres but not for the MRJ packings.
The latter packings are isostatic, which means that the number of constraints by spheres in contact matches exactly the number of degrees of freedom.
For an infinite system, this results in a mean contact number $\bar{z}=6$.
For a finite deformable simulation box, additional constraints have to be taken into account, but carefully identifying the contacts between spheres still yields $\bar{z} \approx 6.00$.
For more details; see Ref.~\cite{AtkinsonEtAl2013}, whose MRJ configurations we here analyze.
Because $\bar{z}>0$ for the MRJ sphere packings, the slope of $F^*_{sv}(r)$ does not vanish at $r=0$ in contrast to the equilibrium hard spheres.

Even more prominent are the differences between $F^*_{ss}(r)$ for the MRJ and equilibrium hard-sphere packings.
There, the functional value $F^*_{ss}(0) := \lim_{r\rightarrow 0} F^*_{ss}(r)$ can be related to the same mean contact number $\bar{z}$:
\begin{align}
  F^*_{ss}(0) = \frac{s}{4D} \cdot \bar{z}.
  \label{eq:value-Fss}
\end{align}
Equations~\eqref{eq:slope-Fsv} and \eqref{eq:value-Fss} provide estimates of the mean number of contacts based on the correlation functions.
In principle, the exact formulas for the correlation functions of hard sphere packings, which are derived here, allow for accurate results.
However, if there are numerical inaccuracies, \eg, in the positions of the spheres or their radii, or if the functions are evaluated only at finite radial distances, slight deviations can appear.

While for the equilibrium hard spheres $F^*_{ss}(0)=0$, the functional values remain strictly positive for the MRJ packings.
Moreover, in contrast to the smooth curves for the equilibrium hard spheres, $F^*_{ss}(r)$ of the MRJ packings is continuous but not differentiable at $r=D$ and $r=2D$.
These discontinuities in the derivative arise because of the contacts between the spheres.
A rigorous connection to the mean number of contacts can be derived by expressing the surface-surface correlation function by the pair-correlation function of the sphere centers~\cite{Torquato1986}.

{The Dirac delta contribution from the spheres at contact to the total correlation function $h(r)=g_2(r)-1$ is
\begin{align}
  h_2^{(c)}(r) = \frac{\bar{z} \cdot \delta(r-D)}{\rho \cdot 4\pi D^2}.
  \label{eq:contr-spheres-at-contact}
\end{align}
Evaluating the Fourier transforms in~\cite{Torquato1986}, we obtain their contribution to the surface-surface correlation function:
\begin{align}
  F_{ss}^{(c)}(r) := \bar{z} \cdot \frac{\rho D\pi}{4} \cdot\begin{cases}
    1, & r \leq D,\\
    \frac{2D-r}{r}, & D < r \leq 2D,\\
    0, & 2D < r.
  \end{cases}
  \label{eq:F_ss_contact}
\end{align}
It is continuous but not differentiable at $r=D$ and $r=2D$.
Note that the contribution from spheres that are not at contact is smooth.
Therefore, the discontinuities in the derivative of $F^*_{ss}$ stem only from spheres that are at contact with each other.}

Similarly to $F_{ss}$, we can also compute the contributions $F_{sv}^{(c)}$ and $S_{2}^{(c)}$ of the spheres in contact to the surface-void and two-point correlation function:
\begin{widetext}
\begin{align}
  F_{sv}^{(c)}(r) &:= \bar{z} \cdot \frac{\rho\pi}{4} \cdot\begin{cases}
    \frac{3Dr-2r^2}{6}, & r \leq D,\\
    \frac{2r^3-9Dr^2+12D^2r-4D^3}{6r}, & D < r \leq 2D,\\
    0, & 2D < r.
  \end{cases}\\
  S_{2}^{(c)}(r) &:= \bar{z} \cdot \frac{\rho\pi}{120D} \cdot\begin{cases}
    r^2\cdot(r^2-5Dr+5D^2), & r \leq D,\\
    -r^4+5Dr^3-5D^2r^2 -10D^3r+20D^4-\frac{8}{r}D^5, & D < r \leq 2D,\\
    0, & 2D < r.
  \end{cases}
\end{align}
\end{widetext}
They are continuous and differentiable.
For $F_{sv}^{(c)}$ at $r=D$ and $S_{2}^{(c)}$ at $r=D$ and $r=2D$, even the second derivatives exists.
However, the second derivative of $F_{sv}^{(c)}$ is discontinuous at $r=2D$ and so are the third derivatives of $F_{sv}^{(c)}$ at $r=D$ and $S_{2}^{(c)}$ at $r=D$ and $r=2D$.


\begin{thebibliography}{89}%
\makeatletter
\providecommand \@ifxundefined [1]{%
 \@ifx{#1\undefined}
}%
\providecommand \@ifnum [1]{%
 \ifnum #1\expandafter \@firstoftwo
 \else \expandafter \@secondoftwo
 \fi
}%
\providecommand \@ifx [1]{%
 \ifx #1\expandafter \@firstoftwo
 \else \expandafter \@secondoftwo
 \fi
}%
\providecommand \natexlab [1]{#1}%
\providecommand \enquote  [1]{``#1''}%
\providecommand \bibnamefont  [1]{#1}%
\providecommand \bibfnamefont [1]{#1}%
\providecommand \citenamefont [1]{#1}%
\providecommand \href@noop [0]{\@secondoftwo}%
\providecommand \href [0]{\begingroup \@sanitize@url \@href}%
\providecommand \@href[1]{\@@startlink{#1}\@@href}%
\providecommand \@@href[1]{\endgroup#1\@@endlink}%
\providecommand \@sanitize@url [0]{\catcode `\\12\catcode `\$12\catcode
  `\&12\catcode `\#12\catcode `\^12\catcode `\_12\catcode `\%12\relax}%
\providecommand \@@startlink[1]{}%
\providecommand \@@endlink[0]{}%
\providecommand \url  [0]{\begingroup\@sanitize@url \@url }%
\providecommand \@url [1]{\endgroup\@href {#1}{\urlprefix }}%
\providecommand \urlprefix  [0]{URL }%
\providecommand \Eprint [0]{\href }%
\providecommand \doibase [0]{http://dx.doi.org/}%
\providecommand \selectlanguage [0]{\@gobble}%
\providecommand \bibinfo  [0]{\@secondoftwo}%
\providecommand \bibfield  [0]{\@secondoftwo}%
\providecommand \translation [1]{[#1]}%
\providecommand \BibitemOpen [0]{}%
\providecommand \bibitemStop [0]{}%
\providecommand \bibitemNoStop [0]{.\EOS\space}%
\providecommand \EOS [0]{\spacefactor3000\relax}%
\providecommand \BibitemShut  [1]{\csname bibitem#1\endcsname}%
\let\auto@bib@innerbib\@empty
\bibitem [{\citenamefont {Henley}(1986)}]{Henley1986}%
  \BibitemOpen
  \bibfield  {author} {\bibinfo {author} {\bibfnamefont {C.~L.}\ \bibnamefont
  {Henley}},\ }\href {\doibase 10.1103/PhysRevB.34.797} {\bibfield  {journal}
  {\bibinfo  {journal} {Phys. Rev. B}\ }\textbf {\bibinfo {volume} {34}},\
  \bibinfo {pages} {797} (\bibinfo {year} {1986})}\BibitemShut {NoStop}%
\bibitem [{\citenamefont {Hales}(2005)}]{Hales2005}%
  \BibitemOpen
  \bibfield  {author} {\bibinfo {author} {\bibfnamefont {T.~C.}\ \bibnamefont
  {Hales}},\ }\href {\doibase 10.4007/annals.2005.162.1065} {\bibfield
  {journal} {\bibinfo  {journal} {Ann. Math.}\ }\textbf {\bibinfo {volume}
  {162}},\ \bibinfo {pages} {1065} (\bibinfo {year} {2005})}\BibitemShut
  {NoStop}%
\bibitem [{\citenamefont {Finney}(1977)}]{Finney1997}%
  \BibitemOpen
  \bibfield  {author} {\bibinfo {author} {\bibfnamefont {J.~L.}\ \bibnamefont
  {Finney}},\ }\href {\doibase 10.1038/266309a0} {\bibfield  {journal}
  {\bibinfo  {journal} {Nature}\ }\textbf {\bibinfo {volume} {266}},\ \bibinfo
  {pages} {309 } (\bibinfo {year} {1977})}\BibitemShut {NoStop}%
\bibitem [{\citenamefont {Zallen}(1998)}]{Zallen1998}%
  \BibitemOpen
  \bibfield  {author} {\bibinfo {author} {\bibfnamefont {R.}~\bibnamefont
  {Zallen}},\ }\href@noop {} {\emph {\bibinfo {title} {The Physics of Amorphous
  Solids}}},\ A Wiley-Interscience publication\ (\bibinfo  {publisher}
  {Wiley},\ \bibinfo {year} {1998})\BibitemShut {NoStop}%
\bibitem [{\citenamefont {Bertrand}\ \emph {et~al.}(2016)\citenamefont
  {Bertrand}, \citenamefont {Behringer}, \citenamefont {Chakraborty},
  \citenamefont {O'Hern},\ and\ \citenamefont
  {Shattuck}}]{bertrand_protocol_2016}%
  \BibitemOpen
  \bibfield  {author} {\bibinfo {author} {\bibfnamefont {T.}~\bibnamefont
  {Bertrand}}, \bibinfo {author} {\bibfnamefont {R.~P.}\ \bibnamefont
  {Behringer}}, \bibinfo {author} {\bibfnamefont {B.}~\bibnamefont
  {Chakraborty}}, \bibinfo {author} {\bibfnamefont {C.~S.}\ \bibnamefont
  {O'Hern}}, \ and\ \bibinfo {author} {\bibfnamefont {M.~D.}\ \bibnamefont
  {Shattuck}},\ }\href {\doibase 10.1103/PhysRevE.93.012901} {\bibfield
  {journal} {\bibinfo  {journal} {Phys. Rev. E}\ }\textbf {\bibinfo {volume}
  {93}},\ \bibinfo {pages} {012901} (\bibinfo {year} {2016})}\BibitemShut
  {NoStop}%
\bibitem [{\citenamefont {Torquato}\ \emph {et~al.}(2000)\citenamefont
  {Torquato}, \citenamefont {Truskett},\ and\ \citenamefont
  {Debenedetti}}]{TorquatoEtAl2000PhysRevLetRCPvsMRJ}%
  \BibitemOpen
  \bibfield  {author} {\bibinfo {author} {\bibfnamefont {S.}~\bibnamefont
  {Torquato}}, \bibinfo {author} {\bibfnamefont {T.~M.}\ \bibnamefont
  {Truskett}}, \ and\ \bibinfo {author} {\bibfnamefont {P.~G.}\ \bibnamefont
  {Debenedetti}},\ }\href {\doibase 10.1103/PhysRevLett.84.2064} {\bibfield
  {journal} {\bibinfo  {journal} {Phys. Rev. Lett.}\ }\textbf {\bibinfo
  {volume} {84}},\ \bibinfo {pages} {2064} (\bibinfo {year}
  {2000})}\BibitemShut {NoStop}%
\bibitem [{\citenamefont {Torquato}\ and\ \citenamefont
  {Stillinger}(2010)}]{TorquatoStillinger2010RevModPhys}%
  \BibitemOpen
  \bibfield  {author} {\bibinfo {author} {\bibfnamefont {S.}~\bibnamefont
  {Torquato}}\ and\ \bibinfo {author} {\bibfnamefont {F.~H.}\ \bibnamefont
  {Stillinger}},\ }\href {\doibase 10.1103/RevModPhys.82.2633} {\bibfield
  {journal} {\bibinfo  {journal} {Rev. Mod. Phys.}\ }\textbf {\bibinfo {volume}
  {82}},\ \bibinfo {pages} {2633} (\bibinfo {year} {2010})}\BibitemShut
  {NoStop}%
\bibitem [{\citenamefont {O'Hern}\ \emph {et~al.}(2002)\citenamefont {O'Hern},
  \citenamefont {Langer}, \citenamefont {Liu},\ and\ \citenamefont
  {Nagel}}]{ohern_PRL_2002}%
  \BibitemOpen
  \bibfield  {author} {\bibinfo {author} {\bibfnamefont {C.~S.}\ \bibnamefont
  {O'Hern}}, \bibinfo {author} {\bibfnamefont {S.~A.}\ \bibnamefont {Langer}},
  \bibinfo {author} {\bibfnamefont {A.~J.}\ \bibnamefont {Liu}}, \ and\
  \bibinfo {author} {\bibfnamefont {S.~R.}\ \bibnamefont {Nagel}},\ }\href
  {\doibase 10.1103/PhysRevLett.88.075507} {\bibfield  {journal} {\bibinfo
  {journal} {Phys. Rev. Lett.}\ }\textbf {\bibinfo {volume} {88}},\ \bibinfo
  {pages} {075507} (\bibinfo {year} {2002})}\BibitemShut {NoStop}%
\bibitem [{\citenamefont {Karayiannis}\ and\ \citenamefont
  {Laso}(2008)}]{Karayiannis_PRL_2008}%
  \BibitemOpen
  \bibfield  {author} {\bibinfo {author} {\bibfnamefont {N.~C.}\ \bibnamefont
  {Karayiannis}}\ and\ \bibinfo {author} {\bibfnamefont {M.}~\bibnamefont
  {Laso}},\ }\href {\doibase 10.1103/PhysRevLett.100.050602} {\bibfield
  {journal} {\bibinfo  {journal} {Phys. Rev. Lett.}\ }\textbf {\bibinfo
  {volume} {100}},\ \bibinfo {pages} {050602} (\bibinfo {year}
  {2008})}\BibitemShut {NoStop}%
\bibitem [{\citenamefont {Xu}\ and\ \citenamefont {Rice}(2011)}]{XuRice2011}%
  \BibitemOpen
  \bibfield  {author} {\bibinfo {author} {\bibfnamefont {X.}~\bibnamefont
  {Xu}}\ and\ \bibinfo {author} {\bibfnamefont {S.~A.}\ \bibnamefont {Rice}},\
  }\href {\doibase 10.1103/PhysRevE.83.021120} {\bibfield  {journal} {\bibinfo
  {journal} {Phys. Rev. E}\ }\textbf {\bibinfo {volume} {83}},\ \bibinfo
  {pages} {021120} (\bibinfo {year} {2011})}\BibitemShut {NoStop}%
\bibitem [{\citenamefont {Ozawa}\ \emph {et~al.}(2012)\citenamefont {Ozawa},
  \citenamefont {Kuroiwa}, \citenamefont {Ikeda},\ and\ \citenamefont
  {Miyazaki}}]{Ozawa_etal_2012}%
  \BibitemOpen
  \bibfield  {author} {\bibinfo {author} {\bibfnamefont {M.}~\bibnamefont
  {Ozawa}}, \bibinfo {author} {\bibfnamefont {T.}~\bibnamefont {Kuroiwa}},
  \bibinfo {author} {\bibfnamefont {A.}~\bibnamefont {Ikeda}}, \ and\ \bibinfo
  {author} {\bibfnamefont {K.}~\bibnamefont {Miyazaki}},\ }\href {\doibase
  10.1103/PhysRevLett.109.205701} {\bibfield  {journal} {\bibinfo  {journal}
  {Phys. Rev. Lett.}\ }\textbf {\bibinfo {volume} {109}},\ \bibinfo {pages}
  {205701} (\bibinfo {year} {2012})}\BibitemShut {NoStop}%
\bibitem [{\citenamefont {Baranau}\ \emph {et~al.}(2013)\citenamefont
  {Baranau}, \citenamefont {Hlushkou}, \citenamefont {Khirevich},\ and\
  \citenamefont {Tallarek}}]{Baranau_etal_2013}%
  \BibitemOpen
  \bibfield  {author} {\bibinfo {author} {\bibfnamefont {V.}~\bibnamefont
  {Baranau}}, \bibinfo {author} {\bibfnamefont {D.}~\bibnamefont {Hlushkou}},
  \bibinfo {author} {\bibfnamefont {S.}~\bibnamefont {Khirevich}}, \ and\
  \bibinfo {author} {\bibfnamefont {U.}~\bibnamefont {Tallarek}},\ }\href
  {\doibase 10.1039/C3SM27374A} {\bibfield  {journal} {\bibinfo  {journal}
  {Soft Matter}\ }\textbf {\bibinfo {volume} {9}},\ \bibinfo {pages} {3361}
  (\bibinfo {year} {2013})}\BibitemShut {NoStop}%
\bibitem [{\citenamefont {Tian}\ \emph {et~al.}(2015)\citenamefont {Tian},
  \citenamefont {Xu}, \citenamefont {Jiao},\ and\ \citenamefont
  {Torquato}}]{tian_geometric-structure_2015}%
  \BibitemOpen
  \bibfield  {author} {\bibinfo {author} {\bibfnamefont {J.}~\bibnamefont
  {Tian}}, \bibinfo {author} {\bibfnamefont {Y.}~\bibnamefont {Xu}}, \bibinfo
  {author} {\bibfnamefont {Y.}~\bibnamefont {Jiao}}, \ and\ \bibinfo {author}
  {\bibfnamefont {S.}~\bibnamefont {Torquato}},\ }\href {\doibase
  10.1038/srep16722} {\bibfield  {journal} {\bibinfo  {journal} {Scientific
  Reports}\ }\textbf {\bibinfo {volume} {5}},\ \bibinfo {pages} {16722}
  (\bibinfo {year} {2015})}\BibitemShut {NoStop}%
\bibitem [{\citenamefont {Ramola}\ and\ \citenamefont
  {Chakraborty}(2016)}]{ramola_disordered_2016}%
  \BibitemOpen
  \bibfield  {author} {\bibinfo {author} {\bibfnamefont {K.}~\bibnamefont
  {Ramola}}\ and\ \bibinfo {author} {\bibfnamefont {B.}~\bibnamefont
  {Chakraborty}},\ }\href {http://arxiv.org/abs/1604.06148} {\bibfield
  {journal} {\bibinfo  {journal} {arXiv:1604.06148}\ } (\bibinfo {year}
  {2016})}\BibitemShut {NoStop}%
\bibitem [{\citenamefont {Bernal}(1960)}]{Bernal1960}%
  \BibitemOpen
  \bibfield  {author} {\bibinfo {author} {\bibfnamefont {J.~D.}\ \bibnamefont
  {Bernal}},\ }\href@noop {} {\bibfield  {journal} {\bibinfo  {journal}
  {Nature}\ }\textbf {\bibinfo {volume} {185}},\ \bibinfo {pages} {68}
  (\bibinfo {year} {1960})}\BibitemShut {NoStop}%
\bibitem [{\citenamefont {Scott}\ and\ \citenamefont
  {Kilgour}(1969)}]{ScottKilgour1969}%
  \BibitemOpen
  \bibfield  {author} {\bibinfo {author} {\bibfnamefont {G.~D.}\ \bibnamefont
  {Scott}}\ and\ \bibinfo {author} {\bibfnamefont {D.~M.}\ \bibnamefont
  {Kilgour}},\ }\href {http://stacks.iop.org/0022-3727/2/i=6/a=311} {\bibfield
  {journal} {\bibinfo  {journal} {J. Phys. D: Appl. Phys.}\ }\textbf {\bibinfo
  {volume} {2}},\ \bibinfo {pages} {863} (\bibinfo {year} {1969})}\BibitemShut
  {NoStop}%
\bibitem [{\citenamefont {Finney}(1970)}]{Finney1970}%
  \BibitemOpen
  \bibfield  {author} {\bibinfo {author} {\bibfnamefont {J.~L.}\ \bibnamefont
  {Finney}},\ }\href {\doibase 10.1098/rspa.1970.0189} {\bibfield  {journal}
  {\bibinfo  {journal} {P. Roy. Soc. A-Math. Phy.}\ }\textbf {\bibinfo {volume}
  {319}},\ \bibinfo {pages} {479} (\bibinfo {year} {1970})}\BibitemShut
  {NoStop}%
\bibitem [{\citenamefont {Berryman}(1983{\natexlab{a}})}]{Berryman1983}%
  \BibitemOpen
  \bibfield  {author} {\bibinfo {author} {\bibfnamefont {J.~G.}\ \bibnamefont
  {Berryman}},\ }\href {\doibase 10.1103/PhysRevA.27.1053} {\bibfield
  {journal} {\bibinfo  {journal} {Phys. Rev. A}\ }\textbf {\bibinfo {volume}
  {27}},\ \bibinfo {pages} {1053} (\bibinfo {year}
  {1983}{\natexlab{a}})}\BibitemShut {NoStop}%
\bibitem [{\citenamefont {O’Hern}\ \emph {et~al.}(2003)\citenamefont
  {O’Hern}, \citenamefont {Silbert}, \citenamefont {Liu},\ and\ \citenamefont
  {Nagel}}]{HernEtAl2003}%
  \BibitemOpen
  \bibfield  {author} {\bibinfo {author} {\bibfnamefont {C.~S.}\ \bibnamefont
  {O’Hern}}, \bibinfo {author} {\bibfnamefont {L.~E.}\ \bibnamefont
  {Silbert}}, \bibinfo {author} {\bibfnamefont {A.~J.}\ \bibnamefont {Liu}}, \
  and\ \bibinfo {author} {\bibfnamefont {S.~R.}\ \bibnamefont {Nagel}},\
  }\href@noop {} {\bibfield  {journal} {\bibinfo  {journal} {Phys. Rev. E}\
  }\textbf {\bibinfo {volume} {68}},\ \bibinfo {pages} {011306} (\bibinfo
  {year} {2003})}\BibitemShut {NoStop}%
\bibitem [{\citenamefont {Aste}(2005)}]{Aste2005}%
  \BibitemOpen
  \bibfield  {author} {\bibinfo {author} {\bibfnamefont {T.}~\bibnamefont
  {Aste}},\ }\href {\doibase 10.1088/0953-8984/17/24/001} {\bibfield  {journal}
  {\bibinfo  {journal} {J. Phys.-Condens. Mat.}\ }\textbf {\bibinfo {volume}
  {17}},\ \bibinfo {pages} {S2361} (\bibinfo {year} {2005})}\BibitemShut
  {NoStop}%
\bibitem [{\citenamefont {Kurita}\ and\ \citenamefont
  {Weeks}(2011)}]{KuritaWeeks2011}%
  \BibitemOpen
  \bibfield  {author} {\bibinfo {author} {\bibfnamefont {R.}~\bibnamefont
  {Kurita}}\ and\ \bibinfo {author} {\bibfnamefont {E.~R.}\ \bibnamefont
  {Weeks}},\ }\href {\doibase 10.1103/PhysRevE.84.030401} {\bibfield  {journal}
  {\bibinfo  {journal} {Phys. Rev. E}\ }\textbf {\bibinfo {volume} {84}},\
  \bibinfo {pages} {030401} (\bibinfo {year} {2011})}\BibitemShut {NoStop}%
\bibitem [{\citenamefont {Berthier}\ \emph {et~al.}(2011)\citenamefont
  {Berthier}, \citenamefont {Chaudhuri}, \citenamefont {Coulais}, \citenamefont
  {Dauchot},\ and\ \citenamefont {Sollich}}]{Berthier_etal_2011}%
  \BibitemOpen
  \bibfield  {author} {\bibinfo {author} {\bibfnamefont {L.}~\bibnamefont
  {Berthier}}, \bibinfo {author} {\bibfnamefont {P.}~\bibnamefont {Chaudhuri}},
  \bibinfo {author} {\bibfnamefont {C.}~\bibnamefont {Coulais}}, \bibinfo
  {author} {\bibfnamefont {O.}~\bibnamefont {Dauchot}}, \ and\ \bibinfo
  {author} {\bibfnamefont {P.}~\bibnamefont {Sollich}},\ }\href {\doibase
  10.1103/PhysRevLett.106.120601} {\bibfield  {journal} {\bibinfo  {journal}
  {Phys. Rev. Lett.}\ }\textbf {\bibinfo {volume} {106}},\ \bibinfo {pages}
  {120601} (\bibinfo {year} {2011})}\BibitemShut {NoStop}%
\bibitem [{\citenamefont {Kapfer}\ \emph {et~al.}(2012)\citenamefont {Kapfer},
  \citenamefont {Mickel}, \citenamefont {Mecke},\ and\ \citenamefont
  {Schr\"oder-Turk}}]{KapferEtAl2012}%
  \BibitemOpen
  \bibfield  {author} {\bibinfo {author} {\bibfnamefont {S.~C.}\ \bibnamefont
  {Kapfer}}, \bibinfo {author} {\bibfnamefont {W.}~\bibnamefont {Mickel}},
  \bibinfo {author} {\bibfnamefont {K.}~\bibnamefont {Mecke}}, \ and\ \bibinfo
  {author} {\bibfnamefont {G.~E.}\ \bibnamefont {Schr\"oder-Turk}},\ }\href
  {\doibase 10.1103/PhysRevE.85.030301} {\bibfield  {journal} {\bibinfo
  {journal} {Phys. Rev. E}\ }\textbf {\bibinfo {volume} {85}},\ \bibinfo
  {pages} {030301} (\bibinfo {year} {2012})}\BibitemShut {NoStop}%
\bibitem [{\citenamefont {Atkinson}\ \emph {et~al.}(2013)\citenamefont
  {Atkinson}, \citenamefont {Stillinger},\ and\ \citenamefont
  {Torquato}}]{AtkinsonEtAl2013}%
  \BibitemOpen
  \bibfield  {author} {\bibinfo {author} {\bibfnamefont {S.}~\bibnamefont
  {Atkinson}}, \bibinfo {author} {\bibfnamefont {F.~H.}\ \bibnamefont
  {Stillinger}}, \ and\ \bibinfo {author} {\bibfnamefont {S.}~\bibnamefont
  {Torquato}},\ }\href {\doibase 10.1103/PhysRevE.88.062208} {\bibfield
  {journal} {\bibinfo  {journal} {Phys. Rev. E}\ }\textbf {\bibinfo {volume}
  {88}},\ \bibinfo {pages} {062208} (\bibinfo {year} {2013})}\BibitemShut
  {NoStop}%
\bibitem [{\citenamefont {Kansal}\ \emph {et~al.}(2002)\citenamefont {Kansal},
  \citenamefont {Torquato},\ and\ \citenamefont {Stillinger}}]{KansalEtAl2002}%
  \BibitemOpen
  \bibfield  {author} {\bibinfo {author} {\bibfnamefont {A.~R.}\ \bibnamefont
  {Kansal}}, \bibinfo {author} {\bibfnamefont {S.}~\bibnamefont {Torquato}}, \
  and\ \bibinfo {author} {\bibfnamefont {F.~H.}\ \bibnamefont {Stillinger}},\
  }\href {\doibase 10.1103/PhysRevE.66.041109} {\bibfield  {journal} {\bibinfo
  {journal} {Phys. Rev. E}\ }\textbf {\bibinfo {volume} {66}},\ \bibinfo
  {pages} {041109} (\bibinfo {year} {2002})}\BibitemShut {NoStop}%
\bibitem [{\citenamefont {Atkinson}\ \emph {et~al.}(2014)\citenamefont
  {Atkinson}, \citenamefont {Stillinger},\ and\ \citenamefont
  {Torquato}}]{AtkinsonEtAl2014}%
  \BibitemOpen
  \bibfield  {author} {\bibinfo {author} {\bibfnamefont {S.}~\bibnamefont
  {Atkinson}}, \bibinfo {author} {\bibfnamefont {F.~H.}\ \bibnamefont
  {Stillinger}}, \ and\ \bibinfo {author} {\bibfnamefont {S.}~\bibnamefont
  {Torquato}},\ }\href@noop {} {\bibfield  {journal} {\bibinfo  {journal}
  {Proc. Natl. Acad. Sci. USA}\ }\textbf {\bibinfo {volume} {111}},\ \bibinfo
  {pages} {18436} (\bibinfo {year} {2014})}\BibitemShut {NoStop}%
\bibitem [{\citenamefont {Torquato}\ and\ \citenamefont
  {Stillinger}(2003)}]{TorquatoStillinger2003}%
  \BibitemOpen
  \bibfield  {author} {\bibinfo {author} {\bibfnamefont {S.}~\bibnamefont
  {Torquato}}\ and\ \bibinfo {author} {\bibfnamefont {F.~H.}\ \bibnamefont
  {Stillinger}},\ }\href {\doibase 10.1103/PhysRevE.68.041113} {\bibfield
  {journal} {\bibinfo  {journal} {Phys. Rev. E}\ }\textbf {\bibinfo {volume}
  {68}},\ \bibinfo {pages} {041113} (\bibinfo {year} {2003})}\BibitemShut
  {NoStop}%
\bibitem [{\citenamefont {Zachary}\ and\ \citenamefont
  {Torquato}(2009)}]{ZacharyTorquato2009}%
  \BibitemOpen
  \bibfield  {author} {\bibinfo {author} {\bibfnamefont {C.~E.}\ \bibnamefont
  {Zachary}}\ and\ \bibinfo {author} {\bibfnamefont {S.}~\bibnamefont
  {Torquato}},\ }\href {http://stacks.iop.org/1742-5468/2009/i=12/a=P12015}
  {\bibfield  {journal} {\bibinfo  {journal} {J. Stat. Mech.}\ }\textbf
  {\bibinfo {volume} {2009}},\ \bibinfo {pages} {P12015} (\bibinfo {year}
  {2009})}\BibitemShut {NoStop}%
\bibitem [{\citenamefont {Donev}\ \emph {et~al.}(2005)\citenamefont {Donev},
  \citenamefont {Stillinger},\ and\ \citenamefont {Torquato}}]{DonevEtAl2005}%
  \BibitemOpen
  \bibfield  {author} {\bibinfo {author} {\bibfnamefont {A.}~\bibnamefont
  {Donev}}, \bibinfo {author} {\bibfnamefont {F.~H.}\ \bibnamefont
  {Stillinger}}, \ and\ \bibinfo {author} {\bibfnamefont {S.}~\bibnamefont
  {Torquato}},\ }\href {\doibase 10.1103/PhysRevLett.95.090604} {\bibfield
  {journal} {\bibinfo  {journal} {Phys. Rev. Lett.}\ }\textbf {\bibinfo
  {volume} {95}},\ \bibinfo {pages} {090604} (\bibinfo {year}
  {2005})}\BibitemShut {NoStop}%
\bibitem [{\citenamefont {Zachary}\ \emph
  {et~al.}(2011{\natexlab{a}})\citenamefont {Zachary}, \citenamefont {Jiao},\
  and\ \citenamefont {Torquato}}]{ZacharyJiaoTorquato2011}%
  \BibitemOpen
  \bibfield  {author} {\bibinfo {author} {\bibfnamefont {C.~E.}\ \bibnamefont
  {Zachary}}, \bibinfo {author} {\bibfnamefont {Y.}~\bibnamefont {Jiao}}, \
  and\ \bibinfo {author} {\bibfnamefont {S.}~\bibnamefont {Torquato}},\
  }\href@noop {} {\bibfield  {journal} {\bibinfo  {journal} {Phys. Rev. Lett.}\
  }\textbf {\bibinfo {volume} {106}},\ \bibinfo {pages} {178001} (\bibinfo
  {year} {2011}{\natexlab{a}})}\BibitemShut {NoStop}%
\bibitem [{\citenamefont {Hopkins}\ \emph {et~al.}(2012)\citenamefont
  {Hopkins}, \citenamefont {Stillinger},\ and\ \citenamefont
  {Torquato}}]{HopkinsStillingerTorquato2012}%
  \BibitemOpen
  \bibfield  {author} {\bibinfo {author} {\bibfnamefont {A.~B.}\ \bibnamefont
  {Hopkins}}, \bibinfo {author} {\bibfnamefont {F.~H.}\ \bibnamefont
  {Stillinger}}, \ and\ \bibinfo {author} {\bibfnamefont {S.}~\bibnamefont
  {Torquato}},\ }\href {\doibase 10.1103/PhysRevE.86.021505} {\bibfield
  {journal} {\bibinfo  {journal} {Phys. Rev. E}\ }\textbf {\bibinfo {volume}
  {86}},\ \bibinfo {pages} {021505} (\bibinfo {year} {2012})}\BibitemShut
  {NoStop}%
\bibitem [{\citenamefont {Klatt}\ and\ \citenamefont
  {Torquato}(2014)}]{KlattTorquato2014}%
  \BibitemOpen
  \bibfield  {author} {\bibinfo {author} {\bibfnamefont {M.~A.}\ \bibnamefont
  {Klatt}}\ and\ \bibinfo {author} {\bibfnamefont {S.}~\bibnamefont
  {Torquato}},\ }\href@noop {} {\bibfield  {journal} {\bibinfo  {journal}
  {Phys. Rev. E}\ }\textbf {\bibinfo {volume} {90}},\ \bibinfo {pages} {052120}
  (\bibinfo {year} {2014})}\BibitemShut {NoStop}%
\bibitem [{\citenamefont {Prager}(1961)}]{Prager1961}%
  \BibitemOpen
  \bibfield  {author} {\bibinfo {author} {\bibfnamefont {S.}~\bibnamefont
  {Prager}},\ }\href {\doibase http://dx.doi.org/10.1063/1.1706246} {\bibfield
  {journal} {\bibinfo  {journal} {{Phys. Fluids}}\ }\textbf {\bibinfo {volume}
  {4}},\ \bibinfo {pages} {1477} (\bibinfo {year} {1961})}\BibitemShut
  {NoStop}%
\bibitem [{\citenamefont {Rubinstein}\ and\ \citenamefont
  {Torquato}(1988)}]{RubinsteinTorquato1988}%
  \BibitemOpen
  \bibfield  {author} {\bibinfo {author} {\bibfnamefont {J.}~\bibnamefont
  {Rubinstein}}\ and\ \bibinfo {author} {\bibfnamefont {S.}~\bibnamefont
  {Torquato}},\ }\href {\doibase http://dx.doi.org/10.1063/1.454474} {\bibfield
   {journal} {\bibinfo  {journal} {J. Chem. Phys.}\ }\textbf {\bibinfo {volume}
  {88}},\ \bibinfo {pages} {6372} (\bibinfo {year} {1988})}\BibitemShut
  {NoStop}%
\bibitem [{\citenamefont {Rubinstein}\ and\ \citenamefont
  {Torquato}(1989)}]{RubinsteinTorquato1989JFM}%
  \BibitemOpen
  \bibfield  {author} {\bibinfo {author} {\bibfnamefont {J.}~\bibnamefont
  {Rubinstein}}\ and\ \bibinfo {author} {\bibfnamefont {S.}~\bibnamefont
  {Torquato}},\ }\href {\doibase 10.1017/S0022112089002211} {\bibfield
  {journal} {\bibinfo  {journal} {J. Fluid Mech.}\ }\textbf {\bibinfo {volume}
  {206}},\ \bibinfo {pages} {25} (\bibinfo {year} {1989})}\BibitemShut
  {NoStop}%
\bibitem [{\citenamefont {{Torquato, S. and Yeong, C. L.
  Y.}}(1997)}]{TorquatoYeong1997}%
  \BibitemOpen
  \bibfield  {author} {\bibinfo {author} {\bibnamefont {{Torquato, S. and
  Yeong, C. L. Y.}}},\ }\href {\doibase http://dx.doi.org/10.1063/1.473941}
  {\bibfield  {journal} {\bibinfo  {journal} {{J. Chem. Phys.}}\ }\textbf
  {\bibinfo {volume} {106}},\ \bibinfo {pages} {8814} (\bibinfo {year}
  {1997})}\BibitemShut {NoStop}%
\bibitem [{\citenamefont {Prager}(1963)}]{Prager1963}%
  \BibitemOpen
  \bibfield  {author} {\bibinfo {author} {\bibfnamefont {S.}~\bibnamefont
  {Prager}},\ }\href {\doibase http://dx.doi.org/10.1016/0009-2509(63)87003-7}
  {\bibfield  {journal} {\bibinfo  {journal} {Chem. Eng. Sci.}\ }\textbf
  {\bibinfo {volume} {18}},\ \bibinfo {pages} {227} (\bibinfo {year}
  {1963})}\BibitemShut {NoStop}%
\bibitem [{\citenamefont {Torquato}\ and\ \citenamefont
  {Avellaneda}(1991)}]{TorquatoAvellaneda1991}%
  \BibitemOpen
  \bibfield  {author} {\bibinfo {author} {\bibfnamefont {S.}~\bibnamefont
  {Torquato}}\ and\ \bibinfo {author} {\bibfnamefont {M.}~\bibnamefont
  {Avellaneda}},\ }\href {\doibase http://dx.doi.org/10.1063/1.461519}
  {\bibfield  {journal} {\bibinfo  {journal} {J. Chem. Phys.}\ }\textbf
  {\bibinfo {volume} {95}},\ \bibinfo {pages} {6477} (\bibinfo {year}
  {1991})}\BibitemShut {NoStop}%
\bibitem [{\citenamefont {Torquato}(2002)}]{Torquato2002}%
  \BibitemOpen
  \bibfield  {author} {\bibinfo {author} {\bibfnamefont {S.}~\bibnamefont
  {Torquato}},\ }\href@noop {} {\emph {\bibinfo {title} {Random Heterogeneous
  Materials: Microstructure and Macroscopic Properties}}},\ Interdisciplinary
  Applied Mathematics\ (\bibinfo  {publisher} {Springer},\ \bibinfo {year}
  {2002})\BibitemShut {NoStop}%
\bibitem [{\citenamefont {Beran}(1968)}]{Beran1968}%
  \BibitemOpen
  \bibfield  {author} {\bibinfo {author} {\bibfnamefont {M.~J.}\ \bibnamefont
  {Beran}},\ }\href {https://books.google.de/books?id=vgs\_AAAAIAAJ} {\emph
  {\bibinfo {title} {Statistical continuum theories}}}\ (\bibinfo  {publisher}
  {Interscience Publishers},\ \bibinfo {year} {1968})\BibitemShut {NoStop}%
\bibitem [{\citenamefont {Torquato}(1985)}]{Torquato1985}%
  \BibitemOpen
  \bibfield  {author} {\bibinfo {author} {\bibfnamefont {S.}~\bibnamefont
  {Torquato}},\ }\href {\doibase http://dx.doi.org/10.1063/1.335593} {\bibfield
   {journal} {\bibinfo  {journal} {J. Appl. Phys.}\ }\textbf {\bibinfo {volume}
  {58}},\ \bibinfo {pages} {3790} (\bibinfo {year} {1985})}\BibitemShut
  {NoStop}%
\bibitem [{\citenamefont {Sen}\ and\ \citenamefont
  {Torquato}(1989)}]{SenTorquato1989}%
  \BibitemOpen
  \bibfield  {author} {\bibinfo {author} {\bibfnamefont {A.~K.}\ \bibnamefont
  {Sen}}\ and\ \bibinfo {author} {\bibfnamefont {S.}~\bibnamefont {Torquato}},\
  }\href {\doibase 10.1103/PhysRevB.39.4504} {\bibfield  {journal} {\bibinfo
  {journal} {Phys. Rev. B}\ }\textbf {\bibinfo {volume} {39}},\ \bibinfo
  {pages} {4504} (\bibinfo {year} {1989})}\BibitemShut {NoStop}%
\bibitem [{\citenamefont {Rechtsman}\ and\ \citenamefont
  {Torquato}(2008)}]{RechtsmanTorquato2008}%
  \BibitemOpen
  \bibfield  {author} {\bibinfo {author} {\bibfnamefont {M.~C.}\ \bibnamefont
  {Rechtsman}}\ and\ \bibinfo {author} {\bibfnamefont {S.}~\bibnamefont
  {Torquato}},\ }\href@noop {} {\bibfield  {journal} {\bibinfo  {journal} {{J.
  Appl. Phys.}}\ }\textbf {\bibinfo {volume} {103}},\ \bibinfo {eid} {084901}
  (\bibinfo {year} {2008})}\BibitemShut {NoStop}%
\bibitem [{\citenamefont {Torquato}\ and\ \citenamefont
  {Rubinstein}(1989)}]{TorquatoRubinstein1989}%
  \BibitemOpen
  \bibfield  {author} {\bibinfo {author} {\bibfnamefont {S.}~\bibnamefont
  {Torquato}}\ and\ \bibinfo {author} {\bibfnamefont {J.}~\bibnamefont
  {Rubinstein}},\ }\href {\doibase http://dx.doi.org/10.1063/1.456655}
  {\bibfield  {journal} {\bibinfo  {journal} {J. Chem. Phys.}\ }\textbf
  {\bibinfo {volume} {90}},\ \bibinfo {pages} {1644} (\bibinfo {year}
  {1989})}\BibitemShut {NoStop}%
\bibitem [{\citenamefont {Doi}(1976)}]{Doi1976}%
  \BibitemOpen
  \bibfield  {author} {\bibinfo {author} {\bibfnamefont {M.}~\bibnamefont
  {Doi}},\ }\href {\doibase 10.1143/JPSJ.40.567} {\bibfield  {journal}
  {\bibinfo  {journal} {J. Phys. Soc. Jpn.}\ }\textbf {\bibinfo {volume}
  {40}},\ \bibinfo {pages} {567} (\bibinfo {year} {1976})}\BibitemShut
  {NoStop}%
\bibitem [{\citenamefont {Torquato}(1991)}]{Torquato1991Review}%
  \BibitemOpen
  \bibfield  {author} {\bibinfo {author} {\bibfnamefont {S.}~\bibnamefont
  {Torquato}},\ }\href {\doibase doi: 10.1115/1.3119494} {\bibfield  {journal}
  {\bibinfo  {journal} {Appl. Mech. Rev.}\ }\textbf {\bibinfo {volume} {44}},\
  \bibinfo {pages} {37} (\bibinfo {year} {1991})}\BibitemShut {NoStop}%
\bibitem [{\citenamefont {Torquato}\ and\ \citenamefont
  {Lado}(1986)}]{TorquatoLado1986}%
  \BibitemOpen
  \bibfield  {author} {\bibinfo {author} {\bibfnamefont {S.}~\bibnamefont
  {Torquato}}\ and\ \bibinfo {author} {\bibfnamefont {F.}~\bibnamefont
  {Lado}},\ }\href {\doibase 10.1103/PhysRevB.33.6428} {\bibfield  {journal}
  {\bibinfo  {journal} {Phys. Rev. B}\ }\textbf {\bibinfo {volume} {33}},\
  \bibinfo {pages} {6428} (\bibinfo {year} {1986})}\BibitemShut {NoStop}%
\bibitem [{Note1()}]{Note1}%
  \BibitemOpen
  \bibinfo {note} {The spectral density is the Fourier transform of the
  autocovariance function.}\BibitemShut {Stop}%
\bibitem [{\citenamefont {Dreyfus}\ \emph {et~al.}(2015)\citenamefont
  {Dreyfus}, \citenamefont {Xu}, \citenamefont {Still}, \citenamefont {Hough},
  \citenamefont {Yodh},\ and\ \citenamefont {Torquato}}]{DreyfusEtAl2015}%
  \BibitemOpen
  \bibfield  {author} {\bibinfo {author} {\bibfnamefont {R.}~\bibnamefont
  {Dreyfus}}, \bibinfo {author} {\bibfnamefont {Y.}~\bibnamefont {Xu}},
  \bibinfo {author} {\bibfnamefont {T.}~\bibnamefont {Still}}, \bibinfo
  {author} {\bibfnamefont {L.~A.}\ \bibnamefont {Hough}}, \bibinfo {author}
  {\bibfnamefont {A.~G.}\ \bibnamefont {Yodh}}, \ and\ \bibinfo {author}
  {\bibfnamefont {S.}~\bibnamefont {Torquato}},\ }\href {\doibase
  10.1103/PhysRevE.91.012302} {\bibfield  {journal} {\bibinfo  {journal} {Phys.
  Rev. E}\ }\textbf {\bibinfo {volume} {91}},\ \bibinfo {pages} {012302}
  (\bibinfo {year} {2015})}\BibitemShut {NoStop}%
\bibitem [{\citenamefont {Quintanilla}\ and\ \citenamefont
  {Torquato}(1997)}]{quintanilla_local_1997}%
  \BibitemOpen
  \bibfield  {author} {\bibinfo {author} {\bibfnamefont {J.}~\bibnamefont
  {Quintanilla}}\ and\ \bibinfo {author} {\bibfnamefont {S.}~\bibnamefont
  {Torquato}},\ }\href {\doibase 10.1063/1.473414} {\bibfield  {journal}
  {\bibinfo  {journal} {J. Chem. Phys.}\ }\textbf {\bibinfo {volume} {106}},\
  \bibinfo {pages} {2741} (\bibinfo {year} {1997})}\BibitemShut {NoStop}%
\bibitem [{\citenamefont {Debye}\ and\ \citenamefont
  {Bueche}(1949)}]{debye_scattering_1949}%
  \BibitemOpen
  \bibfield  {author} {\bibinfo {author} {\bibfnamefont {P.}~\bibnamefont
  {Debye}}\ and\ \bibinfo {author} {\bibfnamefont {A.~M.}\ \bibnamefont
  {Bueche}},\ }\href {\doibase 10.1063/1.1698419} {\bibfield  {journal}
  {\bibinfo  {journal} {J. Appl. Phys.}\ }\textbf {\bibinfo {volume} {20}},\
  \bibinfo {pages} {518} (\bibinfo {year} {1949})}\BibitemShut {NoStop}%
\bibitem [{\citenamefont {Torquato}\ and\ \citenamefont
  {Stell}(1985)}]{torquato_microstructure_1985}%
  \BibitemOpen
  \bibfield  {author} {\bibinfo {author} {\bibfnamefont {S.}~\bibnamefont
  {Torquato}}\ and\ \bibinfo {author} {\bibfnamefont {G.}~\bibnamefont
  {Stell}},\ }\href {\doibase 10.1063/1.448475} {\bibfield  {journal} {\bibinfo
   {journal} {J. Chem. Phys.}\ }\textbf {\bibinfo {volume} {82}},\ \bibinfo
  {pages} {980} (\bibinfo {year} {1985})}\BibitemShut {NoStop}%
\bibitem [{Note2()}]{Note2}%
  \BibitemOpen
  \bibinfo {note} {The pore-size probability density function $P(\delta )$ for
  pores in a phase formed by hard spheres with radius $R$ is only nonzero for
  $\delta < R$. There, it is always (independent of the arrangement of the
  spheres) given by $P(\delta )=3(R-\delta )^2/R^3$~\cite
  {Prager1963}.}\BibitemShut {Stop}%
\bibitem [{\citenamefont {Chiu}\ \emph {et~al.}(2013)\citenamefont {Chiu},
  \citenamefont {Stoyan}, \citenamefont {Kendall},\ and\ \citenamefont
  {Mecke}}]{ChiuEtAl2013}%
  \BibitemOpen
  \bibfield  {author} {\bibinfo {author} {\bibfnamefont {S.}~\bibnamefont
  {Chiu}}, \bibinfo {author} {\bibfnamefont {D.}~\bibnamefont {Stoyan}},
  \bibinfo {author} {\bibfnamefont {W.}~\bibnamefont {Kendall}}, \ and\
  \bibinfo {author} {\bibfnamefont {J.}~\bibnamefont {Mecke}},\ }\href@noop {}
  {\emph {\bibinfo {title} {Stochastic Geometry and Its Applications}}},\ Wiley
  Series in Probability and Statistics\ (\bibinfo  {publisher} {Wiley},\
  \bibinfo {year} {2013})\BibitemShut {NoStop}%
\bibitem [{\citenamefont {Hug}\ \emph {et~al.}(2002)\citenamefont {Hug},
  \citenamefont {Last},\ and\ \citenamefont {Weil}}]{hug_survey_2002}%
  \BibitemOpen
  \bibfield  {author} {\bibinfo {author} {\bibfnamefont {D.}~\bibnamefont
  {Hug}}, \bibinfo {author} {\bibfnamefont {G.}~\bibnamefont {Last}}, \ and\
  \bibinfo {author} {\bibfnamefont {W.}~\bibnamefont {Weil}},\ }in\ \href@noop
  {} {\emph {\bibinfo {booktitle} {Morphology of {Condensed} {Matter}:
  {Physics} and {Geometry} of {Spatially} {Complex} {Systems}}}},\ \bibinfo
  {editor} {edited by\ \bibinfo {editor} {\bibfnamefont {K.}~\bibnamefont
  {Mecke}}\ and\ \bibinfo {editor} {\bibfnamefont {D.}~\bibnamefont {Stoyan}}}\
  (\bibinfo  {publisher} {Springer},\ \bibinfo {address} {Berlin, Heidelberg},\
  \bibinfo {year} {2002})\ pp.\ \bibinfo {pages} {317--357}\BibitemShut
  {NoStop}%
\bibitem [{footnote-correlations()}]{footnote-correlations}%
  \BibitemOpen
  \bibinfo {note} {For generalizations to higher-order contact distributions
  and their relation to correlation functions; see Refs.~\cite{Torquato2002,
  ballani_second-order_2006, ballani_surface_2007,
  ballani_multiple-point_2011}.}\BibitemShut {Stop}%
\bibitem [{\citenamefont {Last}\ and\ \citenamefont
  {Penrose}()}]{LastPenrose16}%
  \BibitemOpen
  \bibfield  {author} {\bibinfo {author} {\bibfnamefont {G.}~\bibnamefont
  {Last}}\ and\ \bibinfo {author} {\bibfnamefont {M.~D.}\ \bibnamefont
  {Penrose}},\ }\href@noop {} {\emph {\bibinfo {title} {Lectures on the
  {P}oisson {P}rocess}}}\ (\bibinfo  {publisher} {to appear in Cambridge
  University Press, 2016})\ \bibinfo {note}
  {\url{http://www.math.kit.edu/stoch/~last/seite/lehrbuch_poissonp/de}}\BibitemShut
  {NoStop}%
\bibitem [{\citenamefont {Zachary}\ and\ \citenamefont
  {Torquato}(2011)}]{ZacharyTorquato2011}%
  \BibitemOpen
  \bibfield  {author} {\bibinfo {author} {\bibfnamefont {C.~E.}\ \bibnamefont
  {Zachary}}\ and\ \bibinfo {author} {\bibfnamefont {S.}~\bibnamefont
  {Torquato}},\ }\href {\doibase 10.1103/PhysRevE.84.056102} {\bibfield
  {journal} {\bibinfo  {journal} {Phys. Rev. E}\ }\textbf {\bibinfo {volume}
  {84}},\ \bibinfo {pages} {056102} (\bibinfo {year} {2011})}\BibitemShut
  {NoStop}%
\bibitem [{\citenamefont {Torquato}(1986)}]{Torquato1986}%
  \BibitemOpen
  \bibfield  {author} {\bibinfo {author} {\bibfnamefont {S.}~\bibnamefont
  {Torquato}},\ }\href {\doibase http://dx.doi.org/10.1063/1.451783} {\bibfield
   {journal} {\bibinfo  {journal} {J. Chem. Phys.}\ }\textbf {\bibinfo {volume}
  {85}},\ \bibinfo {pages} {4622} (\bibinfo {year} {1986})}\BibitemShut
  {NoStop}%
\bibitem [{\citenamefont {Verlet}\ and\ \citenamefont
  {Weis}(1972)}]{VerletWeis1972}%
  \BibitemOpen
  \bibfield  {author} {\bibinfo {author} {\bibfnamefont {L.}~\bibnamefont
  {Verlet}}\ and\ \bibinfo {author} {\bibfnamefont {J.-J.}\ \bibnamefont
  {Weis}},\ }\href {\doibase 10.1103/PhysRevA.5.939} {\bibfield  {journal}
  {\bibinfo  {journal} {Phys. Rev. A}\ }\textbf {\bibinfo {volume} {5}},\
  \bibinfo {pages} {939} (\bibinfo {year} {1972})}\BibitemShut {NoStop}%
\bibitem [{\citenamefont
  {Berryman}(1983{\natexlab{b}})}]{berryman_computing_1983}%
  \BibitemOpen
  \bibfield  {author} {\bibinfo {author} {\bibfnamefont {J.~G.}\ \bibnamefont
  {Berryman}},\ }\href {\doibase 10.1016/0021-9991(83)90021-9} {\bibfield
  {journal} {\bibinfo  {journal} {J. Comput. Phys.}\ }\textbf {\bibinfo
  {volume} {52}},\ \bibinfo {pages} {142} (\bibinfo {year}
  {1983}{\natexlab{b}})}\BibitemShut {NoStop}%
\bibitem [{\citenamefont {Debye}\ \emph {et~al.}(1957)\citenamefont {Debye},
  \citenamefont {Anderson},\ and\ \citenamefont
  {Brumberger}}]{debye_scattering_1957}%
  \BibitemOpen
  \bibfield  {author} {\bibinfo {author} {\bibfnamefont {P.}~\bibnamefont
  {Debye}}, \bibinfo {author} {\bibfnamefont {H.~R.}\ \bibnamefont {Anderson}},
  \ and\ \bibinfo {author} {\bibfnamefont {H.}~\bibnamefont {Brumberger}},\
  }\href {\doibase 10.1063/1.1722830} {\bibfield  {journal} {\bibinfo
  {journal} {J. Appl. Phys.}\ }\textbf {\bibinfo {volume} {28}},\ \bibinfo
  {pages} {679} (\bibinfo {year} {1957})}\BibitemShut {NoStop}%
\bibitem [{footnote-limit-two-point()}]{footnote-limit-two-point}%
  \BibitemOpen
  \bibinfo {note} {This relationship can also be generalized to anisotropic
  media~\cite{berryman_relationship_1987}. Moreover, the proportionality holds
  in any dimension $d$; see
  Ref.~\cite[][Eq.~(2.34)]{Torquato2002}.}\BibitemShut {Stop}%
\bibitem [{\citenamefont {Seaton}\ and\ \citenamefont
  {Glandt}(1986)}]{SeatonGlandt1986}%
  \BibitemOpen
  \bibfield  {author} {\bibinfo {author} {\bibfnamefont {N.~A.}\ \bibnamefont
  {Seaton}}\ and\ \bibinfo {author} {\bibfnamefont {E.~D.}\ \bibnamefont
  {Glandt}},\ }\href {\doibase http://dx.doi.org/10.1063/1.451667} {\bibfield
  {journal} {\bibinfo  {journal} {J. Chem. Phys.}\ }\textbf {\bibinfo {volume}
  {85}},\ \bibinfo {pages} {5262} (\bibinfo {year} {1986})}\BibitemShut
  {NoStop}%
\bibitem [{supplementary_materia()}]{supplementary_material}%
  \BibitemOpen
  \bibinfo {note} {See Supplemental Material at [URL will be inserted by
  publisher] for extended versions of
  Tabs.~\ref{tab:correlations-equi}--\ref{tab:spectral-density}.}\BibitemShut
  {Stop}%
\bibitem [{\citenamefont {Torquato}(2016{\natexlab{a}})}]{Torquato2016b}%
  \BibitemOpen
  \bibfield  {author} {\bibinfo {author} {\bibfnamefont {S.}~\bibnamefont
  {Torquato}},\ }\href {\doibase 10.1103/PhysRevE.94.022122} {\bibfield
  {journal} {\bibinfo  {journal} {Phys. Rev. E}\ }\textbf {\bibinfo {volume}
  {94}},\ \bibinfo {pages} {022122} (\bibinfo {year}
  {2016}{\natexlab{a}})}\BibitemShut {NoStop}%
\bibitem [{Note3()}]{Note3}%
  \BibitemOpen
  \bibinfo {note} {For three dimensions, $\protect \qopname \relax
  m{lim}_{k\rightarrow 0}\protect \mathaccentV {tilde}07E{m}(k) = \pi
  D^3/6$.}\BibitemShut {Stop}%
\bibitem [{\citenamefont {Zachary}\ \emph
  {et~al.}(2011{\natexlab{b}})\citenamefont {Zachary}, \citenamefont {Jiao},\
  and\ \citenamefont {Torquato}}]{zachary_hyperuniformity_2011}%
  \BibitemOpen
  \bibfield  {author} {\bibinfo {author} {\bibfnamefont {C.~E.}\ \bibnamefont
  {Zachary}}, \bibinfo {author} {\bibfnamefont {Y.}~\bibnamefont {Jiao}}, \
  and\ \bibinfo {author} {\bibfnamefont {S.}~\bibnamefont {Torquato}},\ }\href
  {\doibase 10.1103/PhysRevE.83.051308} {\bibfield  {journal} {\bibinfo
  {journal} {Phys. Rev. E}\ }\textbf {\bibinfo {volume} {83}},\ \bibinfo
  {pages} {051308} (\bibinfo {year} {2011}{\natexlab{b}})}\BibitemShut
  {NoStop}%
\bibitem [{Note4()}]{Note4}%
  \BibitemOpen
  \bibinfo {note} {Below this minimal value of $k$, the Fourier transform
  strongly depends on the cutoff.}\BibitemShut {Stop}%
\bibitem [{\citenamefont {Torquato}(2016{\natexlab{b}})}]{Torquato2016a}%
  \BibitemOpen
  \bibfield  {author} {\bibinfo {author} {\bibfnamefont {S.}~\bibnamefont
  {Torquato}},\ }\href {http://stacks.iop.org/0953-8984/28/i=41/a=414012}
  {\bibfield  {journal} {\bibinfo  {journal} {J. Phys. Condens. Mat.}\ }\textbf
  {\bibinfo {volume} {28}},\ \bibinfo {pages} {414012} (\bibinfo {year}
  {2016}{\natexlab{b}})}\BibitemShut {NoStop}%
\bibitem [{\citenamefont {Hansen}\ and\ \citenamefont
  {McDonald}(2006)}]{HansenMcDonald2006}%
  \BibitemOpen
  \bibfield  {author} {\bibinfo {author} {\bibfnamefont {J.~P.}\ \bibnamefont
  {Hansen}}\ and\ \bibinfo {author} {\bibfnamefont {I.~R.}\ \bibnamefont
  {McDonald}},\ }\href@noop {} {\emph {\bibinfo {title} {Theory of Simple
  Liquids}}}\ (\bibinfo  {publisher} {Elsevier Science},\ \bibinfo {year}
  {2006})\BibitemShut {NoStop}%
\bibitem [{\citenamefont {Atkinson}\ \emph {et~al.}(2016)\citenamefont
  {Atkinson}, \citenamefont {Zhang}, \citenamefont {Hopkins},\ and\
  \citenamefont {Torquato}}]{AtkinsonEtAl2016}%
  \BibitemOpen
  \bibfield  {author} {\bibinfo {author} {\bibfnamefont {S.}~\bibnamefont
  {Atkinson}}, \bibinfo {author} {\bibfnamefont {G.}~\bibnamefont {Zhang}},
  \bibinfo {author} {\bibfnamefont {A.~B.}\ \bibnamefont {Hopkins}}, \ and\
  \bibinfo {author} {\bibfnamefont {S.}~\bibnamefont {Torquato}},\ }\href@noop
  {} {\bibfield  {journal} {\bibinfo  {journal} {Phys. Rev. E}\ }\textbf
  {\bibinfo {volume} {94}},\ \bibinfo {pages} {012902} (\bibinfo {year}
  {2016})},\ \bibinfo {note} {this paper carefully studies the link between
  hyperuniformity and jamming in MRJ packings and the ``critical slowing down"
  associated with any numerical packing protocol on approach to
  jamming.}\BibitemShut {Stop}%
\bibitem [{\citenamefont {Torquato}(2010)}]{Torquato2010reformulation}%
  \BibitemOpen
  \bibfield  {author} {\bibinfo {author} {\bibfnamefont {S.}~\bibnamefont
  {Torquato}},\ }\href {\doibase 10.1103/PhysRevE.82.056109} {\bibfield
  {journal} {\bibinfo  {journal} {Phys. Rev. E}\ }\textbf {\bibinfo {volume}
  {82}},\ \bibinfo {pages} {056109} (\bibinfo {year} {2010})}\BibitemShut
  {NoStop}%
\bibitem [{\citenamefont {Torquato}(1995)}]{torquato_nearest-neighbor_1995}%
  \BibitemOpen
  \bibfield  {author} {\bibinfo {author} {\bibfnamefont {S.}~\bibnamefont
  {Torquato}},\ }\href
  {http://journals.aps.org/pre/abstract/10.1103/PhysRevE.51.3170} {\bibfield
  {journal} {\bibinfo  {journal} {Phys. Rev. E}\ }\textbf {\bibinfo {volume}
  {51}},\ \bibinfo {pages} {3170} (\bibinfo {year} {1995})}\BibitemShut
  {NoStop}%
\bibitem [{\citenamefont {Rintoul}\ and\ \citenamefont
  {Torquato}(1998)}]{rintoul_hard-sphere_1998}%
  \BibitemOpen
  \bibfield  {author} {\bibinfo {author} {\bibfnamefont {M.~D.}\ \bibnamefont
  {Rintoul}}\ and\ \bibinfo {author} {\bibfnamefont {S.}~\bibnamefont
  {Torquato}},\ }\href
  {http://journals.aps.org/pre/abstract/10.1103/PhysRevE.58.532} {\bibfield
  {journal} {\bibinfo  {journal} {Phys. Rev. E}\ }\textbf {\bibinfo {volume}
  {58}},\ \bibinfo {pages} {532} (\bibinfo {year} {1998})}\BibitemShut
  {NoStop}%
\bibitem [{\citenamefont {Torquato}\ \emph {et~al.}(1990)\citenamefont
  {Torquato}, \citenamefont {Lu},\ and\ \citenamefont
  {Rubinstein}}]{torquato_nearest-neighbor_1990}%
  \BibitemOpen
  \bibfield  {author} {\bibinfo {author} {\bibfnamefont {S.}~\bibnamefont
  {Torquato}}, \bibinfo {author} {\bibfnamefont {B.}~\bibnamefont {Lu}}, \ and\
  \bibinfo {author} {\bibfnamefont {J.}~\bibnamefont {Rubinstein}},\ }\href
  {http://journals.aps.org/pra/abstract/10.1103/PhysRevA.41.2059} {\bibfield
  {journal} {\bibinfo  {journal} {Phys. Rev. A}\ }\textbf {\bibinfo {volume}
  {41}},\ \bibinfo {pages} {2059} (\bibinfo {year} {1990})}\BibitemShut
  {NoStop}%
\bibitem [{\citenamefont {Torquato}\ and\ \citenamefont
  {Jiao}(2010)}]{TorquatoJiao2010}%
  \BibitemOpen
  \bibfield  {author} {\bibinfo {author} {\bibfnamefont {S.}~\bibnamefont
  {Torquato}}\ and\ \bibinfo {author} {\bibfnamefont {Y.}~\bibnamefont
  {Jiao}},\ }\href {\doibase 10.1103/PhysRevE.82.061302} {\bibfield  {journal}
  {\bibinfo  {journal} {Phys. Rev. E}\ }\textbf {\bibinfo {volume} {82}},\
  \bibinfo {pages} {061302} (\bibinfo {year} {2010})}\BibitemShut {NoStop}%
\bibitem [{\citenamefont {Du}\ \emph {et~al.}(1999)\citenamefont {Du},
  \citenamefont {Faber},\ and\ \citenamefont {Gunzburger}}]{DuEtAl1999}%
  \BibitemOpen
  \bibfield  {author} {\bibinfo {author} {\bibfnamefont {Q.}~\bibnamefont
  {Du}}, \bibinfo {author} {\bibfnamefont {V.}~\bibnamefont {Faber}}, \ and\
  \bibinfo {author} {\bibfnamefont {M.}~\bibnamefont {Gunzburger}},\ }\href
  {\doibase 10.1137/S0036144599352836} {\bibfield  {journal} {\bibinfo
  {journal} {SIAM Rev.}\ }\textbf {\bibinfo {volume} {41}},\ \bibinfo {pages}
  {637} (\bibinfo {year} {1999})}\BibitemShut {NoStop}%
\bibitem [{\citenamefont {Bose}(2002)}]{Bose2008}%
  \BibitemOpen
  \bibfield  {author} {\bibinfo {author} {\bibfnamefont {R.}~\bibnamefont
  {Bose}},\ }\href@noop {} {\emph {\bibinfo {title} {Information Theory, Coding
  and Cryptography}}}\ (\bibinfo  {publisher} {McGraw Hill, New Delhi},\
  \bibinfo {year} {2002})\BibitemShut {NoStop}%
\bibitem [{\citenamefont {Conway}\ and\ \citenamefont
  {Sloane}(1998)}]{Conway1998}%
  \BibitemOpen
  \bibfield  {author} {\bibinfo {author} {\bibfnamefont {J.~H.}\ \bibnamefont
  {Conway}}\ and\ \bibinfo {author} {\bibfnamefont {N.~J.~A.}\ \bibnamefont
  {Sloane}},\ }\href@noop {} {\emph {\bibinfo {title} {Sphere Packings,
  Lattices and Groups}}}\ (\bibinfo  {publisher} {Springer-Verlag, New York},\
  \bibinfo {year} {1998})\BibitemShut {NoStop}%
\bibitem [{Note5()}]{Note5}%
  \BibitemOpen
  \bibinfo {note} {The estimated probability density function is the empirical
  histogram weighted by the total number of samples and the bin width. In other
  words, the PDF is a relative frequency histogram weighted by the size of each
  bin.}\BibitemShut {Stop}%
\bibitem [{Note6()}]{Note6}%
  \BibitemOpen
  \bibinfo {note} {In a finite simulation box with periodic boundary
  conditions, the maximal, allowed radius is half of the minimum width of the
  box. Otherwise different representatives of the same sphere might be included
  in one test ball.}\BibitemShut {Stop}%
\bibitem [{\citenamefont {Florescu}\ \emph {et~al.}(2009)\citenamefont
  {Florescu}, \citenamefont {Torquato},\ and\ \citenamefont
  {Steinhardt}}]{florescu_designer_2009}%
  \BibitemOpen
  \bibfield  {author} {\bibinfo {author} {\bibfnamefont {M.}~\bibnamefont
  {Florescu}}, \bibinfo {author} {\bibfnamefont {S.}~\bibnamefont {Torquato}},
  \ and\ \bibinfo {author} {\bibfnamefont {P.~J.}\ \bibnamefont {Steinhardt}},\
  }\href {http://www.pnas.org/content/106/49/20658.short} {\bibfield  {journal}
  {\bibinfo  {journal} {Proc. Natl. Acad. Sci. USA}\ }\textbf {\bibinfo
  {volume} {106}},\ \bibinfo {pages} {20658} (\bibinfo {year}
  {2009})}\BibitemShut {NoStop}%
\bibitem [{\citenamefont {Charnes}\ \emph {et~al.}(1976)\citenamefont
  {Charnes}, \citenamefont {Frome},\ and\ \citenamefont
  {Yu}}]{charnes_equivalence_1976}%
  \BibitemOpen
  \bibfield  {author} {\bibinfo {author} {\bibfnamefont {A.}~\bibnamefont
  {Charnes}}, \bibinfo {author} {\bibfnamefont {E.~L.}\ \bibnamefont {Frome}},
  \ and\ \bibinfo {author} {\bibfnamefont {P.~L.}\ \bibnamefont {Yu}},\ }\href
  {\doibase 10.2307/2285762} {\bibfield  {journal} {\bibinfo  {journal} {J. Am.
  Stat. Assoc.}\ }\textbf {\bibinfo {volume} {71}},\ \bibinfo {pages} {169}
  (\bibinfo {year} {1976})}\BibitemShut {NoStop}%
\bibitem [{\citenamefont {Smith}\ and\ \citenamefont
  {Torquato}(1988)}]{SmithTorquato1988}%
  \BibitemOpen
  \bibfield  {author} {\bibinfo {author} {\bibfnamefont {P.}~\bibnamefont
  {Smith}}\ and\ \bibinfo {author} {\bibfnamefont {S.}~\bibnamefont
  {Torquato}},\ }\href {\doibase
  http://dx.doi.org/10.1016/0021-9991(88)90136-2} {\bibfield  {journal}
  {\bibinfo  {journal} {J. Comp. Phys.}\ }\textbf {\bibinfo {volume} {76}},\
  \bibinfo {pages} {176} (\bibinfo {year} {1988})}\BibitemShut {NoStop}%
\bibitem [{\citenamefont {Ballani}(2006)}]{ballani_second-order_2006}%
  \BibitemOpen
  \bibfield  {author} {\bibinfo {author} {\bibfnamefont {F.}~\bibnamefont
  {Ballani}},\ }\href {\doibase 10.1112/S0025579300000139} {\bibfield
  {journal} {\bibinfo  {journal} {Mathematika}\ }\textbf {\bibinfo {volume}
  {53}},\ \bibinfo {pages} {255} (\bibinfo {year} {2006})}\BibitemShut
  {NoStop}%
\bibitem [{\citenamefont {Ballani}(2007)}]{ballani_surface_2007}%
  \BibitemOpen
  \bibfield  {author} {\bibinfo {author} {\bibfnamefont {F.}~\bibnamefont
  {Ballani}},\ }\href {\doibase 10.1239/aap/1175266466} {\bibfield  {journal}
  {\bibinfo  {journal} {Adv. Appl. Probab.}\ }\textbf {\bibinfo {volume}
  {39}},\ \bibinfo {pages} {1} (\bibinfo {year} {2007})}\BibitemShut {NoStop}%
\bibitem [{\citenamefont {Ballani}(2011)}]{ballani_multiple-point_2011}%
  \BibitemOpen
  \bibfield  {author} {\bibinfo {author} {\bibfnamefont {F.}~\bibnamefont
  {Ballani}},\ }\href {\doibase 10.1002/mana.200810287} {\bibfield  {journal}
  {\bibinfo  {journal} {Math. Nachr.}\ }\textbf {\bibinfo {volume} {284}},\
  \bibinfo {pages} {938} (\bibinfo {year} {2011})}\BibitemShut {NoStop}%
\bibitem [{\citenamefont {Berryman}(1987)}]{berryman_relationship_1987}%
  \BibitemOpen
  \bibfield  {author} {\bibinfo {author} {\bibfnamefont {J.~G.}\ \bibnamefont
  {Berryman}},\ }\href {\doibase 10.1063/1.527804} {\bibfield  {journal}
  {\bibinfo  {journal} {J. Math. Phys.}\ }\textbf {\bibinfo {volume} {28}},\
  \bibinfo {pages} {244} (\bibinfo {year} {1987})}\BibitemShut {NoStop}%
\end{thebibliography}

%

\end{document}